\documentclass{article}
\usepackage[accepted]{icml2025}

\usepackage{microtype}
\usepackage{graphicx}
\usepackage{subfigure}
\usepackage{subcaption}
\usepackage{booktabs} 
\usepackage{algorithmic}
\usepackage{dcolumn}
\usepackage{bm}

\usepackage{hyperref}

\usepackage{amsmath}
\usepackage{amssymb}
\usepackage{mathtools}
\usepackage{amsthm}
\usepackage{algorithm}
\usepackage{tikz,colortbl}
\definecolor{track}{RGB}{245,214,155}
\definecolor{track_border}{RGB}{174,153,114}
\definecolor{shower_border}{RGB}{94,140,127}
\definecolor{shower}{RGB}{121,181,164}
\definecolor{delta}{RGB}{186,74,9}
\definecolor{delta_border}{RGB}{147,59,7}
\definecolor{michel_border}{RGB}{18,65,106}
\definecolor{michel}{RGB}{24,88,144}
\definecolor{led}{RGB}{238, 181, 255}
\definecolor{led_border}{RGB}{212, 161, 227}

\DeclareRobustCommand{\colorsquare}[1]{\tikz{\path[draw=#1_border,fill=#1, thick] (0,0) rectangle (5pt,5pt);}}

\newcommand{\best}[1]{\cellcolor{gray!15}{#1}}
\newcommand{\besto}[1]{\colorbox{gray!15}{{#1}}}

\input{commands}

\theoremstyle{plain}

\theoremstyle{definition}

\theoremstyle{remark}

\icmltitlerunning{Particle Trajectory Representation Learning}

\begin{document}

\twocolumn[
\icmltitle{Particle Trajectory Representation Learning with Masked Point Modeling}



\icmlsetsymbol{equal}{*}

\begin{icmlauthorlist}
\icmlauthor{Sam Young}{stanford,slac}
\icmlauthor{Yeon-jae Jwa}{slac}
\icmlauthor{Kazuhiro Terao}{slac}
\end{icmlauthorlist}

\icmlaffiliation{stanford}{Stanford University, Stanford, CA, USA}
\icmlaffiliation{slac}{SLAC National Accelerator Laboratory, Menlo Park, CA, USA}

\icmlcorrespondingauthor{Sam Young}{youngsam@stanford.edu}

\icmlkeywords{self-supervised learning, high energy physics, neutrino physics, 3D computer vision, point cloud learning, open dataset}

\vskip 0.3in
]



   



\printAffiliationsAndNotice{}  

\begin{abstract}
Effective self-supervised learning (SSL) techniques have been key to unlocking large datasets for representation learning. While many promising methods have been developed using online corpora and captioned photographs, their application to scientific domains, where data encodes highly specialized knowledge, remains a challenge. Liquid Argon Time Projection Chambers (LArTPCs) provide high-resolution 3D imaging for fundamental physics, but analysis of their sparse, complex point cloud data often relies on supervised methods trained on large simulations, introducing potential biases. We introduce the \textbf{Po}int-based \textbf{L}iquid \textbf{Ar}gon \textbf{M}asked \textbf{A}uto\textbf{e}ncoder (PoLAr-MAE), applying masked point modeling to unlabeled LArTPC images using domain-specific volumetric tokenization and energy prediction. We show this SSL approach learns physically meaningful trajectory representations directly from data. This yields remarkable data efficiency: fine-tuning on just 100 labeled events achieves track/shower semantic segmentation performance comparable to the state-of-the-art supervised baseline trained on $>$100,000 events. Furthermore, internal attention maps exhibit emergent instance segmentation of particle trajectories. While challenges remain, particularly for fine-grained features, we make concrete SSL's potential for building a foundation model for LArTPC image analysis capable of serving as a common base for all data reconstruction tasks. To facilitate further progress, we release PILArNet-M, a large dataset of 1M LArTPC events. Project site: \url{https://youngsm.com/polarmae}.
\end{abstract}


\begin{figure}[t]
    \centering
    \includegraphics[width=1\linewidth]{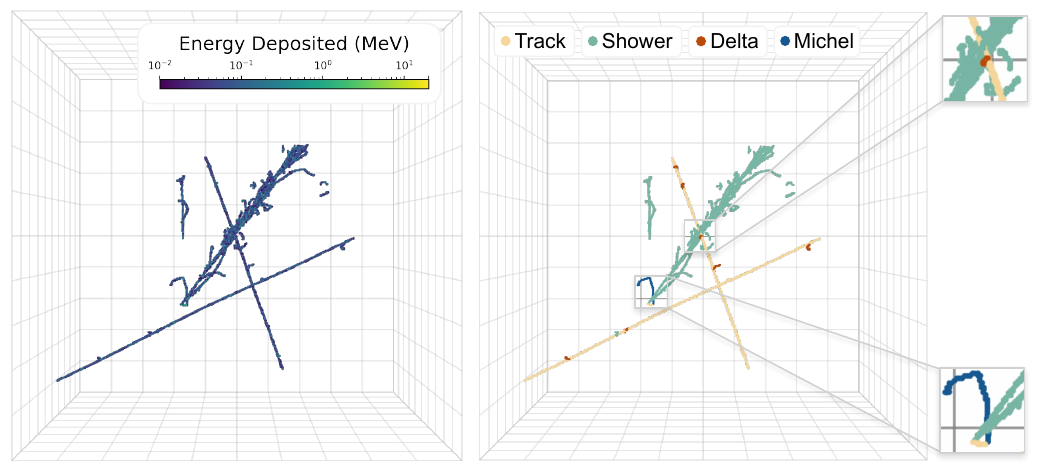}
    \vspace{-10pt}
    \caption{\textit{Illustration of liquid argon time projection chamber (LArTPC) data.} \textbf{(Left.)} High energy particles enter the detector medium, depositing energy along their trajectories. These depositions are measured by a complex particle imaging system. \textbf{(Right.)} A significant part of particle trajectory reconstruction is semantic segmentation, whereby each individual occupied voxel is classified as coming from a track-like particle~\colorsquare{track}, electromagnetic shower~\colorsquare{shower}, delta ray~\colorsquare{delta}, or Michel electron~\colorsquare{michel}. A single event can contain both overlapping trajectories and complex interactions in a small amount of space, as illustrated in the two zoomed in portions of the event.}
    \label{fig:example}
    \vspace{-5mm}
\end{figure}

\section{Introduction}

Liquid Argon Time Projection Chambers (LArTPCs) are a cornerstone technology in modern experimental neutrino physics, enabling detailed studies of neutrino oscillations, interactions, and searches for physics beyond the Standard Model \cite{Abi2020}. As high-resolution imaging detectors \cite{rubbia1977liquid}, LArTPCs capture the three-dimensional trajectories of charged particles produced in neutrino interactions within large volumes of liquid argon. When charged particles traverse the argon, they leave trails of ionization electrons. These electrons drift under a strong electric field and are collected on finely segmented anode planes, allowing for precise reconstruction of particle paths and energy depositions, often achieving millimeter resolution over meter-scale distances (visualized in Fig.~\ref{fig:example}). The resulting data, while globally sparse ($>99\%$ empty voxels), contains intricate ionization patterns corresponding to different particle types. These topologies are critical for event interpretation.

Minimum ionizing particles like muons, protons, and charged pions produce long, relatively straight \textbf{tracks} characterized by dense, linear ionization trails. In contrast, electrons, positrons, and photons initiate \textbf{electromagnetic showers}, which appear as diffuse, often conically expanding structures composed of many short, scattered track segments resulting from pair production and bremsstrahlung. Beyond these primary topologies, two important secondary structures provide additional physics information. \textbf{Michel electrons} arise from the decay of a stopping muon ($\mu \rightarrow e^-~\nu~\bar{\nu}$) and manifest as short (typically few cm), sometimes slightly curved electron tracks originating specifically from the Bragg peak of a muon track. \textbf{Delta rays} are energetic electrons knocked out of Ar atoms by a high-energy charged particle traversing the medium; they appear as short electron tracks branching off directly from the side of a primary particle track. Accurately reconstructing and identifying these distinct track, shower, Michel, and delta ray topologies from the complex LArTPC images is a critical step for achieving the physics goals of current and future neutrino experiments. An example event containing these complex topologies is shown in the right side of Fig.~\ref{fig:example}.

The current state-of-the-art approach for LArTPC data reconstruction leverages deep neural networks trained with supervised learning on large, detailed Monte Carlo simulated datasets, as demonstrated by frameworks such as \textsc{SPINE} \cite{drielsma2021scalableendtoenddeeplearningbaseddata}. This ``simulate $\rightarrow$ supervised train $\rightarrow$ calibrate" paradigm has proven highly effective and is central to ongoing analyses. However, like any approach, it involves certain trade-offs and presents aspects that motivate the exploration of complementary techniques. For instance, developing and validating these supervised models requires significant investment in simulation generation and careful calibration to mitigate potential discrepancies between simulation and real detector data (domain shift) \cite{Acciarri2017, Adams2019}. Models trained for one detector geometry or condition may not be directly transferable, and managing distinct networks for the multitude of reconstruction tasks (segmentation, clustering, particle ID, etc.) can be complex. The computational resources required for large-scale simulation also remain a consideration for next-generation experiments. These factors, alongside the desire for methods that might offer greater adaptability or uncover unexpected features in data, encourage investigation into alternative machine learning strategies.

Self-supervised learning (SSL) offers a promising path forward. By learning representations from the inherent structure of unlabeled data itself, SSL frameworks, often developed as foundation models \cite{bommasani2021opportunities}, have shown remarkable success in computer vision and other scientific fields \cite{Jumper2021, Hou2024, Irwin2022, Ross2022, lanusse2023astroclip, dillmann2024representation, doi:10.1126/science.adi2336, nguyen2023climax, kochkov2023neural, Trinh2024, xin2024deepseek, mccabe2023multiple, herde2024poseidonefficientfoundationmodels} for their robustness, transferability of representations, and out-of-distribution generalization. In this ``pre-train $\rightarrow$ fine-tune'' paradigm, a large model is initially \textit{pre-trained} on a vast corpora of unlabeled data using unsupervised or self-supervised learning to learn efficient representations of the data. Then, the model is \textit{fine-tuned} in a supervised manner to utilize these learned representations to very quickly adapt to any downstream task with a small amount of data. This trivial adaption to very few examples is often described as `few-shot learning' within ML literature.

SSL's application to the unique data characteristics and analysis workflows of LArTPCs remains largely unexplored. This paper presents the first study applying masked modeling directly to raw, unlabeled 3D LArTPC point cloud data. We investigate whether SSL can learn meaningful physical representations that enable highly data-efficient downstream performance.

To this end, we develop the \textbf{Po}int-based \textbf{L}iquid \textbf{Ar}gon \textbf{M}asked \textbf{A}uto\textbf{e}ncoder (PoLAr-MAE). Building upon the Point-MAE framework \cite{pang2022maskedautoencoderspointcloud} designed for generic point clouds, PoLAr-MAE incorporates modifications tailored to LArTPC data: a novel volumetric tokenization method to effectively group sparse ionization points into meaningful patches, and an auxiliary energy prediction task to focus the model on calorimetric information. The model is pretrained on a large dataset of unlabeled simulated LArTPC events using a masked autoencoding objective, i.e., learning to reconstruct masked portions of particle trajectories.

Our results demonstrate that PoLAr-MAE learns powerful representations that give rise to efficient transfer to pixel-level tasks. Fine-tuning PoLAr-MAE for semantic segmentation achieves performance on par with the fully supervised Sparse UResNet \cite{graham20183d, drielsma2021scalableendtoenddeeplearningbaseddata} baseline for identifying dominant track and shower topologies, while requiring drastically less labeled data -- matching the baseline trained on O(100,000) events using only O(100) labeled events for fine-tuning. This highlights the potential of SSL to dramatically reduce the need for large labeled datasets, a significant advantage in the context of complex detector simulations and potential domain shifts -- the so-called `sim2real` gap. Qualitative analysis reveals that specific attention heads in the transformer encoder learn to focus on individual particle instances, effectively performing emergent instance segmentation without explicit supervision. However, we also find that resolving fine-grained, sub-token phenomena like Michel electrons and delta rays remains challenging for the current architecture, indicating clear avenues for future work.

This work establishes the viability and data efficiency of SSL masked modeling for LArTPC data analysis, paving the way for more robust, scalable, and adaptable reconstruction workflows in current and future experiments like DUNE. To facilitate further research and development by the community in this emerging area, we release PILArNet-M, a large simulated LArTPC dataset, containing over one million events and 5.2 billion labeled energy depositions.

In sum, our main contributions are:
\begin{itemize}
    \item The first successful application and adaptation of self-supervised masked modeling for representation learning directly on sparse 3D point cloud data from LArTPCs.
    \item Demonstration that SSL-pretrained representations capture physically meaningful structures, evidenced by emergent instance segmentation in attention maps and extreme data efficiency for track/shower classification.
    \item Introduction and validation of C-NMS, a volumetric tokenization strategy tailored for sparse particle trajectory data, and demonstration of the benefit of an auxiliary energy prediction task.
    \item The public release of PILArNet-M, a large LArTPC simulation dataset, to serve as a benchmark and resource for future ML research in particle physics.
\end{itemize}

\section{Related Work}

\textbf{LArTPC Data Analysis and Reconstruction.} The reconstruction of particle interactions in LArTPCs is a complex task tackled by sophisticated software frameworks. The current state-of-the-art often involves algorithms based on deep neural networks trained in a supervised manner on large Monte Carlo simulations. A prominent example is the \textsc{SPINE} framework \cite{drielsma2021scalableendtoenddeeplearningbaseddata}, which employs a combination of sparse convolutional networks (SCNs) \cite{domine2020scalable, domine2021point} and graph neural networks (GNNs) \cite{drielsma2021clustering} in a multi-task cascade to perform dense segmentation, clustering, and particle identification. \textsc{SPINE}'s Sparse UResNet achieves high performance in semantic segmentation, particularly for track and shower classification (reporting 97.7\% and 99.5\% precision, respectively, when trained on over 125,000 simulated events), and serves as the primary supervised benchmark for evaluating the downstream task performance of our self-supervised model. A key motivation for our work is to investigate whether SSL can result in representations that are naturally separable and require very little fine-tuning to achieve comparable performance to the fully supervised approach.

\textbf{Self-Supervised Learning for Point Clouds.} Our work leverages recent advances in self-supervised learning adapted for 3D point cloud data. Many successful approaches build upon the Transformer architecture \cite{vaswani2017attention}, originally developed for natural language processing and later adapted for computer vision (Vision Transformer, ViT) \cite{dosovitskiy2020image} and point clouds \cite{guo2021pct, zhao2021point}. A powerful SSL paradigm is masked modeling, inspired by its success in language (Bidirectional Encoder Representations from Transformers, BERT) \cite{devlin2018bert} and vision (Masked Auto-encoder, MAE) \cite{he2021maskedautoencodersscalablevision}. This involves masking portions of the input data and training the model to predict the missing content. For point clouds, Point-MAE \cite{pang2022maskedautoencoderspointcloud} and Point-BERT \cite{yu2022pointbertpretraining3dpoint} demonstrated the effectiveness of this approach, typically by dividing the point cloud into local patches, masking a majority of them, and using a Transformer encoder-decoder to reconstruct the masked information. We base our PoLAr-MAE model on the Point-MAE architecture due to its conceptual simplicity and strong performance, adapting its patchification and reconstruction targets for the specific nature of LArTPC data, as detailed in Sec.~\ref{sec:method}. While contrastive learning \cite{chen2020simple, he2020momentum, wilkinson2025contrastivelearningrobustrepresentations} and self-distillation \cite{oquab2024dinov2learningrobustvisual,zeid2023point2vecselfsupervisedrepresentationlearning} methods represent additional major SSL directions, we focus here on masked modeling as a first exploration of SSL for raw LArTPC images.

\textbf{Representation Learning in High Energy Physics.} Self-supervised learning is gaining traction within HEP, although applications often differ significantly based on the physics context (e.g., collider vs. neutrino physics) and the data representation used. For example, \cite{golling2024maskedparticlemodelingsets} proposed masked modeling on sets of reconstructed particles (like jets), predicting properties of masked particles using features learned by a VQ-VAE. R3SL \cite{r3sl} employed contrastive learning where augmentations were generated by re-simulating physics processes, requiring access to simulation internals and ground truth. Other efforts focus on contrastive learning for particle tracking at colliders, often operating on graph representations of detector hits or using supervised objectives \cite{Lieret2024, Miao2024}. Specific to LArTPCs, recent work explored contrastive learning following SimCLR \cite{chen2020simple} aiming for representations robust to detector variations \cite{wilkinson2025contrastivelearningrobustrepresentations}. Interestingly, that study found that using realistic detector variations (like noise levels or wire efficiencies) as the sole source of augmentation provided insufficient signal for learning powerful representations, indicating that stronger augmentations or pretext tasks are likely necessary for LArTPC data. Geometric Deep Learning (GDL), particularly methods incorporating physical symmetries like SE(3)-equivariance \cite{pmlr-v139-satorras21a, fuchs2020se3Transformers3dPointClouds}, represents another important direction for analyzing 3D physics data, although these architectures can be more complex to implement and train. Our work distinguishes itself by applying unsupervised masked modeling -- a strong pretext task involving fine-grained reconstruction -- directly to the raw, high-resolution, sparse 3D energy depositions (point clouds) from LArTPCs, without relying on reconstructed physics objects, simulation interventions, or explicit geometric GDL architectures, aiming to learn fundamental representations of particle trajectories in this specific data modality.

\begin{figure*}[ht!!]
    \centering
    \includegraphics[width=1\linewidth]{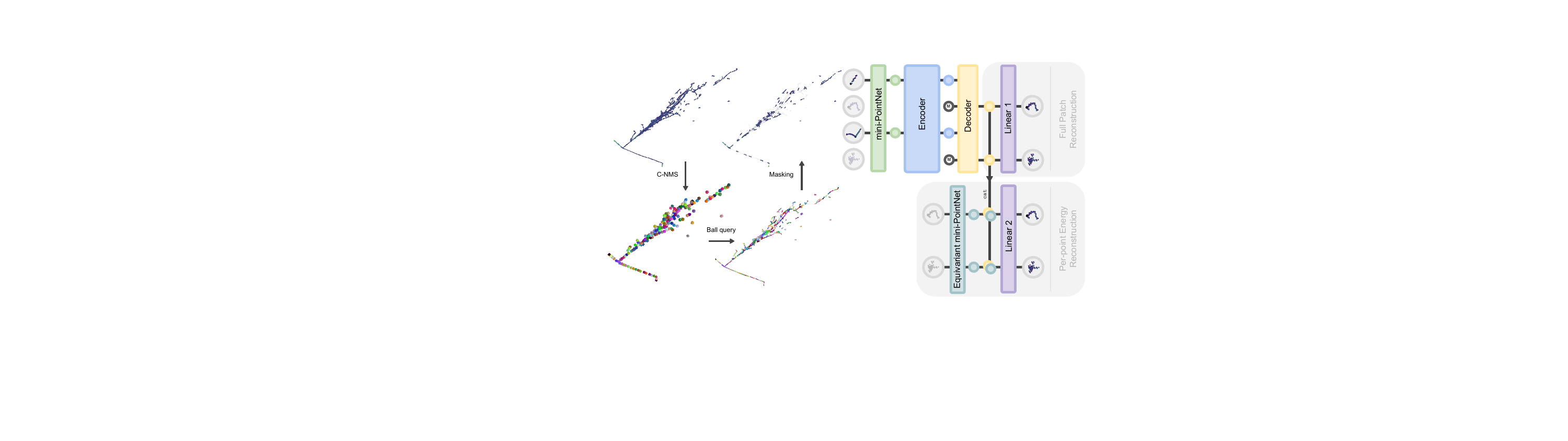}
\caption{\textit{PoLAr-MAE pre-training.} Input point clouds are partitioned into localized patches: seeds are selected via farthest point sampling (FPS), refined by centrality-based non-maximum suppression (C-NMS), and grouped via ball queries. A subset of patches is randomly masked, and visible patches are encoded into tokens using a mini-PointNet (right). The Transformer Encoder captures global relationships between tokens, while the Decoder predicts features of the missing patches using learned masked embeddings added onto learned positional encodings of the masked patch centers. Reconstructed patch coordinates are recovered via a linear projection. Additionally, an auxiliary task infills per-point energies by concatenating decoder outputs with Equivariant mini-PointNet (right) features from unmasked patches, leveraging their permutation-equivariant structure for point-wise regression.}  
\label{fig:arch}
\end{figure*}

\section{Dataset and Evaluation}
\label{sec:data}

In this section, we provide an overview of the simulated LArTPC dataset used for training and evaluation, PILArNet-M, and outline our evaluation strategy. This dataset represents an expansion of the earlier PILArNet dataset \cite{adams2020pilarnetpublicdatasetparticle} through an order-of-magnitude increase in the number of simulated events.

The dataset comprises 1,210,080 simulated LArTPC images generated using standard HEP simulation tools (details in \cite{adams2020pilarnetpublicdatasetparticle}). Each image represents particle interactions within a cubic detector volume of $(2.3~\text{m})^3$, discretized into $768^3$ voxels with a $3~\text{mm}$ pitch.

For each active voxel, the dataset provides the deposited energy (in MeV) along with metadata labels derived from the simulation truth, enabling detailed analysis:
\vspace{-2mm}
\begin{itemize}
    \item \textit{Fragment ID}: A unique identifier (per event) for the smallest contiguous clusters of energy depositions, useful for low-level clustering tasks.
    \item \textit{Group ID}: A unique identifier (per event) aggregating fragments belonging to the same physical particle trajectory (e.g., a single muon track), often corresponding to ``particles" in physics analyses.
    \item \textit{Interaction ID}: A unique identifier (per event) grouping particles originating from the same primary physics interaction vertex.
    \item \textit{Semantic Type}: An integer classifying the particle type responsible for the energy deposition. Categories include electromagnetic showers (ID: 0), minimum ionizing particle tracks (ID: 1), Michel electrons (ID: 2), delta rays (ID: 3), and low-energy depositions often considered background (ID: 4).
\end{itemize}

\noindent This comprehensive labeling supports evaluating various reconstruction tasks, including the semantic segmentation focused on in this work, as well as potential future studies on particle clustering or interaction vertexing. Further details on the dataset composition and statistics can be found in Appendix \ref{app:dataset}.

\subsection{Data Preprocessing} Before input to the network, several preprocessing steps are applied. First, voxels labeled as low-energy depositions (ID: 4) are removed, as these typically represent noise or diffuse background hits not associated with primary particle trajectories \footnote{These spurious deposits provide minimal physics information for the tasks considered here and can often be removed effectively using standard clustering techniques like DBSCAN \cite{10.5555/3001460.3001507}.} and considerably increase the memory usage of our network. We only consider events containing at least 1024 remaining points after this cut. The 3D coordinates ($\textbf{x}_i$) of the remaining points in each event are normalized to lie within a unit sphere centered at the origin. The energy depositions ($E_i$) are log-transformed to handle their large dynamic range and then scaled to the interval $[-1, 1]$:
\begin{equation}
\bar{E}_i = 2\left(\frac{\log_{10}(E_i+\epsilon) - \log_{10}(\epsilon)}{\log_{10}(E_\text{max}+\epsilon) - \log_{10}(\epsilon)}\right) - 1,
\end{equation}
where $\epsilon = 0.01$ MeV and $E_\text{max} = 20$ MeV are chosen based on the observed energy range in the dataset. This normalization ensures consistent input scaling for the network. During training, the only data augmentation applied is random 3D rotation of the entire event point cloud; other geometric augmentations like scaling or geometric jittering are avoided as they may correspond to non-physical transformations of the particle trajectories and energy depositions.


\subsection{Training and Validation Sets} The full dataset contains 1,033,377 events meeting the preprocessing criteria. This set is used for self-supervised pre-training. A held-out set of 10,471 events is reserved for validation and testing. For evaluating the quality of the pre-trained representations (linear probing), a small random subset of ~30 validation events is used. For evaluating the downstream semantic segmentation task after fine-tuning, we train on subsets of the main training set (with varying sizes, down to 100 events) and evaluate performance on the full 10,471-event held-out validation set.

\subsection{Evaluation Metrics}
\label{sec:eval}
We employ two stages of evaluation: one during pre-training, and one during fine-tuning. To assess the semantic information captured by the SSL representations without full fine-tuning, we follow standard practice \cite{pang2022maskedautoencoderspointcloud, he2021maskedautoencodersscalablevision} and train linear Support Vector Machines (SVMs) on the frozen representations (tokens, i.e.,  patches of points) produced by the pre-trained encoder. Since each token represents a group of points that may contain multiple semantic types, we train separate One-vs-Rest (OvR) classifiers to predict the presence of each semantic type within a token. Performance is measured using the $F_1$-score (harmonic mean of precision and recall: $F_1=2/(\text{recall}^{-1}+\text{precision}^{-1})$) for each class. Additionally, after fine-tuning a segmentation head on top of the pre-trained encoder (using varying amounts of labeled data), we evaluate point-wise semantic segmentation performance. We report standard classification metrics, including per-class $F_1$-scores and precision, allowing direct comparison with the fully supervised Sparse UResNet baseline. Results are presented in Sec.\ref{sec:results}, with further detailed results in Appendix~\ref{app:segsem}.

\section{Method}
\label{sec:method}

Figure \ref{fig:arch} outlines PoLAr-MAE's overall architecture and pretraining strategy, adapted from Point-MAE with an added energy reconstruction task. The LArTPC image is converted into a point cloud and processed through a modified grouping module. Some groups are masked, while visible ones are embedded into latent discrete tokens. A Vision Transformer (ViT) \cite{ranftl2021vision}-based autoencoder, with a heavy (i.e., many parameters) encoder and light (i.e., comparatively less parameters) decoder, reconstructs full patches. For energy reconstruction, a permutation-equivariant embedding module processes 3D point positions, and a linear head predicts per-point energy using decoded masked embeddings and encoded 3D positions.

\subsection{Tokenization of Particle Trajectories}
\label{sec:tok}

\subsubsection{Patch Grouping}

\begin{figure}[t]
    \centering
    \includegraphics[width=0.8\linewidth]{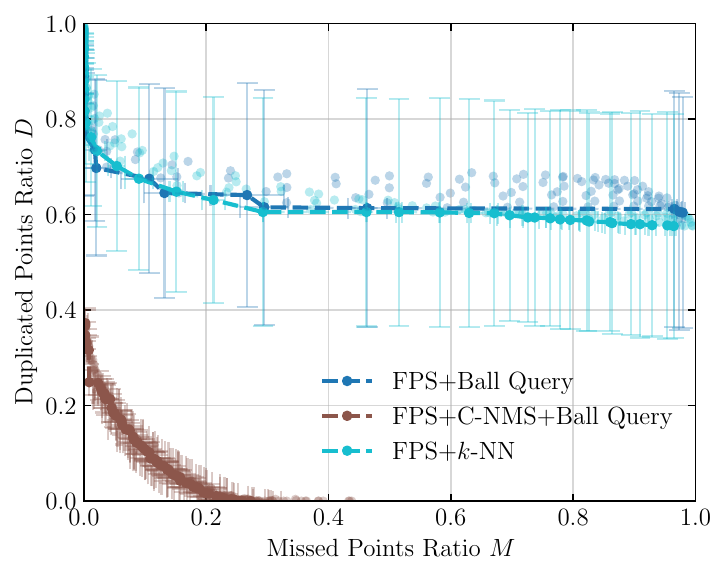}
    \caption{\textit{Pareto frontiers for each grouping method.} This plot evaluates three methods: FPS+$k$-NN, FPS+Ball Query, and FPS+C-NMS+Ball Query, where lower values on both axes indicate better performance. Points along each curve represent optimal parameter configurations for minimizing point duplication (y-axis) and missed coverage (x-axis), and are normalized to percentage scales (0.0 = 0\%, 1.0 = 100\%). Error bars represent 1 standard deviation across 32 LArTPC events. Further analysis of this plot is given in Appendix \ref{app:cnms}.}
\label{fig:pareto}
\vspace{-3mm}
\end{figure}

The common method for grouping point clouds into smaller patches involves farthest point sampling (FPS) -- an iterative algorithm that selects points that are farthest away from the previously selected points -- to sample group centers, followed by a $k$-nearest neighbors ($k$-NN) or ball query to find nearby neighbors for inclusion in the patch. However, in our dataset, where point density varies in regions containing signal, traditional FPS combined with $k$-NN or ball queries proves insufficient. Specifically, these approaches result in either an excessive number of ungrouped points or excessive overlap between local patches (see Figure \ref{fig:pareto}). 

To address this issue, we introduce a volumetric sampling method, \textbf{centrality-based non-maximum suppression (C-NMS)}, which extends the conventional NMS approach used in computer vision to operate on spherical regions. C-NMS enables efficient sampling of the minimal number of patches while maintaining low levels of ungrouped points and controlled overlap between patches. The method proceeds as follows:

\begin{enumerate}
\itemsep0em 
\item A large set of candidate group centers is sampled, e.g., via FPS or by taking the entire point cloud.
\item  Each group center is treated as the center of a sphere with a fixed radius $r$.
\item Using a greedy NMS algorithm, overlapping spheres are iteratively removed until no two remaining spheres exceed a predefined overlap factor $f$, defined as the percentage of the sphere diameter. \footnote{This is to say that $f=1$ means that sphere centers must be at least 1 entire diameter away from each other, i.e., with no overlap. A value of $f=0$ means that spheres are allowed to overlap completely.}
\item Finally, points for each group are sampled via a ball query using the same radius $r$.
\end{enumerate}

The resulting groups form a variable number $G$ of point groups, each containing a variable number $K$ of points. This approach approximates the effect of voxelizing the volume spanned by all points with a voxel pitch equal to the sphere diameter, while inherently reducing aliasing artifacts. Unlike FPS+$k$-NN, which lacks control over patch overlap, C-NMS ensures a tunable overlap between patches. In traditional FPS+\{$k$-NN, ball query\}, researchers often employ a fixed number of patches and a fixed number of points per patch. By contrast, C-NMS dynamically determines both the number of tokens $G$ and the number of points per token $K$ based on two parameters: the sphere radius $r$ and the overlap factor $f$. This method, which is highly parallelizable, has been implemented as a custom CUDA kernel for extremely fast processing. Additional commentary, further information, and ablation studies on this approach can be found in Appendix \ref{app:cnms}.

After grouping via C-NMS, each patch is normalized by subtracting the patch center:
$
\mathbf{x}_i \rightarrow \mathbf{x}_i - \mathbf{c},
$
where $\mathbf{x}_i$ is the original point position's first three dimensions, and $\mathbf{c}$ is the patch center, defined by the sphere centroid returned by C-NMS. Following PointNeXT \cite{qian2022pointnext}, we scale each normalized point by the sphere radius $r$, such that all patch points' positional coordinates lie within $[-1,1]^3$:
$
\mathbf{x}_i \rightarrow \frac{\mathbf{x}_i}{r}.
$ Note that we only apply these normalizations to the coordinates of each point, and not the energy value.

\subsubsection{Patch Encoding} 

After grouping, each patch group is encoded into a single latent vector using a permutation-invariant mini-PointNet \cite{qi2017pointnet}. The mini-PointNet is a small MLP network that first uplifts each individual point in each group to a higher latent dimension and then performs max-pooling along the channel dimension, so as to ``collapse" all individual point embeddings into a single latent vector. Since groups may contain less than $K$ points per patch, we pad each group with copies of its first point to avoid propagating padding artifacts. The mini-PointNet is invariant to input permutations, and as such is also invariant to this form of padding.

The patch embeddings from the mini-PointNet form the input sequence to a vanilla Transformer encoder \cite{vaswani2017attention}, which computes global contextual information across all patch tokens. To preserve spatial information lost during grouping, positional embeddings are added to the patch tokens. Each positional embedding \cite{vaswani2017attention} is learned via a three-layer MLP and corresponds to a mapping from the 3D coordinates of the patch center to some latent vector. We omit the final LayerNorm \cite{ba2016layernormalization} operation in the Transformer encoder, as it resulted in better representations.

\subsection{Patch-based masked event modeling}

To facilitate effective self-supervised pretraining for LArTPC data, we employ a masked autoencoder framework. A masking module selects a random subset of patch tokens, leaving the remaining unmasked tokens to be processed through the patch embedding module and Transformer encoder, producing contextually enriched embeddings. A shallow decoder then takes the encoded visible tokens along with a duplicated number of placeholder learnable mask tokens. Positional embeddings corresponding to the spatial position of the patch centers are added to both visible and masked token embeddings, enabling the decoder to reconstruct the latent representations of the masked tokens given spatial cues. This process is shown in Fig. \ref{fig:arch}.

The Point-MAE framework takes the decoded mask tokens and applies a simple linear layer to predict the original masked point groups. Formally, for each masked token, the decoder outputs $\textbf{x}_g \in \mathbb{R}^{K_\text{max} \times (3+1)}$ predicted points, where $K_\text{max}$ is the maximum possible number of points in any group. To handle variable group sizes, we take only the first $K$ points for each masked group, where $K$ corresponds to the number of points in the ground truth group. The reconstruction is evaluated using the Chamfer Distance ($d_{CD}$) loss between the predicted points $\{\mathbf{x}_g^{\text{pred}}\}$ and the original points $\{\mathbf{x}_g^{\text{true}}\}$ of the masked patches, which is defined as

\begin{equation}
d_{CD}(S_1, S_2) = \sum_{x\in S_1} \min_{y\in S_2} ||x-y||_2^2 + \sum_{y\in S_2}\min_{x\in S_1} ||x-y||^2_2. 
\end{equation}

\noindent This loss is the sum of the minimum distance between each point in one set $S_1$ with any point in another $S_2$ plus its permutation, and is permutation invariant to point ordering.

To further facilitate trajectory representation learning, we introduce an auxiliary energy reconstruction task, whereby the network learns to predict per-point energies given their individual 3D positions. This is a natural extension, as the energy deposition over the length of a given trajectory $(\frac{dE}{dx})$ is an important discriminator for particle identification (PID) in any LArTPC analysis. However, permutation-invariant PointNets struggle with per-point predictions in variable-sized groups due to unordered processing. We address this with an \textit{Equivariant Mini-PointNet} that imposes structured order via sinusoidal positional encodings \cite{vaswani2017attention} processed through a learnable MLP. Unlike the original, positional embeddings corresponding to the (arbitrary) tensor ordering of points within each patch are injected after the first shared MLP layer, intentionally breaking permutation invariance while retaining the aforementioned padding invariance. A final linear projection layer concatenates positional masked tokens with the predicted mask tokens from the decoder, predicting $K_\text{max}$ energy values per group via the $L_2$ loss. This enables precise per-point energy reconstruction while handling variable group sizes.

\subsection{Pre-training}

We train our model on the training set described in Sec.~\ref{sec:data} using an effective batch size of 512 across 4 NVIDIA A100 GPUs for 500,000 iterations, corresponding to about 60 epochs. We use a grouping radius of 5 voxels (15 mm) with an optimized C-NMS overlap fraction $f=0.73$, mask 60\% of input tokens randomly, and adopt ViT-S-like specifications \cite{ranftl2021vision} for the encoder/decoder. Groups are capped at 32 points, with farthest point sampling (FPS) \cite{qi2017pointnet++} applied when exceeding this limit. With our dataset, this results in about 400 groups per event (Figure \ref{fig:group_lens}). Full architectural details and secondary hyperparameters are provided in Appendix~\ref{app:hyperparameters}.

\begin{figure}
    \centering
    \includegraphics[width=0.9\linewidth]{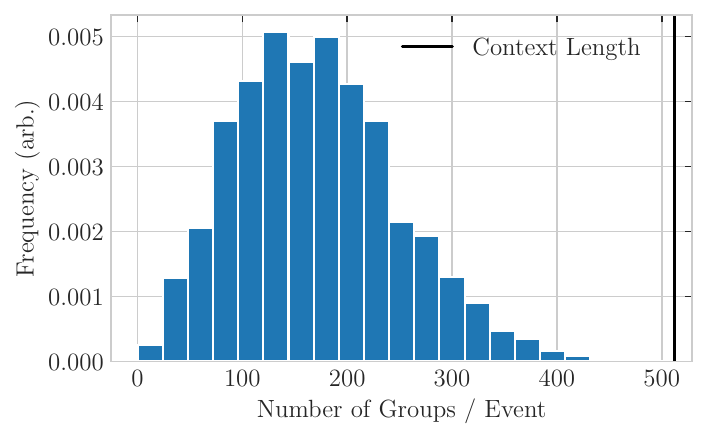}
    \vspace{-15pt}
    \caption{\textit{Per-event grouping statistics.} We set the context length of the model to be 512 groups, which leaves plenty of room.}
    \label{fig:group_lens}
\end{figure}

\section{Results}
\label{sec:results}

We evaluate the PoLAr-MAE framework using quantitative metrics to assess the quality of its pre-trained representations and its performance on downstream semantic segmentation. Additionally, we provide qualitative visualizations of the learned features and model attention maps.

\subsection{Quantitative Results}

\subsubsection{Linear Token Classification}

\setlength{\tabcolsep}{2pt}  

Table \ref{tab:svm} provides SVM F-scores (F\textsubscript{1}), defined as the harmonic mean of the precision and recall of the classifier (i.e., $F_1=\frac{2}{\text{recall}^{-1}+\text{precision}^{-1}})$ at the end of training for both the Point-MAE and PoLAr-MAE. We give metrics for Point-MAE from two training runs: one with 100,000 training steps ($\sim$2 epochs), and one with 500,000 training steps ($\sim$60 epochs). These metrics make clear that the model definitively learns the semantic meaning of different types of trajectories without explicitly providing particle ID information. By just pre-training, the SVM can be taught to classify tokens as containing tracks \colorsquare{track} and showers \colorsquare{shower} with F-scores of 99.4\% and 97.7\%, respectively. Particle types that often span one to two tokens are much less discernible, and as such are harder, but not impossible, to discern, having F-scores of 51.8\% for Michels \colorsquare{michel} and 44.0\% for delta rays \colorsquare{delta}. Interestingly, the model's representations of patches contain tracks and showers become easily linearly separable just within around 10 epochs of training.

\begin{table}[t]
\centering
\caption{\textit{SVM Validation Results.} We present final SVM validation F\textsubscript{1} scores for each class—where F\textsubscript{1} is the harmonic mean of precision and recall—and an overall mean across classes. \textbf{Bold} entries highlight the best results.}
\vskip 0.15in
\label{tab:svm}
\begin{tabular}{llcccccc}
\toprule
& & & \multicolumn{4}{c}{Per-class F\textsubscript{1}}\\
\cmidrule(lr){4-7}
Method & Steps & F\textsubscript{1} & Track & Shower & Delta & Michel \\
 &  &  & \colorsquare{track} & \colorsquare{shower} & \colorsquare{delta} & \colorsquare{michel} \\
\midrule
{Point-MAE}  & 100k & 0.680 & 0.990 & 0.970 & 0.436 & 0.325 \\
{Point-MAE}  & 500k & 0.719 & 0.993 & 0.967 & \best{0.572} & 0.342 \\
{PoLAr-MAE}    & 500k & \best{0.732} & \best{0.994} & \best{0.977} & 0.518 &\best{ 0.440} \\
\bottomrule
\end{tabular}
\end{table}

\subsubsection{Semantic Segmentation}

\begin{figure}[t!]
    \centering
    \includegraphics[width=1\linewidth]{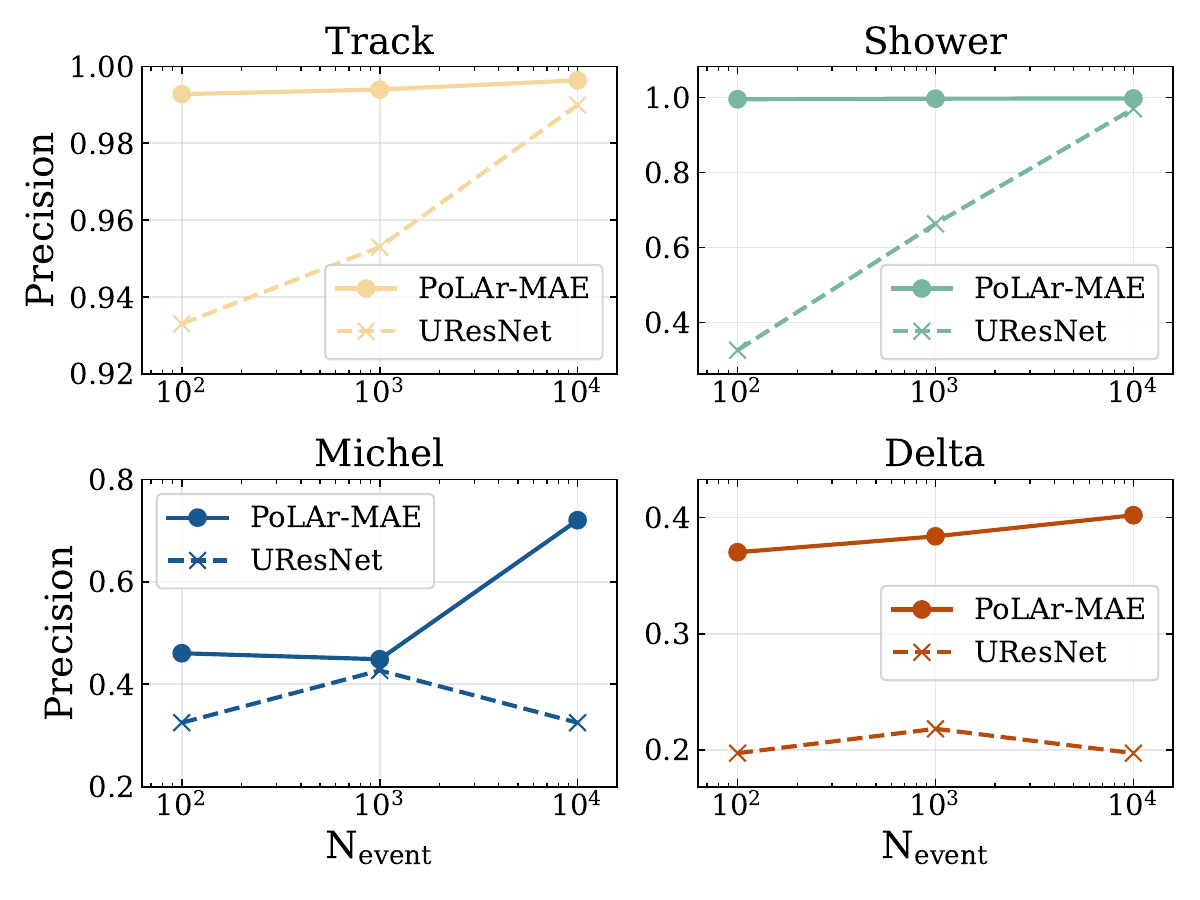}
    \caption{\textit{Data efficiency of the pre-trained model.} We fine-tune PoLAr-MAE on small labeled datasets consisting of 100, 1,000, and 10,000 events and compare precision results to the supervised UResNet baseline when trained on the same number of events. Despite being fine-tuned on just 100 events, PoLAr-MAE reaches $>0.99\%$ precision for track/shower disambiguation, and outperforms the supervised approach in all three dataset sizes.}
    \label{fig:spine_comparison_quad}
\end{figure}

\setlength{\tabcolsep}{2pt}  
\begin{figure}[t!]
    \centering
    \begin{tabular}{c@{\hspace{2mm}}*{3}{c}@{}} 
    & \textsc{Event 1} & \textsc{Event 2} & \textsc{Event 3} \\
    \rotatebox[origin=l]{90}{\textsc{\small ~~~~~Truth}} & 
    \includegraphics[width=0.3\linewidth]{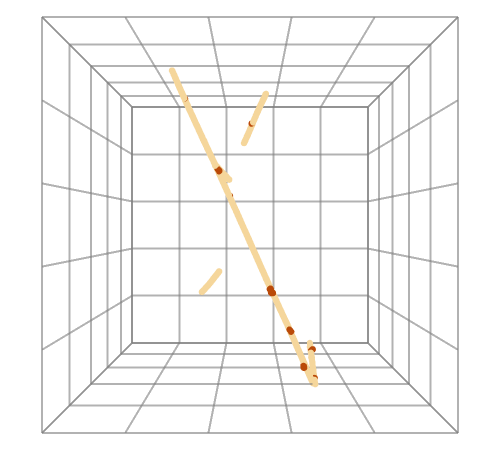}&
    \includegraphics[width=0.3\linewidth]{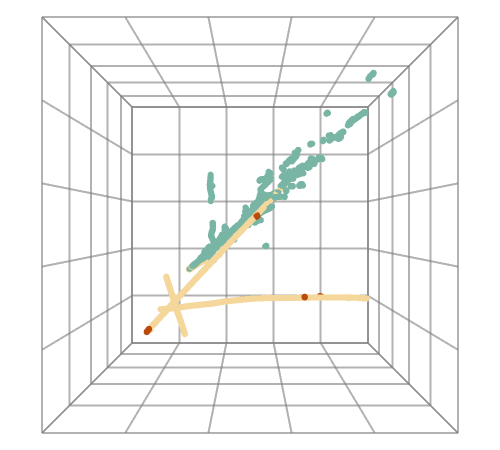} &
    \includegraphics[width=0.3\linewidth]{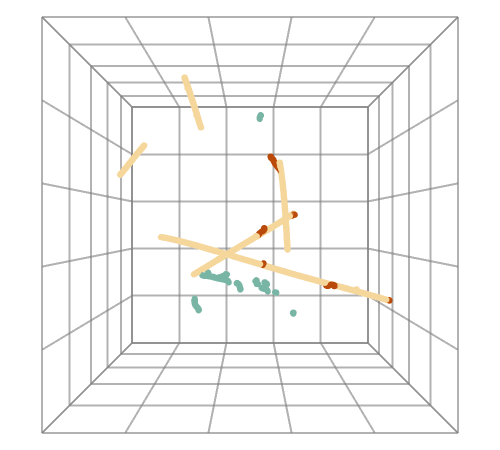} \\
    
    \rotatebox[origin=l]{90}{\textsc{\small ~~~~~PEFT}} & 
    \includegraphics[width=0.3\linewidth]{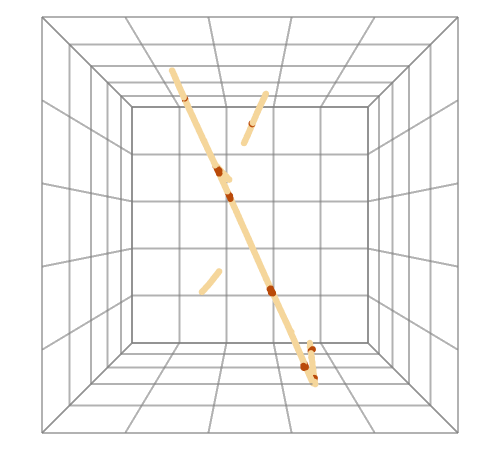} &
    \includegraphics[width=0.3\linewidth]{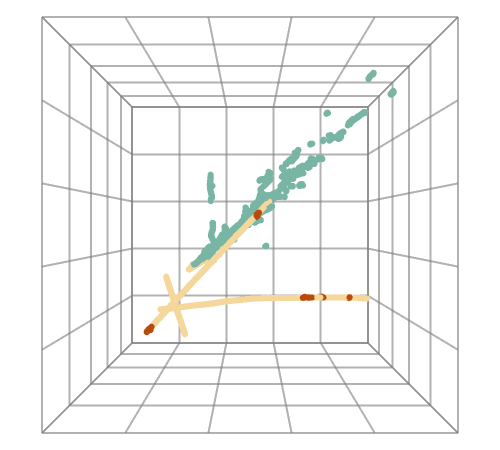} &
    \includegraphics[width=0.3\linewidth]{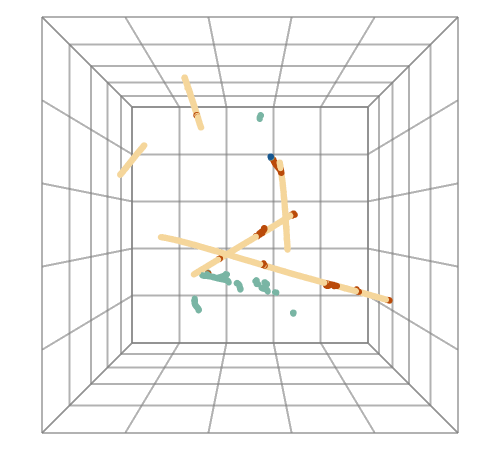} \\

    \rotatebox[origin=l]{90}{\textsc{\small ~~~~~~FFT}} & 
    \includegraphics[width=0.3\linewidth]{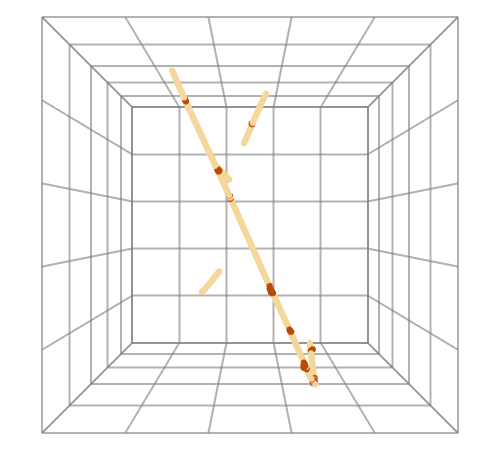} &
    \includegraphics[width=0.3\linewidth]{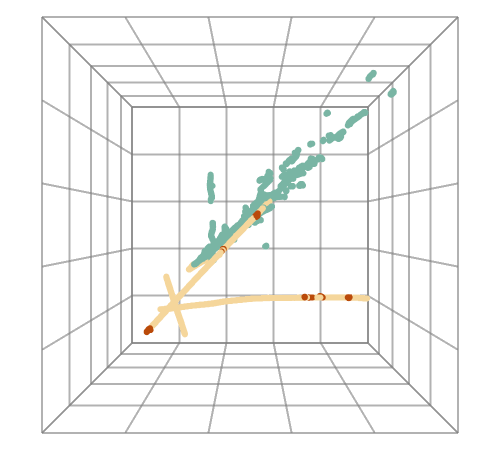} &
    \includegraphics[width=0.3\linewidth]{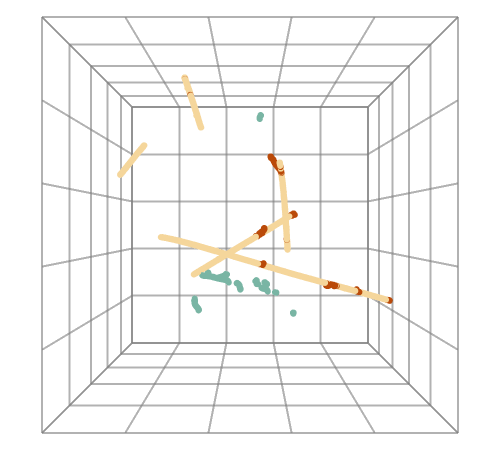} \\
    \end{tabular}
    \makebox[\linewidth][r]{%
        \includegraphics[width=0.5\linewidth]{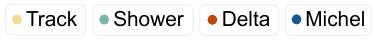}}
    \caption{\textit{Visualization of finetuned PoLAr-MAE models on the semantic segmentation task.} Classes are one of \{track \colorsquare{track}, shower \colorsquare{shower}, delta ray \colorsquare{delta}, Michel \colorsquare{michel}\}. Strong performance for the PoLAr-MAE models trained with parameter-efficient fine-tuning (PEFT) methods and via full fine-tuning (FFT) are shown. Most of the error results from incorrectly classifying spatially small semantic types, e.g., delta rays alongside tracks. The finetuned Point-MAE predictions are visually similar to that of the PoLAr-MAE models, and are not pictured. Additional examples (with predictions from the Point-MAE models) can be found in Appendix \ref{app:segsem}. Best viewed zoomed in.}
    \label{fig:segsem_examples}
    \vspace{-5mm}
\end{figure}

To evaluate the effectiveness of the representations learned by the PoLAr-MAE encoder, we fine-tune the model to perform dense classification of individual voxels as one of the four particle types: tracks, showers, delta rays, and Michel.  We train simple feature upsampling and classification heads to predict per-point classes, following PointNet++ \cite{qi2017pointnet++} and Point-MAE \cite{pang2022maskedautoencoderspointcloud}. We fine-tune both PoLAr-MAE and Point-MAE pretrained models on 10,000, 1,000, and 100 events, which are 10x, 100x, and 1000x less than the number of events used to train the supervised baseline UResNet \cite{drielsma2021scalableendtoenddeeplearningbaseddata}. We opt to use the Focal loss \cite{lin2018focallossdenseobject}, which is defined as

\begin{equation}
\text{FL}(p_t) = -\alpha_t(1-p_t)^\gamma \log{p_t},
\end{equation}

\noindent where $p_t\in [0,1]$ is the model's estimated probability for the ground truth class, $\alpha_t$ is the inverse frequency of the true label within the entire dataset, and $\gamma=2$. The Focal loss, an extension of standard cross entropy loss, focuses the relative loss towards harder (low confidence) examples by downweighting the loss if the model is very confident in its prediction. We apply weights $\alpha_t$ to compensate for the large class imbalance between tracks/showers and Michels/deltas.

We perform full fine-tuning (optimizing all parameters of the model) as well as parameter-efficient fine-tuning (PEFT), i.e., freezing all components of the model except the relatively tiny feature upscaling and segmentation heads, to evaluate how well the representations the encoder learns actually are.

\subsubsection{Data efficient fine-tuning}

To further evaluate the effectiveness of the representations learned by the PoLAr-MAE encoder, we evaluate the model when fine-tuned for semantic segmentation on an extremely small number of labeled events -- 100 and 1000, and 10,000. To demonstrate the power of initializing the segmentation model with pre-trained weights, we compare segmentation results to SPINE's UResNet baseline trained from scratch on the same amounts of data. As shown in Figure \ref{fig:spine_comparison_quad}, PoLAr-MAE achieves a precision of 0.995 for showers and 0.993 for tracks after being fine-tuned on just 100 LArTPC events. This compares favorably to the fully supervised UResNet, which when trained on 100 events yields a precision of 0.326 for showers and 0.933 for tracks. Despite comparatively lower precision scores on delta ray and Michel electron classification overall, PoLAr-MAE generally outperforms or matches the fully supervised UResNet across these categories at all three smaller data sizes, notably for Michels with 10,000 events.

\begin{table}[t]
\centering
\caption{\textit{Semantic segmentation results.} We present F\textsubscript{1} scores for semantic segmentation of each class within the validation set -- where F\textsubscript{1} is the harmonic mean of precision and recall -- and an overall mean across classes. The best results are \besto{highlighted}.}
\vskip 0.15in
\label{tab:segsem}
\begin{tabular}{llcccc}
\toprule
& & \multicolumn{4}{c}{Per-class F\textsubscript{1}}\\
\cmidrule(lr){3-6}
Model & F\textsubscript{1} & Track & Shower & Delta & Michel \\
&  & \colorsquare{track} & \colorsquare{shower} & \colorsquare{delta} & \colorsquare{michel} \\
\midrule
\textsc{Point-MAE PEFT}  & 0.772 & \best{0.965} & 0.983 & \best{0.569} & 0.572 \\
\textsc{PoLAr-MAE PEFT}   & 0.798 & 0.961 & 0.990  & 0.542 & 0.698 \\
\midrule
\textsc{Point-MAE FFT} & 0.831 & 0.963 & \best{0.994} & 0.561 & 0.807 \\
\textsc{PoLAr-MAE FFT} & \best{0.837} & 0.964 & \best{0.994} & \best{0.569} & \best{0.823} \\
\bottomrule
\end{tabular}
\end{table}

\begin{figure}[t]
    \centering
    \includegraphics[width=0.9\linewidth]{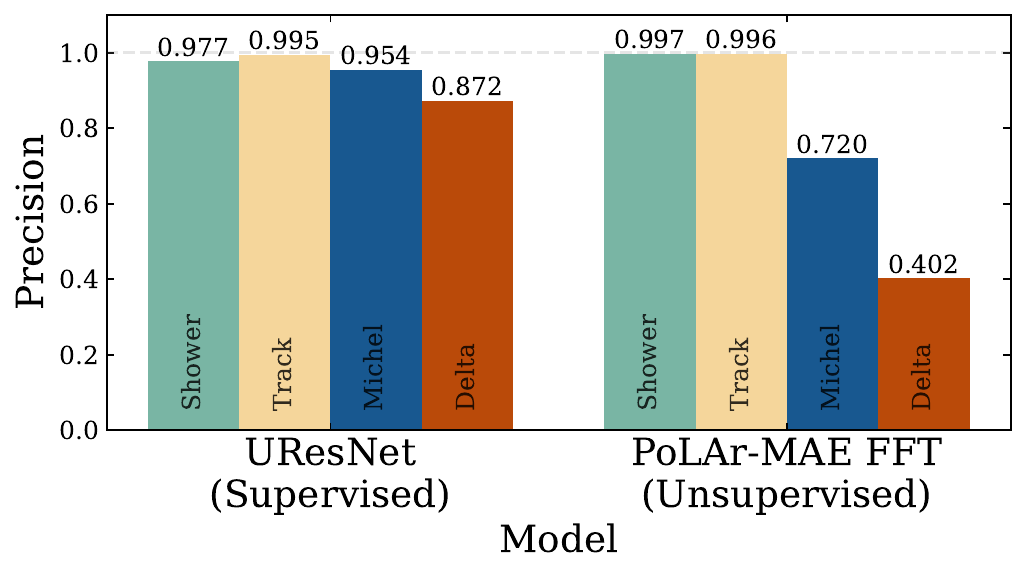}
    \vspace{-10pt}
    \caption{\textit{Comparison to the fully supervised approach.} We compare PoLAr-MAE fine-tuned on 10,000 events to reported average classification precision values of SPINE trained for 100,000 events \cite{drielsma2021scalableendtoenddeeplearningbaseddata}. Notably, our fully fine-tuned PoLAr-MAE model outperforms the Sparse UResNet-based SPINE on track and shower classification while training on 10x less labeled data.}
    \label{fig:spine_comparison}
\end{figure}

Table \ref{tab:segsem} shows results on semantic segmentation across the two pretraining strategies and the two fine-tuning strategies using F\textsubscript{1} scores as evaluation metrics. Overall, our fine-tuned models perform exceptionally well, especially given these models were fine-tuned with just 10,000 events. PoLAr-MAE overall outperforms or is comparable to Point-MAE in the FFT models, however there is a clear trade-off in Michel/delta classification with the PEFT models. Qualitative examples of semantic segmentation predictions are shown in Fig. \ref{fig:segsem_examples}. A direct comparison of precision values reported in \cite{drielsma2021scalableendtoenddeeplearningbaseddata} are shown in Figure \ref{fig:spine_comparison}. Critically, our model outperforms the fully supervised approach for shower and track classification by 2\% and 0.1\%, respectively, while performing fairly worse for Michel and delta-ray classification. A more detailed comparison between models can be found in Appendix \ref{app:segsem}.

\subsection{Qualitative Results}

\subsubsection{PCA of Patch Features}

To get a qualitative understanding of what these representations of trajectories work out, we plot out the center of each token in an event and color it according to a casting of its latent vectors to RGB space using principal component analysis (PCA). To ensure that the colors more directly relate to the underlying semantic meaning of individual tokens, we regress out any spatial bias by training a simple affine model to predict embeddings from their patch centers. A full explanation of this process is described in Appendix \ref{app:spatial_debiasing}.

Shown in Figure \ref{fig:latent_comparison}, the pretrained models clearly shows an understanding of semantics between tracks, showers, deltas, and Michels. A comparison to the latent representations of a model initialized with random weights (also de-biased) is shown for comparison. Indeed, even trajectories that cross very close to one another, despite being very spatially close, have different representations. Tokens along a single particle trajectory are shown to have similar representations. Particle trajectories that all come from the same vertex also differ in representations. Additional examples can be found in the appendix.

\begin{figure}
    \begin{tabular}{c@{\hspace{2mm}}*{3}{c}@{}} 
    & \textsc{Event 1} & \textsc{Event 2} & \textsc{Event 3} \\
    \rotatebox[origin=l]{90}{\textsc{~~~~Random}} & 
    \includegraphics[width=0.3\linewidth]{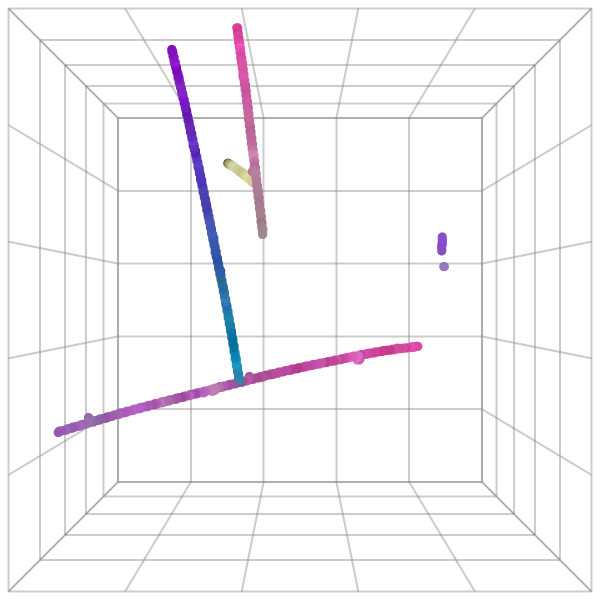 }&
    \includegraphics[width=0.3\linewidth]{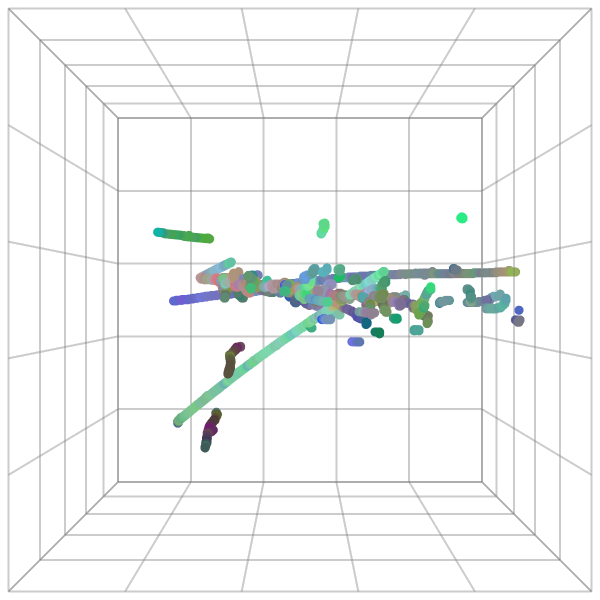} &
    \includegraphics[width=0.3\linewidth]{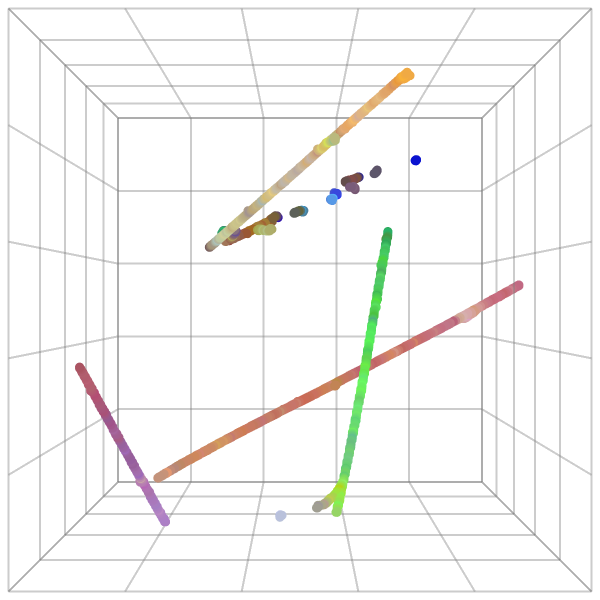} \\
    
    
    \rotatebox[origin=l]{90}{\textsc{~PoLAr-MAE}} & 
    \includegraphics[width=0.3\linewidth]{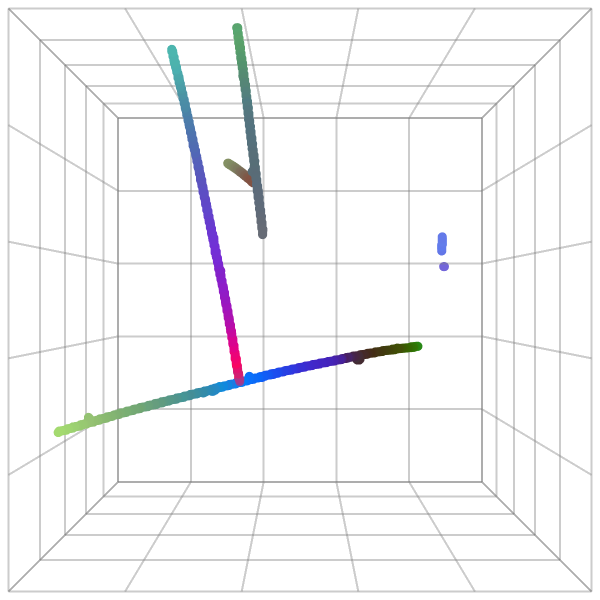} &
    \includegraphics[width=0.3\linewidth]{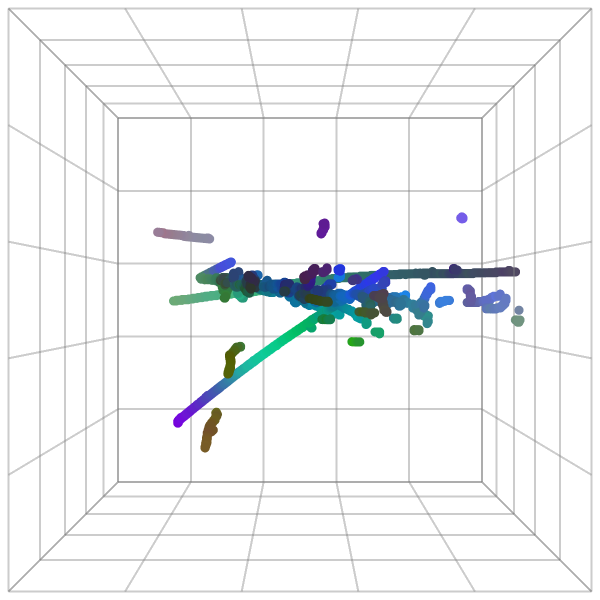} &
    \includegraphics[width=0.3\linewidth]{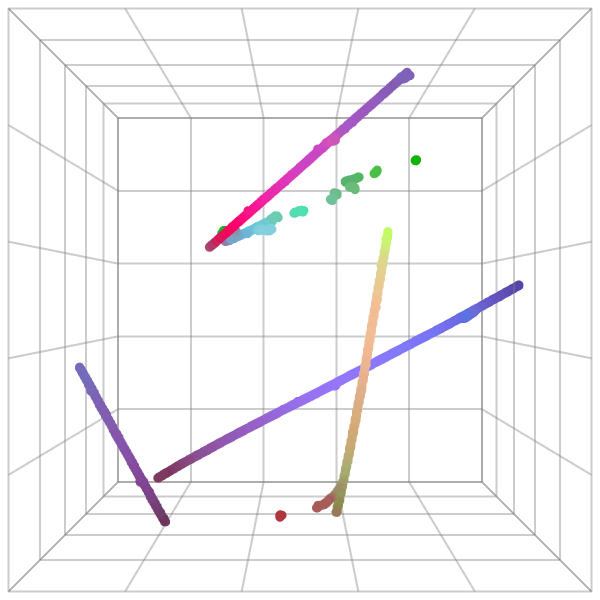} \\
    \end{tabular}
    \caption{\textit{Visualization of spatially debiased learned representations after pretraining.} We use PCA to project the learned representations into RGB space after spatial debiasing. The randomly initialized model displays a strong positional bias, while the Point-MAE model shows a clearer bias towards individual trajectories. The Point-MAE model contains visually similar embeddings as PoLAr-MAE, and is not pictured. Best viewed zoomed in. Additional examples (with embeddings for Point-MAE) as well as the debiasiing procedure can be found in Appendix \ref{app:pretraining}.}
    \label{fig:latent_comparison}
    \vspace{-2mm}
\end{figure}

\begin{figure*}[ht!!]
    \centering
    \includegraphics[width=0.85\linewidth]{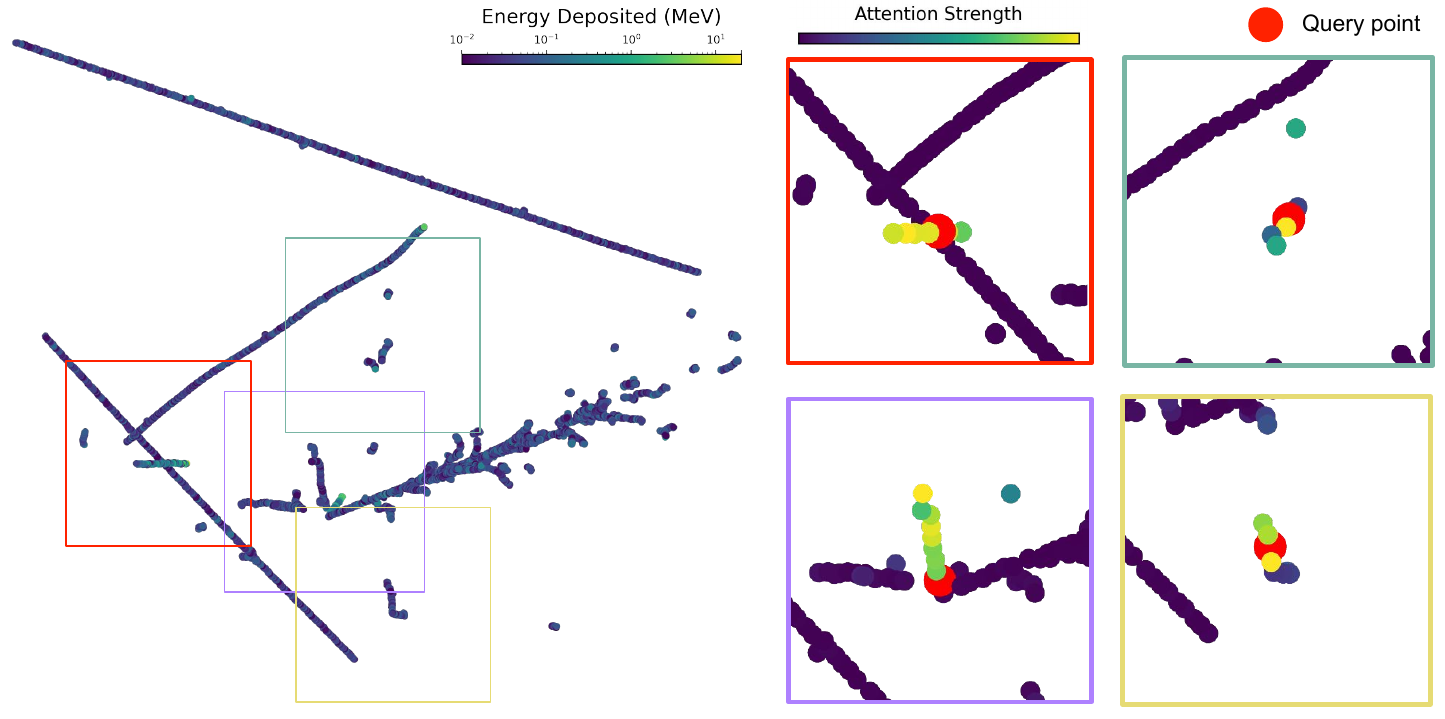}
    \includegraphics[width=0.83\linewidth]{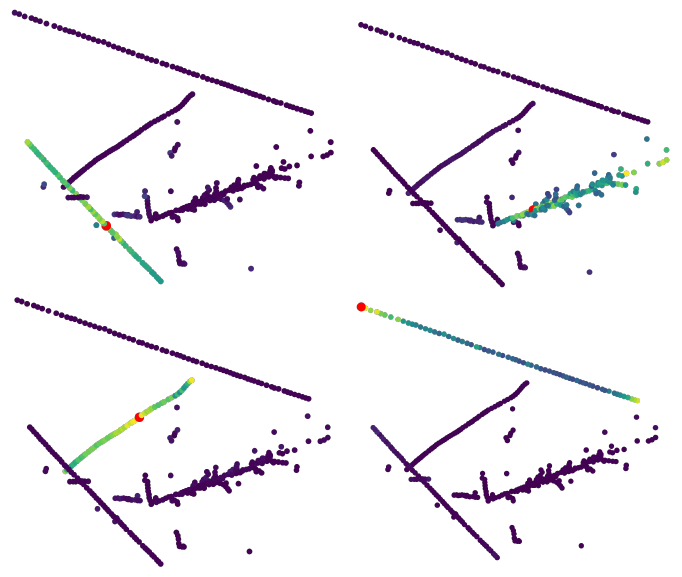}
    \caption{\textit{Visualization of attention maps from the 6th head of the 5th encoder layer.} Top row: full event used as input (left) and zoomed-in regions showing attention focusing on small trajectories (right). Bottom four panels: attention patterns when querying tokens (red dots) on longer tracks and a shower. Attention largely conforms to individual particle instances. Color intensity reflects the strength of attention contributed by other tokens to the queried token.}
    \label{fig:attn_map}
\end{figure*}

\subsubsection{Attention Maps}
\label{sec:attention_maps}

To gain qualitative insights into the representations learned by PoLAr-MAE, particularly how the model processes spatial relationships between different parts of an event, we visualize the attention mechanisms within the Transformer encoder. Figure \ref{fig:attn_map} showcases these attention patterns from the 6th attention head in the 5th layer of the encoder. In these visualizations, a specific token (the ``query point," marked in red) is selected, and all other tokens in the point cloud are colored according to the attention strength they contribute \textit{to} this queried token. High attention (yellow) indicates tokens that the model deems most relevant to understanding or contextualizing the queried token.

The figure is structured to demonstrate this phenomenon across different scales and particle types. The top row first presents the full input event alongside several zoomed-in regions. These magnified views highlight how the attention mechanism can delineate even small, distinct trajectories, such as short tracks or Michel electrons (e.g., top right panels, within colored bounding boxes). When querying tokens on longer tracks or within electromagnetic showers (bottom four panels of Fig.~\ref{fig:attn_map}, the attention patterns consistently concentrate along the queried particle's trajectory or within the characteristic spread of the shower. 

This consistent focusing of attention along individual particle trajectories, whether they are short and isolated or long and part of a complex event, shows that the self-supervised pre-training task encourages the model to learn not just local geometric features but also a more high-level understanding of particle contiguity and separation. This intrinsic grouping capability is a highly desirable property for downstream tasks, as it suggests the learned representations are inherently structured to distinguish between different particle instances.

\section{Future Work and Discussion}

This work presents the first successful implementation and adaptation of a self-supervised, masked autoencoding framework, PoLAr-MAE, for representation learning directly on the sparse, 3-D point cloud data from Liquid Argon Time Projection Chambers (LArTPCs). Our results demonstrate that by learning to reconstruct masked portions of particle trajectories, the model develops a robust and transferable understanding of the underlying physics without relying on any labeled examples. These learned representations are highly effective, proving amenable to abstract downstream tasks and enabling remarkable data efficiency.

Successfully transferring SSL paradigms from computer vision to the HEP domain often necessitates domain-specific adaptations. Our development of the Centrality-Based Non-Maximum Suppression (C-NMS) tokenization serves as an example of a modification. Unlike traditional grouping methods common in computer vision, C-NMS is designed to dynamically adjust to the characteristic spatial density variations within TPC point clouds, ensuring minimal token overlap and maximized coverage of particle trajectories. Such adaptations are crucial when applying existing architectures to new scientific domains with distinct data characteristics.

The high $F_1$ scores achieved for track (0.994) and shower (0.977) token classification with linear SVMs validate the efficacy of our SSL approach, while the semantic segmentation results show a successful transfer to dense classification. These metrics are on par with state-of-the-art supervised frameworks, such as the Sparse UResNet trained on $>100,000$ events for SPINE \cite{drielsma2021scalableendtoenddeeplearningbaseddata}, despite not relying on any labeled examples. This demonstrates that SSL frameworks can robustly capture the complex geometries and energy depositions characteristic of TPC data. However, the suboptimal modeling of the highly-infrequent particles, such as Michel electrons and delta rays, indicates a limitation in our current framework. 

Qualitative insights from the model's internal mechanisms further validate the learned representations and offer an understanding of how these results are achieved. Visualizations of the learned representations via spatially de-biased PCA (i.e., Fig.~\ref{fig:latent_comparison}), when compared to the randomly initialized network, show smooth gradients and a clear bias towards individual particle instances. Additionally, visualizations of attention maps from the attention head in the 5th encoder layer (as shown in Figure~\ref{fig:attn_map}) reveal that PoLAr-MAE learns to selectively focus on voxels belonging to the same particle trajectory. These phenomena are observed for both small, distinct trajectories like Michel electrons and short tracks, as well as for more extended structures such as long tracks and electromagnetic showers. This intrinsic grouping capability is a highly desirable property, and suggests that the learned representations are inherently structured to distinguish between different particle instances. We emphasize that this property is emergent, learned solely from the pretext task of predicting the masked content of LArTPC images from a partial view.

Further architectural innovations are crucial for resolving fine-grained sub-token structures and enhancing model efficiency, for current fixed-resolution, token-based ViT-like models like PoLAr-MAE and Point-MAE can be limited. TPC data is inherently multi-scale with both localized (centimeter-scale) energy depositions and extended (meter-scale) structures. Hierarchical architectures \cite{wu2024pointtransformerv3simpler, ryali2023hierahierarchicalvisiontransformer, fan2021multiscale, liu2021swin, zhdanov2025erwin} address the parameter and representational inefficiencies of flat ViT designs, which rely on a fixed resolution, fixed channel approach for all layers throughout the network. By starting spatially fine-grained with fewer channels and becoming spatially coarser yet richer in features, these models enable more efficient parameter use and robust multi-scale feature learning. The success of such hierarchical designs in vision tasks requiring finely detailed per-pixel classification, for example by the Hiera backbone \cite{ryali2023hierahierarchicalvisiontransformer} in Meta's Segment Anything Model 2 \cite{ravi2025sam}, clearly underscores their potential. Additionally, point-native architectures, such as Point Transformers \cite{zhao2021point, wu2022pointtransformerv2grouped, wu2024pointtransformerv3simpler} and Erwin \cite{zhdanov2025erwin}, offer a promising direction by operating directly on individual point features. We therefore see the exploration of point-native and hierarchical models as a critical next step in advancing LArTPC foundation model research.

Exploring alternative SSL paradigms, such as contrastive learning \cite{xie2020pointcontrastunsupervisedpretraining3d, wilkinson2025contrastivelearningrobustrepresentations}, student-teacher self-distillation for view consistency \cite{wu2025sonataselfsupervisedlearningreliable} and masked latent prediction \cite{zeid2023point2vecselfsupervisedrepresentationlearning}, local-global view consistency \cite{wu2025sonataselfsupervisedlearningreliable} may also offer benefits and should be explored in future work. 

Clearly, there are many unexplored avenues to pursue in the quest for creating a LArTPC foundation model; in light of this, we hope that the introduction of the PILArNet-M dataset into the community will spur additional research in this domain.

Throughout the development of this work, many changes were introduced to enforce the learning of better representations. We expand on several modifications we tried that did not result in any meaningful change in token semantics, in Appendix \ref{app:failures}.

\label{sec:discussion}

\section*{Acknowledgements}

This work is supported by the U.S. Department of Energy, Office of Science, and Office of High Energy Physics under Contract No. DE-AC02-76SF00515. SY is supported by the Stanford HAI Graduate Fellowship.

\bibliography{main}

\begin{thebibliography}{60}
\providecommand{\natexlab}[1]{#1}
\providecommand{\url}[1]{\texttt{#1}}
\expandafter\ifx\csname urlstyle\endcsname\relax
  \providecommand{\doi}[1]{doi: #1}\else
  \providecommand{\doi}{doi: \begingroup \urlstyle{rm}\Url}\fi

\bibitem[Abi et~al.(2020)Abi, Acciarri, Acero, Adamov, Adams, Adinolfi, Ahmad, Ahmed, and Alion]{Abi2020}
Abi, B., Acciarri, R., Acero, M., Adamov, G., Adams, D., Adinolfi, M., Ahmad, Z., Ahmed, J., and Alion, T.
\newblock Volume iv. the dune far detector single-phase technology.
\newblock \emph{Journal of Instrumentation}, 15\penalty0 (08):\penalty0 T08010–T08010, August 2020.
\newblock ISSN 1748-0221.
\newblock \doi{10.1088/1748-0221/15/08/t08010}.
\newblock URL \url{http://dx.doi.org/10.1088/1748-0221/15/08/T08010}.

\bibitem[Acciarri et~al.(2017)Acciarri, Adams, An, Asaadi, Auger, Bagby, Baller, Barr, Bass, Bay, Bishai, Blake, Bolton, Bugel, Camilleri, Caratelli, Carls, Fernandez, Cavanna, Chen, Church, Cianci, Collin, Conrad, Convery, Crespo-Anadón, Del~Tutto, Devitt, Dytman, Eberly, Ereditato, Sanchez, Esquivel, Fleming, Foreman, Furmanski, Garvey, Genty, Goeldi, Gollapinni, Graf, Gramellini, Greenlee, Grosso, Guenette, Hackenburg, Hamilton, Hen, Hewes, Hill, Ho, Horton-Smith, James, de~Vries, Jen, Jiang, Johnson, Jones, Joshi, Jostlein, Kaleko, Karagiorgi, Ketchum, Kirby, Kirby, Kobilarcik, Kreslo, Laube, Li, Lister, Littlejohn, Lockwitz, Lorca, Louis, Luethi, Lundberg, Luo, Marchionni, Mariani, Marshall, Caicedo, Meddage, Miceli, Mills, Moon, Mooney, Moore, Mousseau, Murrells, Naples, Nienaber, Nowak, Palamara, Paolone, Papavassiliou, Pate, Pavlovic, Porzio, Pulliam, Qian, Raaf, Rafique, Rochester, von Rohr, Russell, Schmitz, Schukraft, Seligman, Shaevitz, Sinclair, Snider, Soderberg, Söldner-Rembold, Soleti,
  Spentzouris, Spitz, St.~John, Strauss, Szelc, Tagg, Terao, Thomson, Toups, Tsai, Tufanli, Usher, Van~de Water, Viren, Weber, Weston, Wickremasinghe, Wolbers, Wongjirad, Woodruff, Yang, Zeller, Zennamo, and Zhang]{Acciarri2017}
Acciarri, R., Adams, C., An, R., Asaadi, J., Auger, M., Bagby, L., Baller, B., Barr, G., Bass, M., Bay, F., Bishai, M., Blake, A., Bolton, T., Bugel, L., Camilleri, L., Caratelli, D., Carls, B., Fernandez, R.~C., Cavanna, F., Chen, H., Church, E., Cianci, D., Collin, G., Conrad, J., Convery, M., Crespo-Anadón, J., Del~Tutto, M., Devitt, D., Dytman, S., Eberly, B., Ereditato, A., Sanchez, L.~E., Esquivel, J., Fleming, B., Foreman, W., Furmanski, A., Garvey, G., Genty, V., Goeldi, D., Gollapinni, S., Graf, N., Gramellini, E., Greenlee, H., Grosso, R., Guenette, R., Hackenburg, A., Hamilton, P., Hen, O., Hewes, J., Hill, C., Ho, J., Horton-Smith, G., James, C., de~Vries, J.~J., Jen, C.-M., Jiang, L., Johnson, R., Jones, B., Joshi, J., Jostlein, H., Kaleko, D., Karagiorgi, G., Ketchum, W., Kirby, B., Kirby, M., Kobilarcik, T., Kreslo, I., Laube, A., Li, Y., Lister, A., Littlejohn, B., Lockwitz, S., Lorca, D., Louis, W., Luethi, M., Lundberg, B., Luo, X., Marchionni, A., Mariani, C., Marshall, J., Caicedo, D.~M.,
  Meddage, V., Miceli, T., Mills, G., Moon, J., Mooney, M., Moore, C., Mousseau, J., Murrells, R., Naples, D., Nienaber, P., Nowak, J., Palamara, O., Paolone, V., Papavassiliou, V., Pate, S., Pavlovic, Z., Porzio, D., Pulliam, G., Qian, X., Raaf, J., Rafique, A., Rochester, L., von Rohr, C.~R., Russell, B., Schmitz, D., Schukraft, A., Seligman, W., Shaevitz, M., Sinclair, J., Snider, E., Soderberg, M., Söldner-Rembold, S., Soleti, S., Spentzouris, P., Spitz, J., St.~John, J., Strauss, T., Szelc, A., Tagg, N., Terao, K., Thomson, M., Toups, M., Tsai, Y.-T., Tufanli, S., Usher, T., Van~de Water, R., Viren, B., Weber, M., Weston, J., Wickremasinghe, D., Wolbers, S., Wongjirad, T., Woodruff, K., Yang, T., Zeller, G., Zennamo, J., and Zhang, C.
\newblock Convolutional neural networks applied to neutrino events in a liquid argon time projection chamber.
\newblock \emph{Journal of Instrumentation}, 12\penalty0 (03):\penalty0 P03011, mar 2017.
\newblock \doi{10.1088/1748-0221/12/03/P03011}.
\newblock URL \url{https://dx.doi.org/10.1088/1748-0221/12/03/P03011}.

\bibitem[Adams et~al.(2019)Adams, Alrashed, An, Anthony, Asaadi, Ashkenazi, Auger, Balasubramanian, Baller, Barnes, Barr, Bass, Bay, Bhat, Bhattacharya, Bishai, Blake, Bolton, Camilleri, Caratelli, Caro~Terrazas, Carr, Castillo~Fernandez, Cavanna, Cerati, Chen, Church, Cianci, Cohen, Collin, Conrad, Convery, Cooper-Troendle, Crespo-Anad\'on, Del~Tutto, Devitt, Diaz, Duffy, Dytman, Eberly, Ereditato, Escudero~Sanchez, Esquivel, Evans, Fadeeva, Fitzpatrick, Fleming, Franco, Furmanski, Garcia-Gamez, Genty, Goeldi, Gollapinni, Goodwin, Gramellini, Greenlee, Grosso, Guenette, Guzowski, Hackenburg, Hamilton, Hen, Hewes, Hill, Horton-Smith, Hourlier, Huang, James, Jan~de Vries, Ji, Jiang, Johnson, Joshi, Jostlein, Jwa, Karagiorgi, Ketchum, Kirby, Kirby, Kobilarcik, Kreslo, Lepetic, Li, Lister, Littlejohn, Lockwitz, Lorca, Louis, Luethi, Lundberg, Luo, Marchionni, Marcocci, Mariani, Marshall, Martin-Albo, Martinez~Caicedo, Mastbaum, Meddage, Mettler, Mistry, Mogan, Moon, Mooney, Moore, Mousseau, Murphy, Murrells,
  Naples, Nienaber, Nowak, Palamara, Pandey, Paolone, Papadopoulou, Papavassiliou, Pate, Pavlovic, Piasetzky, Porzio, Pulliam, Qian, Raaf, Rafique, Ren, Rochester, Ross-Lonergan, Rudolf~von Rohr, Russell, Scanavini, Schmitz, Schukraft, Seligman, Shaevitz, Sharankova, Sinclair, Smith, Snider, Soderberg, S\"oldner-Rembold, Soleti, Spentzouris, Spitz, St.~John, Strauss, Sutton, Sword-Fehlberg, Szelc, Tagg, Tang, Terao, Thomson, Thornton, Toups, Tsai, Tufanli, Usher, Van De~Pontseele, Van~de Water, Viren, Weber, Wei, Wickremasinghe, Wierman, Williams, Wolbers, Wongjirad, Woodruff, Yang, Yarbrough, Yates, Zeller, Zennamo, and Zhang]{Adams2019}
Adams, C., Alrashed, M., An, R., Anthony, J., Asaadi, J., Ashkenazi, A., Auger, M., Balasubramanian, S., Baller, B., Barnes, C., Barr, G., Bass, M., Bay, F., Bhat, A., Bhattacharya, K., Bishai, M., Blake, A., Bolton, T., Camilleri, L., Caratelli, D., Caro~Terrazas, I., Carr, R., Castillo~Fernandez, R., Cavanna, F., Cerati, G., Chen, Y., Church, E., Cianci, D., Cohen, E.~O., Collin, G.~H., Conrad, J.~M., Convery, M., Cooper-Troendle, L., Crespo-Anad\'on, J.~I., Del~Tutto, M., Devitt, A., Diaz, A., Duffy, K., Dytman, S., Eberly, B., Ereditato, A., Escudero~Sanchez, L., Esquivel, J., Evans, J.~J., Fadeeva, A.~A., Fitzpatrick, R.~S., Fleming, B.~T., Franco, D., Furmanski, A.~P., Garcia-Gamez, D., Genty, V., Goeldi, D., Gollapinni, S., Goodwin, O., Gramellini, E., Greenlee, H., Grosso, R., Guenette, R., Guzowski, P., Hackenburg, A., Hamilton, P., Hen, O., Hewes, J., Hill, C., Horton-Smith, G.~A., Hourlier, A., Huang, E.-C., James, C., Jan~de Vries, J., Ji, X., Jiang, L., Johnson, R.~A., Joshi, J., Jostlein, H.,
  Jwa, Y.-J., Karagiorgi, G., Ketchum, W., Kirby, B., Kirby, M., Kobilarcik, T., Kreslo, I., Lepetic, I., Li, Y., Lister, A., Littlejohn, B.~R., Lockwitz, S., Lorca, D., Louis, W.~C., Luethi, M., Lundberg, B., Luo, X., Marchionni, A., Marcocci, S., Mariani, C., Marshall, J., Martin-Albo, J., Martinez~Caicedo, D.~A., Mastbaum, A., Meddage, V., Mettler, T., Mistry, K., Mogan, A., Moon, J., Mooney, M., Moore, C.~D., Mousseau, J., Murphy, M., Murrells, R., Naples, D., Nienaber, P., Nowak, J., Palamara, O., Pandey, V., Paolone, V., Papadopoulou, A., Papavassiliou, V., Pate, S.~F., Pavlovic, Z., Piasetzky, E., Porzio, D., Pulliam, G., Qian, X., Raaf, J.~L., Rafique, A., Ren, L., Rochester, L., Ross-Lonergan, M., Rudolf~von Rohr, C., Russell, B., Scanavini, G., Schmitz, D.~W., Schukraft, A., Seligman, W., Shaevitz, M.~H., Sharankova, R., Sinclair, J., Smith, A., Snider, E.~L., Soderberg, M., S\"oldner-Rembold, S., Soleti, S.~R., Spentzouris, P., Spitz, J., St.~John, J., Strauss, T., Sutton, K., Sword-Fehlberg, S.,
  Szelc, A.~M., Tagg, N., Tang, W., Terao, K., Thomson, M., Thornton, R.~T., Toups, M., Tsai, Y.-T., Tufanli, S., Usher, T., Van De~Pontseele, W., Van~de Water, R.~G., Viren, B., Weber, M., Wei, H., Wickremasinghe, D.~A., Wierman, K., Williams, Z., Wolbers, S., Wongjirad, T., Woodruff, K., Yang, T., Yarbrough, G., Yates, L.~E., Zeller, G.~P., Zennamo, J., and Zhang, C.
\newblock Deep neural network for pixel-level electromagnetic particle identification in the microboone liquid argon time projection chamber.
\newblock \emph{Phys. Rev. D}, 99:\penalty0 092001, May 2019.
\newblock \doi{10.1103/PhysRevD.99.092001}.
\newblock URL \url{https://link.aps.org/doi/10.1103/PhysRevD.99.092001}.

\bibitem[Adams et~al.(2020)Adams, Terao, and Wongjirad]{adams2020pilarnetpublicdatasetparticle}
Adams, C., Terao, K., and Wongjirad, T.
\newblock Pilarnet: Public dataset for particle imaging liquid argon detectors in high energy physics, 2020.
\newblock URL \url{https://arxiv.org/abs/2006.01993}.

\bibitem[Ba et~al.(2016)Ba, Kiros, and Hinton]{ba2016layernormalization}
Ba, J.~L., Kiros, J.~R., and Hinton, G.~E.
\newblock Layer normalization, 2016.
\newblock URL \url{https://arxiv.org/abs/1607.06450}.

\bibitem[Bommasani et~al.(2021)Bommasani, Hudson, Adeli, Altman, Arora, von Arx, Bernstein, Bohg, Bosselut, Brunskill, et~al.]{bommasani2021opportunities}
Bommasani, R., Hudson, D.~A., Adeli, E., Altman, R., Arora, S., von Arx, S., Bernstein, M.~S., Bohg, J., Bosselut, A., Brunskill, E., et~al.
\newblock On the opportunities and risks of foundation models.
\newblock \emph{arXiv preprint arXiv:2108.07258}, 2021.

\bibitem[Chen et~al.(2020)Chen, Kornblith, Norouzi, and Hinton]{chen2020simple}
Chen, T., Kornblith, S., Norouzi, M., and Hinton, G.
\newblock A simple framework for contrastive learning of visual representations.
\newblock In \emph{International conference on machine learning}, pp.\  1597--1607. PMLR, 2020.

\bibitem[Devlin(2018)]{devlin2018bert}
Devlin, J.
\newblock Bert: Pre-training of deep bidirectional transformers for language understanding.
\newblock \emph{arXiv preprint arXiv:1810.04805}, 2018.

\bibitem[Dillmann et~al.(2024)Dillmann, Mart{\'\i}nez-Galarza, Soria, Di~Stefano, and Kashyap]{dillmann2024representation}
Dillmann, S., Mart{\'\i}nez-Galarza, R., Soria, R., Di~Stefano, R., and Kashyap, V.~L.
\newblock Representation learning for time-domain high-energy astrophysics: Discovery of extragalactic fast x-ray transient xrt 200515.
\newblock \emph{arXiv preprint arXiv:2412.01150}, 2024.

\bibitem[Domin{\'e} et~al.(2020)Domin{\'e}, Terao, and Collaboration)]{domine2020scalable}
Domin{\'e}, L., Terao, K., and Collaboration), D.
\newblock Scalable deep convolutional neural networks for sparse, locally dense liquid argon time projection chamber data.
\newblock \emph{Physical Review D}, 102\penalty0 (1):\penalty0 012005, 2020.

\bibitem[Domin{\'e} et~al.(2021)Domin{\'e}, de~Soux, Drielsma, Koh, Itay, Lin, Terao, Tsang, Usher, and Collaboration)]{domine2021point}
Domin{\'e}, L., de~Soux, P.~C., Drielsma, F., Koh, D.~H., Itay, R., Lin, Q., Terao, K., Tsang, K.~V., Usher, T.~L., and Collaboration), D.
\newblock Point proposal network for reconstructing 3d particle endpoints with subpixel precision in liquid argon time projection chambers.
\newblock \emph{Physical Review D}, 104\penalty0 (3):\penalty0 032004, 2021.

\bibitem[Dosovitskiy(2020)]{dosovitskiy2020image}
Dosovitskiy, A.
\newblock An image is worth 16x16 words: Transformers for image recognition at scale.
\newblock \emph{arXiv preprint arXiv:2010.11929}, 2020.

\bibitem[Drielsma et~al.(2021{\natexlab{a}})Drielsma, Lin, de~Soux, Domin{\'e}, Itay, Koh, Nelson, Terao, Tsang, Usher, et~al.]{drielsma2021clustering}
Drielsma, F., Lin, Q., de~Soux, P.~C., Domin{\'e}, L., Itay, R., Koh, D.~H., Nelson, B.~J., Terao, K., Tsang, K.~V., Usher, T.~L., et~al.
\newblock Clustering of electromagnetic showers and particle interactions with graph neural networks in liquid argon time projection chambers.
\newblock \emph{Physical Review D}, 104\penalty0 (7):\penalty0 072004, 2021{\natexlab{a}}.

\bibitem[Drielsma et~al.(2021{\natexlab{b}})Drielsma, Terao, Dominé, and Koh]{drielsma2021scalableendtoenddeeplearningbaseddata}
Drielsma, F., Terao, K., Dominé, L., and Koh, D.~H.
\newblock Scalable, end-to-end, deep-learning-based data reconstruction chain for particle imaging detectors, 2021{\natexlab{b}}.
\newblock URL \url{https://arxiv.org/abs/2102.01033}.

\bibitem[Ester et~al.(1996)Ester, Kriegel, Sander, and Xu]{10.5555/3001460.3001507}
Ester, M., Kriegel, H.-P., Sander, J., and Xu, X.
\newblock A density-based algorithm for discovering clusters in large spatial databases with noise.
\newblock In \emph{Proceedings of the Second International Conference on Knowledge Discovery and Data Mining}, KDD'96, pp.\  226–231. AAAI Press, 1996.

\bibitem[Fan et~al.(2021)Fan, Xiong, Mangalam, Li, Yan, Malik, and Feichtenhofer]{fan2021multiscale}
Fan, H., Xiong, B., Mangalam, K., Li, Y., Yan, Z., Malik, J., and Feichtenhofer, C.
\newblock Multiscale vision transformers.
\newblock In \emph{Proceedings of the IEEE/CVF international conference on computer vision}, pp.\  6824--6835, 2021.

\bibitem[Fuchs et~al.(2020)Fuchs, Worrall, Fischer, and Welling]{fuchs2020se3Transformers3dPointClouds}
Fuchs, F., Worrall, D., Fischer, V., and Welling, M.
\newblock Se(3)-transformers: 3d roto-translation equivariant attention networks.
\newblock In Larochelle, H., Ranzato, M., Hadsell, R., Balcan, M., and Lin, H. (eds.), \emph{Advances in Neural Information Processing Systems}, volume~33, pp.\  1970--1981. Curran Associates, Inc., 2020.
\newblock URL \url{https://proceedings.neurips.cc/paper_files/paper/2020/file/15231a7ce4ba789d13b722cc5c955834-Paper.pdf}.

\bibitem[Golling et~al.(2024)Golling, Heinrich, Kagan, Klein, Leigh, Osadchy, and Raine]{golling2024maskedparticlemodelingsets}
Golling, T., Heinrich, L., Kagan, M., Klein, S., Leigh, M., Osadchy, M., and Raine, J.~A.
\newblock Masked particle modeling on sets: Towards self-supervised high energy physics foundation models, 2024.
\newblock URL \url{https://arxiv.org/abs/2401.13537}.

\bibitem[Graham et~al.(2018)Graham, Engelcke, and Van Der~Maaten]{graham20183d}
Graham, B., Engelcke, M., and Van Der~Maaten, L.
\newblock 3d semantic segmentation with submanifold sparse convolutional networks.
\newblock In \emph{Proceedings of the IEEE conference on computer vision and pattern recognition}, pp.\  9224--9232, 2018.

\bibitem[Guo et~al.(2021)Guo, Cai, Liu, Mu, Martin, and Hu]{guo2021pct}
Guo, M.-H., Cai, J.-X., Liu, Z.-N., Mu, T.-J., Martin, R.~R., and Hu, S.-M.
\newblock Pct: Point cloud transformer.
\newblock \emph{Computational Visual Media}, 7:\penalty0 187--199, 2021.

\bibitem[Harris et~al.(2024)Harris, Kagan, Krupa, Maier, and Woodward]{r3sl}
Harris, P., Kagan, M., Krupa, J., Maier, B., and Woodward, N.
\newblock Re-simulation-based self-supervised learning for pre-training foundation models, 2024.
\newblock URL \url{https://arxiv.org/abs/2403.07066}.

\bibitem[He et~al.(2020)He, Fan, Wu, Xie, and Girshick]{he2020momentum}
He, K., Fan, H., Wu, Y., Xie, S., and Girshick, R.
\newblock Momentum contrast for unsupervised visual representation learning.
\newblock In \emph{Proceedings of the IEEE/CVF conference on computer vision and pattern recognition}, pp.\  9729--9738, 2020.

\bibitem[He et~al.(2021)He, Chen, Xie, Li, Dollár, and Girshick]{he2021maskedautoencodersscalablevision}
He, K., Chen, X., Xie, S., Li, Y., Dollár, P., and Girshick, R.
\newblock Masked autoencoders are scalable vision learners, 2021.
\newblock URL \url{https://arxiv.org/abs/2111.06377}.

\bibitem[Herde et~al.(2024)Herde, Raonić, Rohner, Käppeli, Molinaro, de~Bézenac, and Mishra]{herde2024poseidonefficientfoundationmodels}
Herde, M., Raonić, B., Rohner, T., Käppeli, R., Molinaro, R., de~Bézenac, E., and Mishra, S.
\newblock Poseidon: Efficient foundation models for pdes, 2024.
\newblock URL \url{https://arxiv.org/abs/2405.19101}.

\bibitem[Hou et~al.(2024)Hou, He, Fang, Mei, Xu, Wu, Tian, Zhang, Zeng, Gou, Xin, Le, Xia, Zhou, Hui, Pan, Eden, Yang, Han, Shu, Guo, Li, Holmes, Li, and Shi]{Hou2024}
Hou, X., He, Y., Fang, P., Mei, S.-Q., Xu, Z., Wu, W.-C., Tian, J.-H., Zhang, S., Zeng, Z.-Y., Gou, Q.-Y., Xin, G.-Y., Le, S.-J., Xia, Y.-Y., Zhou, Y.-L., Hui, F.-M., Pan, Y.-F., Eden, J.-S., Yang, Z.-H., Han, C., Shu, Y.-L., Guo, D., Li, J., Holmes, E.~C., Li, Z.-R., and Shi, M.
\newblock Using artificial intelligence to document the hidden rna virosphere.
\newblock \emph{Cell}, 187\penalty0 (24):\penalty0 6929--6942.e16, November 2024.
\newblock ISSN 0092-8674.
\newblock \doi{10.1016/j.cell.2024.09.027}.
\newblock URL \url{http://dx.doi.org/10.1016/j.cell.2024.09.027}.

\bibitem[Irwin et~al.(2022)Irwin, Dimitriadis, He, and Bjerrum]{Irwin2022}
Irwin, R., Dimitriadis, S., He, J., and Bjerrum, E.~J.
\newblock Chemformer: a pre-trained transformer for computational chemistry.
\newblock \emph{Machine Learning: Science and Technology}, 3\penalty0 (1):\penalty0 015022, January 2022.
\newblock ISSN 2632-2153.
\newblock \doi{10.1088/2632-2153/ac3ffb}.
\newblock URL \url{http://dx.doi.org/10.1088/2632-2153/ac3ffb}.

\bibitem[Jumper et~al.(2021)Jumper, Evans, Pritzel, Green, Figurnov, Ronneberger, Tunyasuvunakool, Bates, Žídek, Potapenko, Bridgland, Meyer, Kohl, Ballard, Cowie, Romera-Paredes, Nikolov, Jain, Adler, Back, Petersen, Reiman, Clancy, Zielinski, Steinegger, Pacholska, Berghammer, Bodenstein, Silver, Vinyals, Senior, Kavukcuoglu, Kohli, and Hassabis]{Jumper2021}
Jumper, J., Evans, R., Pritzel, A., Green, T., Figurnov, M., Ronneberger, O., Tunyasuvunakool, K., Bates, R., Žídek, A., Potapenko, A., Bridgland, A., Meyer, C., Kohl, S. A.~A., Ballard, A.~J., Cowie, A., Romera-Paredes, B., Nikolov, S., Jain, R., Adler, J., Back, T., Petersen, S., Reiman, D., Clancy, E., Zielinski, M., Steinegger, M., Pacholska, M., Berghammer, T., Bodenstein, S., Silver, D., Vinyals, O., Senior, A.~W., Kavukcuoglu, K., Kohli, P., and Hassabis, D.
\newblock Highly accurate protein structure prediction with alphafold.
\newblock \emph{Nature}, 596\penalty0 (7873):\penalty0 583–589, July 2021.
\newblock ISSN 1476-4687.
\newblock \doi{10.1038/s41586-021-03819-2}.
\newblock URL \url{http://dx.doi.org/10.1038/s41586-021-03819-2}.

\bibitem[Kochkov et~al.(2023)Kochkov, Yuval, Langmore, Norgaard, Smith, Mooers, Lottes, Rasp, D{\"u}ben, Kl{\"o}wer, et~al.]{kochkov2023neural}
Kochkov, D., Yuval, J., Langmore, I., Norgaard, P., Smith, J.~A., Mooers, G., Lottes, J., Rasp, S., D{\"u}ben, P.~D., Kl{\"o}wer, M., et~al.
\newblock Neural general circulation models.
\newblock \emph{CoRR}, 2023.

\bibitem[Lam et~al.(2023)Lam, Sanchez-Gonzalez, Willson, Wirnsberger, Fortunato, Alet, Ravuri, Ewalds, Eaton-Rosen, Hu, Merose, Hoyer, Holland, Vinyals, Stott, Pritzel, Mohamed, and Battaglia]{doi:10.1126/science.adi2336}
Lam, R., Sanchez-Gonzalez, A., Willson, M., Wirnsberger, P., Fortunato, M., Alet, F., Ravuri, S., Ewalds, T., Eaton-Rosen, Z., Hu, W., Merose, A., Hoyer, S., Holland, G., Vinyals, O., Stott, J., Pritzel, A., Mohamed, S., and Battaglia, P.
\newblock Learning skillful medium-range global weather forecasting.
\newblock \emph{Science}, 382\penalty0 (6677):\penalty0 1416--1421, 2023.
\newblock \doi{10.1126/science.adi2336}.
\newblock URL \url{https://www.science.org/doi/abs/10.1126/science.adi2336}.

\bibitem[Lanusse et~al.(2023)Lanusse, Parker, Golkar, Bietti, Cranmer, Eickenberg, Krawezik, McCabe, Ohana, Pettee, et~al.]{lanusse2023astroclip}
Lanusse, F., Parker, L.~H., Golkar, S., Bietti, A., Cranmer, M., Eickenberg, M., Krawezik, G., McCabe, M., Ohana, R., Pettee, M., et~al.
\newblock Astroclip: Cross-modal pre-training for astronomical foundation models.
\newblock In \emph{NeurIPS 2023 AI for Science Workshop}, 2023.

\bibitem[Li et~al.(2024)Li, Madarasingha, and Thilakarathna]{li2024diffpmaediffusionmaskedautoencoders}
Li, Y., Madarasingha, C., and Thilakarathna, K.
\newblock Diffpmae: Diffusion masked autoencoders for point cloud reconstruction, 2024.
\newblock URL \url{https://arxiv.org/abs/2312.03298}.

\bibitem[Lieret \& DeZoort(2024)Lieret and DeZoort]{Lieret2024}
Lieret, K. and DeZoort, G.
\newblock An object condensation pipeline for charged particle tracking at the high luminosity lhc.
\newblock \emph{EPJ Web of Conferences}, 295:\penalty0 09004, 2024.
\newblock ISSN 2100-014X.
\newblock \doi{10.1051/epjconf/202429509004}.
\newblock URL \url{http://dx.doi.org/10.1051/epjconf/202429509004}.

\bibitem[Lin et~al.(2018)Lin, Goyal, Girshick, He, and Dollár]{lin2018focallossdenseobject}
Lin, T.-Y., Goyal, P., Girshick, R., He, K., and Dollár, P.
\newblock Focal loss for dense object detection, 2018.
\newblock URL \url{https://arxiv.org/abs/1708.02002}.

\bibitem[Liu et~al.(2021)Liu, Lin, Cao, Hu, Wei, Zhang, Lin, and Guo]{liu2021swin}
Liu, Z., Lin, Y., Cao, Y., Hu, H., Wei, Y., Zhang, Z., Lin, S., and Guo, B.
\newblock Swin transformer: Hierarchical vision transformer using shifted windows.
\newblock In \emph{Proceedings of the IEEE/CVF international conference on computer vision}, pp.\  10012--10022, 2021.

\bibitem[McCabe et~al.(2023)McCabe, Blancard, Parker, Ohana, Cranmer, Bietti, Eickenberg, Golkar, Krawezik, Lanusse, et~al.]{mccabe2023multiple}
McCabe, M., Blancard, B. R.-S., Parker, L.~H., Ohana, R., Cranmer, M., Bietti, A., Eickenberg, M., Golkar, S., Krawezik, G., Lanusse, F., et~al.
\newblock Multiple physics pretraining for physical surrogate models.
\newblock \emph{arXiv preprint arXiv:2310.02994}, 2023.

\bibitem[Miao et~al.(2024)Miao, Lu, Liu, Duarte, and Li]{Miao2024}
Miao, S., Lu, Z., Liu, M., Duarte, J., and Li, P.
\newblock Locality-sensitive hashing-based efficient point transformer with applications in high-energy physics.
\newblock In \emph{Proceedings of the 41st International Conference on Machine Learning}, ICML'24. JMLR.org, 2024.

\bibitem[Nguyen et~al.(2023)Nguyen, Brandstetter, Kapoor, Gupta, and Grover]{nguyen2023climax}
Nguyen, T., Brandstetter, J., Kapoor, A., Gupta, J.~K., and Grover, A.
\newblock Climax: A foundation model for weather and climate.
\newblock \emph{arXiv preprint arXiv:2301.10343}, 2023.

\bibitem[Oquab et~al.(2024)Oquab, Darcet, Moutakanni, Vo, Szafraniec, Khalidov, Fernandez, Haziza, Massa, El-Nouby, Assran, Ballas, Galuba, Howes, Huang, Li, Misra, Rabbat, Sharma, Synnaeve, Xu, Jegou, Mairal, Labatut, Joulin, and Bojanowski]{oquab2024dinov2learningrobustvisual}
Oquab, M., Darcet, T., Moutakanni, T., Vo, H., Szafraniec, M., Khalidov, V., Fernandez, P., Haziza, D., Massa, F., El-Nouby, A., Assran, M., Ballas, N., Galuba, W., Howes, R., Huang, P.-Y., Li, S.-W., Misra, I., Rabbat, M., Sharma, V., Synnaeve, G., Xu, H., Jegou, H., Mairal, J., Labatut, P., Joulin, A., and Bojanowski, P.
\newblock Dinov2: Learning robust visual features without supervision, 2024.
\newblock URL \url{https://arxiv.org/abs/2304.07193}.

\bibitem[Pang et~al.(2022)Pang, Wang, Tay, Liu, Tian, and Yuan]{pang2022maskedautoencoderspointcloud}
Pang, Y., Wang, W., Tay, F. E.~H., Liu, W., Tian, Y., and Yuan, L.
\newblock Masked autoencoders for point cloud self-supervised learning, 2022.
\newblock URL \url{https://arxiv.org/abs/2203.06604}.

\bibitem[Qi et~al.(2017{\natexlab{a}})Qi, Su, Mo, and Guibas]{qi2017pointnet}
Qi, C.~R., Su, H., Mo, K., and Guibas, L.~J.
\newblock Pointnet: Deep learning on point sets for 3d classification and segmentation.
\newblock In \emph{Proceedings of the IEEE conference on computer vision and pattern recognition}, pp.\  652--660, 2017{\natexlab{a}}.

\bibitem[Qi et~al.(2017{\natexlab{b}})Qi, Yi, Su, and Guibas]{qi2017pointnet++}
Qi, C.~R., Yi, L., Su, H., and Guibas, L.~J.
\newblock Pointnet++: Deep hierarchical feature learning on point sets in a metric space.
\newblock \emph{Advances in neural information processing systems}, 30, 2017{\natexlab{b}}.

\bibitem[Qian et~al.(2022)Qian, Li, Peng, Mai, Hammoud, Elhoseiny, and Ghanem]{qian2022pointnext}
Qian, G., Li, Y., Peng, H., Mai, J., Hammoud, H., Elhoseiny, M., and Ghanem, B.
\newblock Pointnext: Revisiting pointnet++ with improved training and scaling strategies.
\newblock \emph{Advances in neural information processing systems}, 35:\penalty0 23192--23204, 2022.

\bibitem[Ranftl et~al.(2021)Ranftl, Bochkovskiy, and Koltun]{ranftl2021vision}
Ranftl, R., Bochkovskiy, A., and Koltun, V.
\newblock Vision transformers for dense prediction.
\newblock In \emph{Proceedings of the IEEE/CVF international conference on computer vision}, pp.\  12179--12188, 2021.

\bibitem[Ravi et~al.(2025)Ravi, Gabeur, Hu, Hu, Ryali, Ma, Khedr, R{\"a}dle, Rolland, Gustafson, Mintun, Pan, Alwala, Carion, Wu, Girshick, Dollar, and Feichtenhofer]{ravi2025sam}
Ravi, N., Gabeur, V., Hu, Y.-T., Hu, R., Ryali, C., Ma, T., Khedr, H., R{\"a}dle, R., Rolland, C., Gustafson, L., Mintun, E., Pan, J., Alwala, K.~V., Carion, N., Wu, C.-Y., Girshick, R., Dollar, P., and Feichtenhofer, C.
\newblock {SAM} 2: Segment anything in images and videos.
\newblock In \emph{The Thirteenth International Conference on Learning Representations}, 2025.
\newblock URL \url{https://openreview.net/forum?id=Ha6RTeWMd0}.

\bibitem[Ross et~al.(2022)Ross, Belgodere, Chenthamarakshan, Padhi, Mroueh, and Das]{Ross2022}
Ross, J., Belgodere, B., Chenthamarakshan, V., Padhi, I., Mroueh, Y., and Das, P.
\newblock Large-scale chemical language representations capture molecular structure and properties.
\newblock \emph{Nature Machine Intelligence}, 4\penalty0 (12):\penalty0 1256–1264, December 2022.
\newblock ISSN 2522-5839.
\newblock \doi{10.1038/s42256-022-00580-7}.
\newblock URL \url{http://dx.doi.org/10.1038/s42256-022-00580-7}.

\bibitem[Rubbia(1977)]{rubbia1977liquid}
Rubbia, C.
\newblock {The liquid-argon time projection chamber: a new concept for neutrino detectors}.
\newblock Technical report, CERN, Geneva, 1977.
\newblock URL \url{https://cds.cern.ch/record/117852}.

\bibitem[Ryali et~al.(2023)Ryali, Hu, Bolya, Wei, Fan, Huang, Aggarwal, Chowdhury, Poursaeed, Hoffman, Malik, Li, and Feichtenhofer]{ryali2023hierahierarchicalvisiontransformer}
Ryali, C., Hu, Y.-T., Bolya, D., Wei, C., Fan, H., Huang, P.-Y., Aggarwal, V., Chowdhury, A., Poursaeed, O., Hoffman, J., Malik, J., Li, Y., and Feichtenhofer, C.
\newblock Hiera: A hierarchical vision transformer without the bells-and-whistles, 2023.
\newblock URL \url{https://arxiv.org/abs/2306.00989}.

\bibitem[Satorras et~al.(2021)Satorras, Hoogeboom, and Welling]{pmlr-v139-satorras21a}
Satorras, V.~G., Hoogeboom, E., and Welling, M.
\newblock E(n) equivariant graph neural networks.
\newblock In Meila, M. and Zhang, T. (eds.), \emph{Proceedings of the 38th International Conference on Machine Learning}, volume 139 of \emph{Proceedings of Machine Learning Research}, pp.\  9323--9332. PMLR, 18--24 Jul 2021.
\newblock URL \url{https://proceedings.mlr.press/v139/satorras21a.html}.

\bibitem[Trinh et~al.(2024)Trinh, Wu, Le, He, and Luong]{Trinh2024}
Trinh, T.~H., Wu, Y., Le, Q.~V., He, H., and Luong, T.
\newblock Solving olympiad geometry without human demonstrations.
\newblock \emph{Nature}, 625\penalty0 (7995):\penalty0 476–482, January 2024.
\newblock ISSN 1476-4687.
\newblock \doi{10.1038/s41586-023-06747-5}.
\newblock URL \url{http://dx.doi.org/10.1038/s41586-023-06747-5}.

\bibitem[Vaswani(2017)]{vaswani2017attention}
Vaswani, A.
\newblock Attention is all you need.
\newblock \emph{Advances in Neural Information Processing Systems}, 2017.

\bibitem[Wilkinson et~al.(2025)Wilkinson, Radev, and Alonso-Monsalve]{wilkinson2025contrastivelearningrobustrepresentations}
Wilkinson, A., Radev, R., and Alonso-Monsalve, S.
\newblock Contrastive learning for robust representations of neutrino data, 2025.
\newblock URL \url{https://arxiv.org/abs/2502.07724}.

\bibitem[Wu et~al.(2022)Wu, Lao, Jiang, Liu, and Zhao]{wu2022pointtransformerv2grouped}
Wu, X., Lao, Y., Jiang, L., Liu, X., and Zhao, H.
\newblock Point transformer v2: Grouped vector attention and partition-based pooling, 2022.
\newblock URL \url{https://arxiv.org/abs/2210.05666}.

\bibitem[Wu et~al.(2024)Wu, Jiang, Wang, Liu, Liu, Qiao, Ouyang, He, and Zhao]{wu2024pointtransformerv3simpler}
Wu, X., Jiang, L., Wang, P.-S., Liu, Z., Liu, X., Qiao, Y., Ouyang, W., He, T., and Zhao, H.
\newblock Point transformer v3: Simpler, faster, stronger, 2024.
\newblock URL \url{https://arxiv.org/abs/2312.10035}.

\bibitem[Wu et~al.(2025)Wu, DeTone, Frost, Shen, Xie, Yang, Engel, Newcombe, Zhao, and Straub]{wu2025sonataselfsupervisedlearningreliable}
Wu, X., DeTone, D., Frost, D., Shen, T., Xie, C., Yang, N., Engel, J., Newcombe, R., Zhao, H., and Straub, J.
\newblock Sonata: Self-supervised learning of reliable point representations, 2025.
\newblock URL \url{https://arxiv.org/abs/2503.16429}.

\bibitem[Xie et~al.(2020)Xie, Gu, Guo, Qi, Guibas, and Litany]{xie2020pointcontrastunsupervisedpretraining3d}
Xie, S., Gu, J., Guo, D., Qi, C.~R., Guibas, L.~J., and Litany, O.
\newblock Pointcontrast: Unsupervised pre-training for 3d point cloud understanding, 2020.
\newblock URL \url{https://arxiv.org/abs/2007.10985}.

\bibitem[Xin et~al.(2024)Xin, Guo, Shao, Ren, Zhu, Liu, Ruan, Li, and Liang]{xin2024deepseek}
Xin, H., Guo, D., Shao, Z., Ren, Z., Zhu, Q., Liu, B., Ruan, C., Li, W., and Liang, X.
\newblock Deepseek-prover: Advancing theorem proving in llms through large-scale synthetic data.
\newblock \emph{arXiv preprint arXiv:2405.14333}, 2024.

\bibitem[Yu et~al.(2022)Yu, Tang, Rao, Huang, Zhou, and Lu]{yu2022pointbertpretraining3dpoint}
Yu, X., Tang, L., Rao, Y., Huang, T., Zhou, J., and Lu, J.
\newblock Point-bert: Pre-training 3d point cloud transformers with masked point modeling, 2022.
\newblock URL \url{https://arxiv.org/abs/2111.14819}.

\bibitem[Zeid et~al.(2023)Zeid, Schult, Hermans, and Leibe]{zeid2023point2vecselfsupervisedrepresentationlearning}
Zeid, K.~A., Schult, J., Hermans, A., and Leibe, B.
\newblock Point2vec for self-supervised representation learning on point clouds, 2023.
\newblock URL \url{https://arxiv.org/abs/2303.16570}.

\bibitem[Zhao et~al.(2021)Zhao, Jiang, Jia, Torr, and Koltun]{zhao2021point}
Zhao, H., Jiang, L., Jia, J., Torr, P.~H., and Koltun, V.
\newblock Point transformer.
\newblock In \emph{Proceedings of the IEEE/CVF international conference on computer vision}, pp.\  16259--16268, 2021.

\bibitem[Zhdanov et~al.(2025)Zhdanov, Welling, and van~de Meent]{zhdanov2025erwin}
Zhdanov, M., Welling, M., and van~de Meent, J.-W.
\newblock Erwin: A tree-based hierarchical transformer for large-scale physical systems.
\newblock \emph{arXiv preprint arXiv:2502.17019}, 2025.

\end{thebibliography}
\bibliographystyle{icml2025}

\newpage
\appendix
\onecolumn

\section{Dataset}
\label{app:dataset}

\begin{figure*}[t!]
    \centering
    \includegraphics[width=0.375\linewidth]{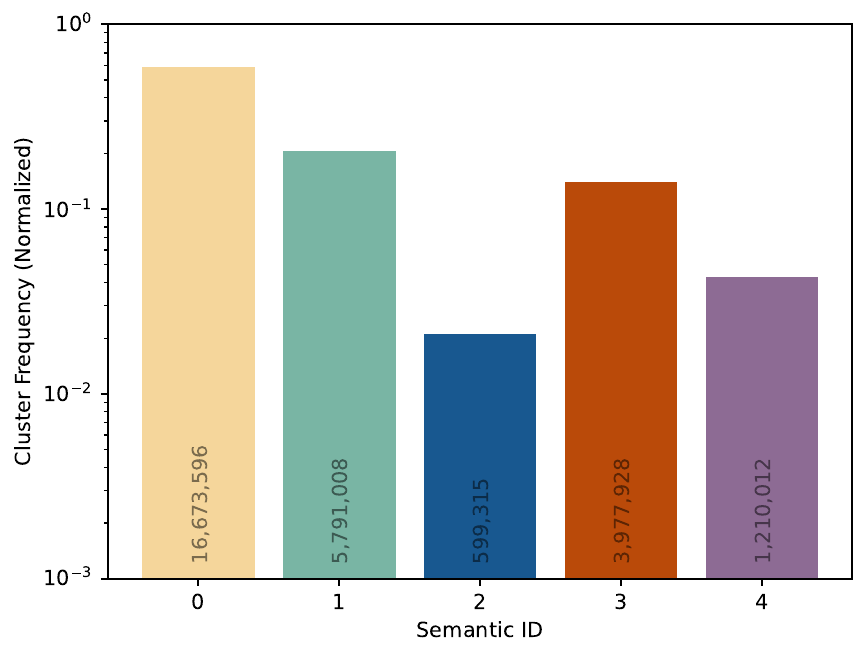}
    \includegraphics[width=0.5\linewidth]{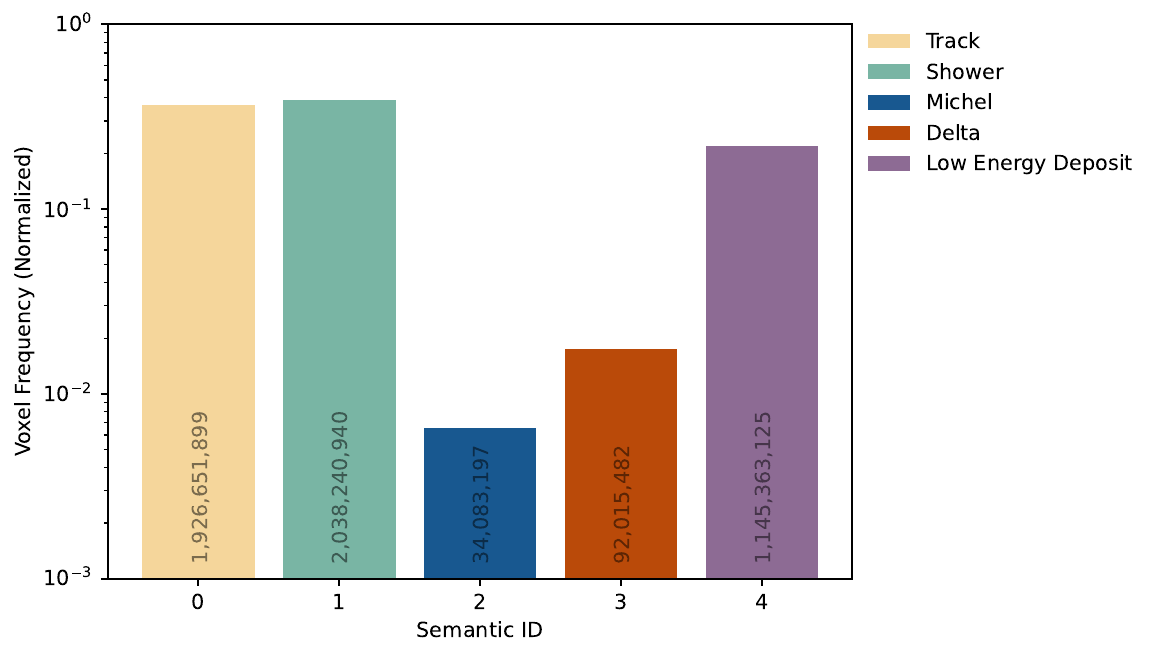}
    \includegraphics[width=0.375\linewidth]{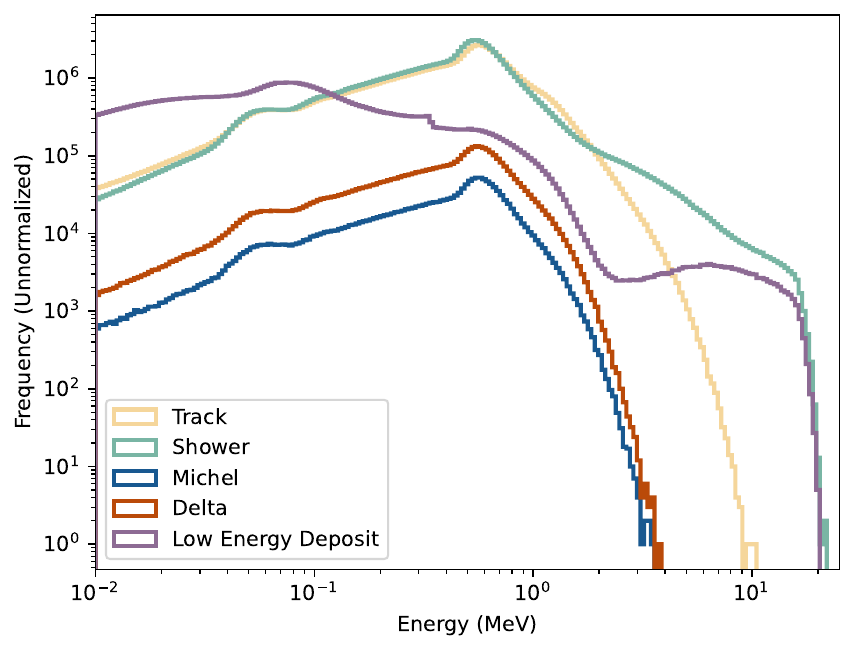}
    \caption{\textit{PILArNet-M statistics.} Starting from top left and going clockwise: cluster frequencies, i.e., how prevalent each particle type is on a cluster (trajectory) basis; voxel frequencies, i.e., the total semantic ID makeup of all voxels in the dataset; voxel energy distributions, i.e., a histogram of the energies of any occupied voxel across all events in the dataset. The number in the bar plots represents the raw number of each quantity found in the dataset, while the bars themselves are normalized to 1. Note that in the energy deposition distribution, energy values below 0.001 MeV are not present.}
    \label{fig:statistics}
\end{figure*}

\begin{figure*}[ht]
    \centering
    \begin{tabular}{c@{\hspace{1mm}}*{8}{c}@{}} 
    & \textsc{Event 4} & \textsc{Event 5} & \textsc{Event 6} & \textsc{Event 7} & \textsc{Event 8} & \textsc{Event 9} & \textsc{Event 10} \\
    \rotatebox[origin=l]{90}{\textsc{~~Fragment}} & 
    \includegraphics[width=0.135\linewidth]{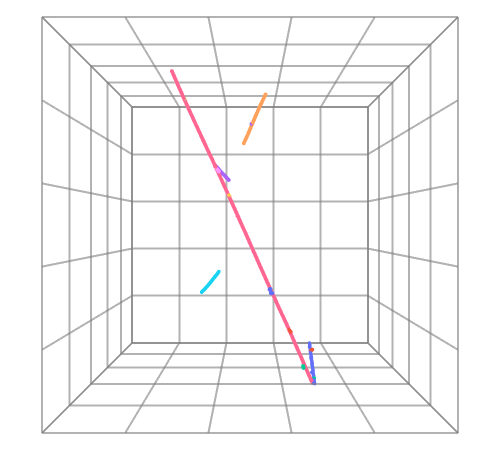}&
    \includegraphics[width=0.135\linewidth]{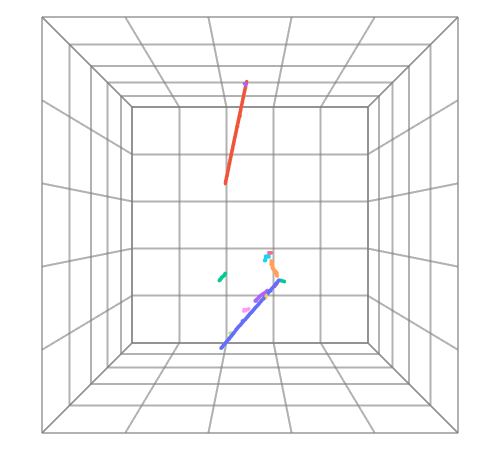} &
    \includegraphics[width=0.135\linewidth]{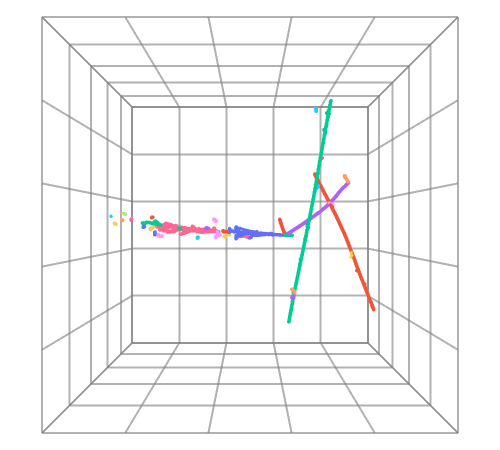} &
    \includegraphics[width=0.135\linewidth]{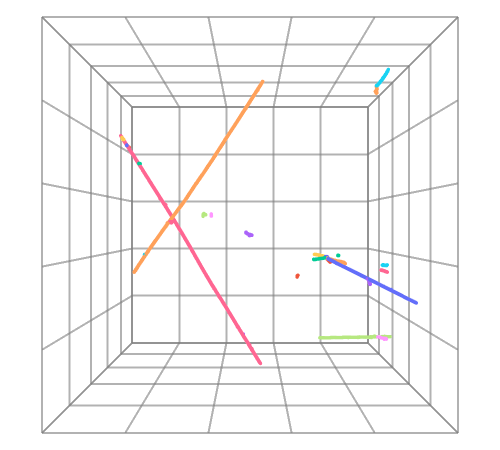} &
    \includegraphics[width=0.135\linewidth]{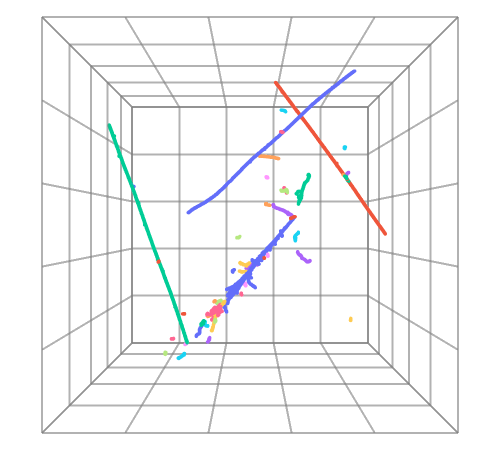} &
    \includegraphics[width=0.135\linewidth]{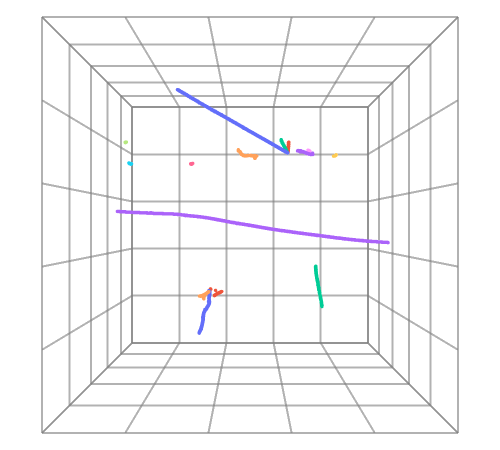} &
    \includegraphics[width=0.135\linewidth]{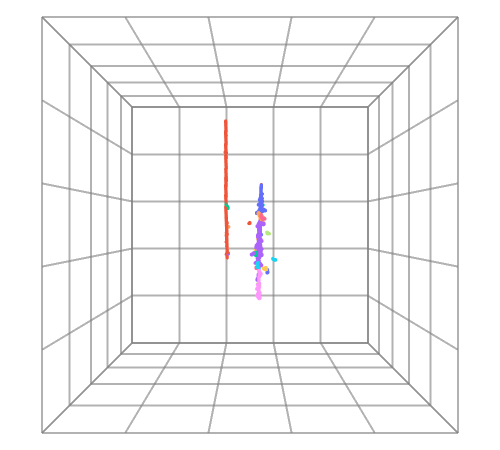} &

    \\
    
    \rotatebox[origin=l]{90}{\textsc{~~~Group}} & 
    \includegraphics[width=0.135\linewidth]{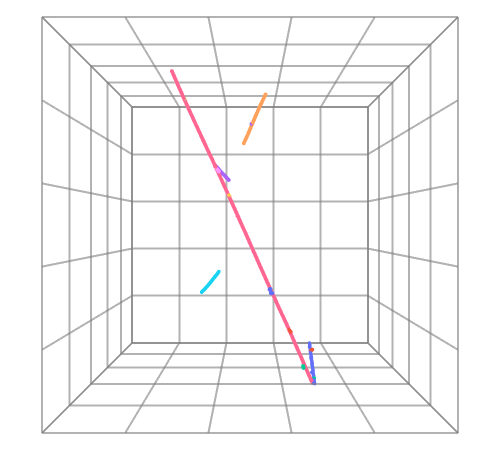}&
    \includegraphics[width=0.135\linewidth]{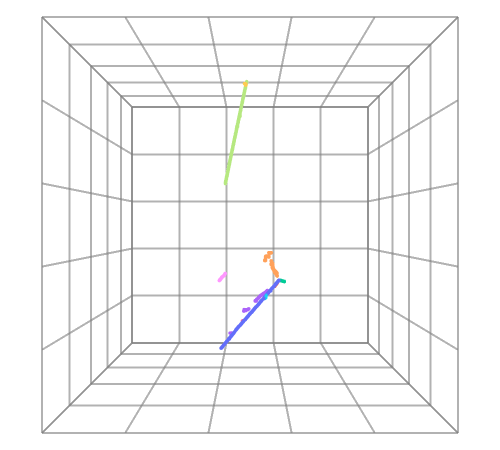} &
    \includegraphics[width=0.135\linewidth]{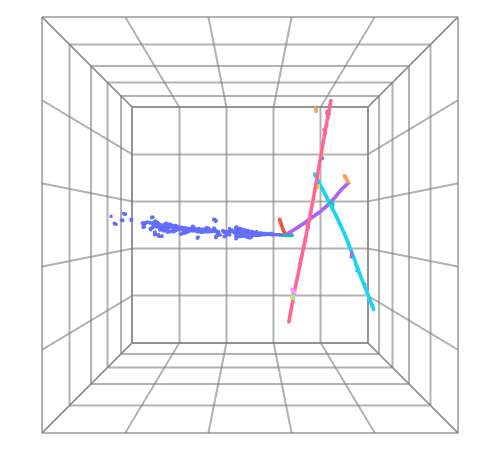} &
    \includegraphics[width=0.135\linewidth]{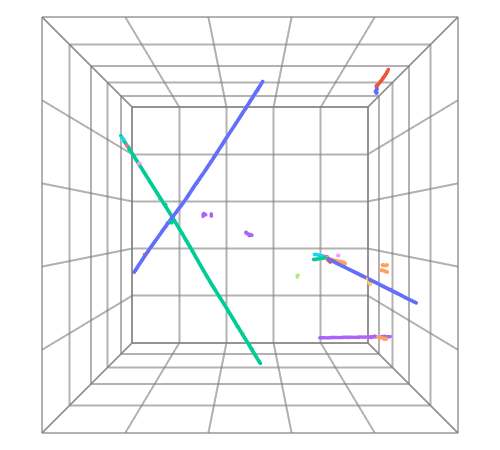} &
    \includegraphics[width=0.135\linewidth]{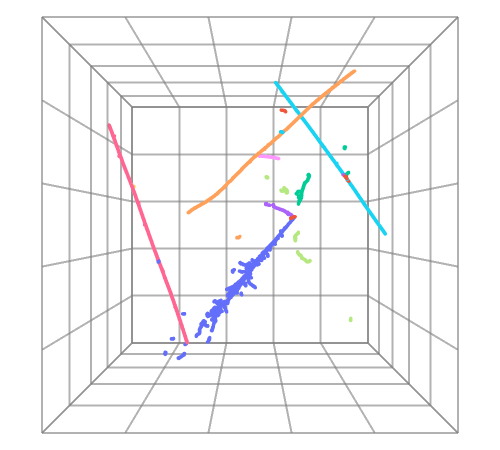} &
    \includegraphics[width=0.135\linewidth]{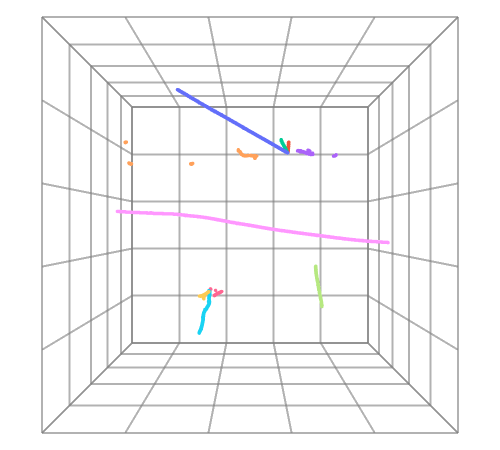} &
    \includegraphics[width=0.135\linewidth]{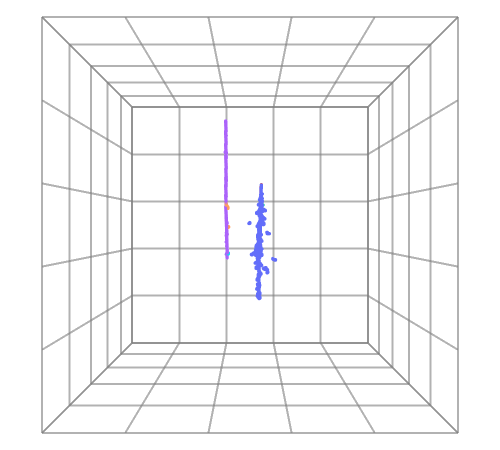} &

    \\
    \rotatebox[origin=l]{90}{\textsc{Interaction}} & 
    \includegraphics[width=0.135\linewidth]{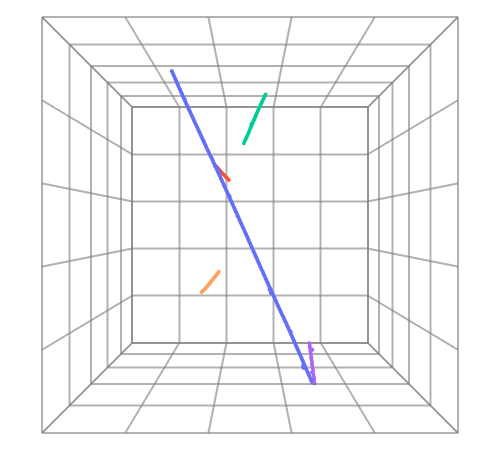}&
    \includegraphics[width=0.135\linewidth]{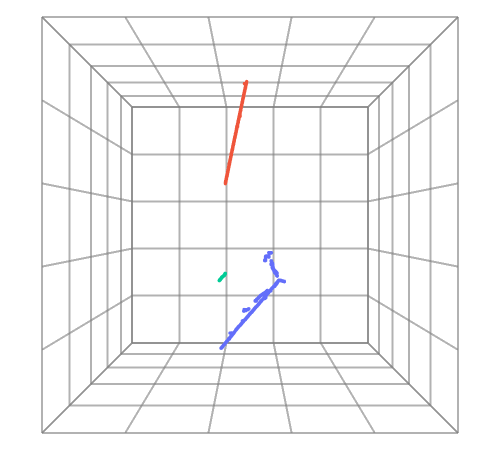} &
    \includegraphics[width=0.135\linewidth]{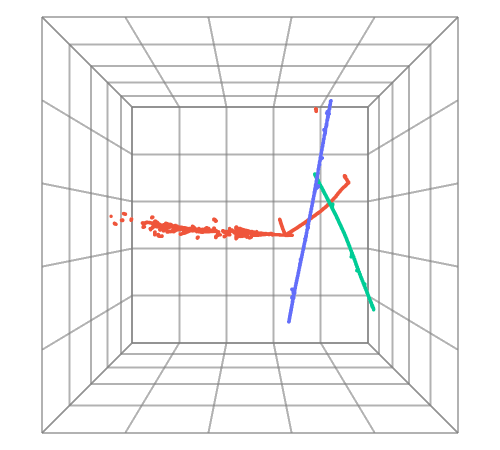} &
    \includegraphics[width=0.135\linewidth]{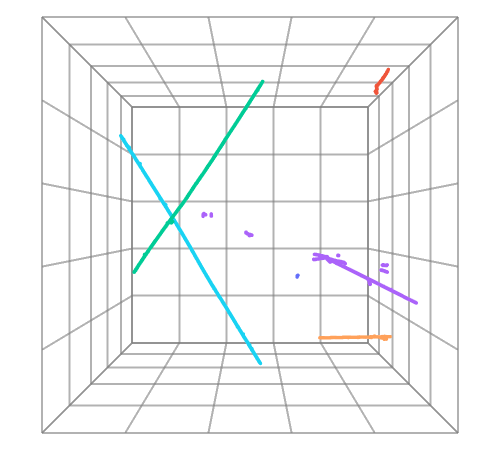} &
    \includegraphics[width=0.135\linewidth]{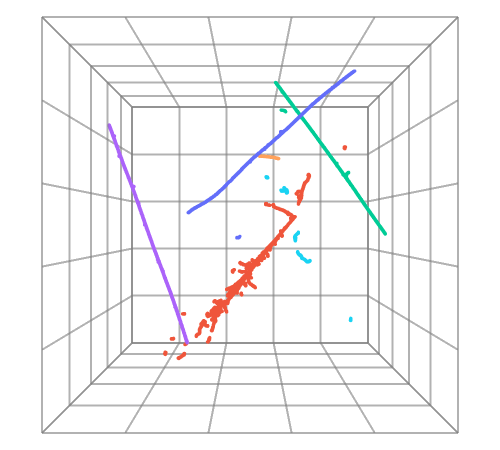} &
    \includegraphics[width=0.135\linewidth]{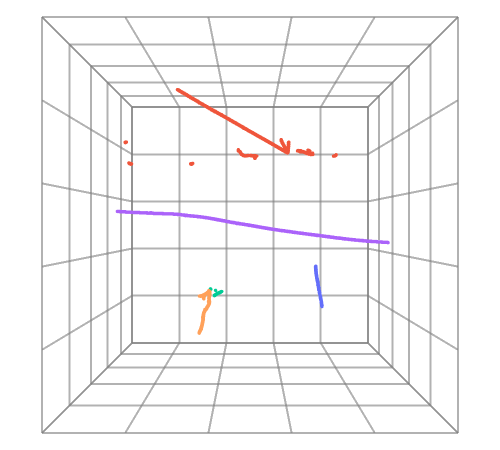} &
    \includegraphics[width=0.135\linewidth]{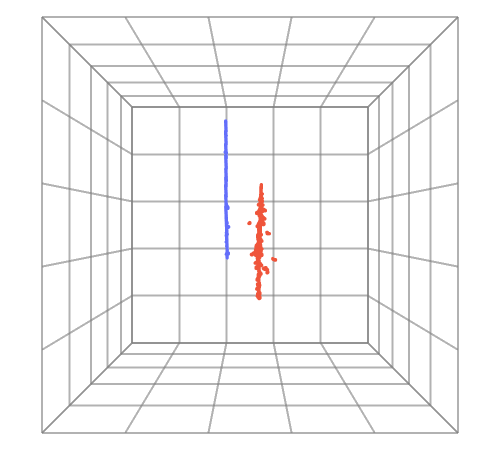} &

    \\
    \rotatebox[origin=l]{90}{\textsc{~~~Semantics}} & 
    \includegraphics[width=0.135\linewidth]{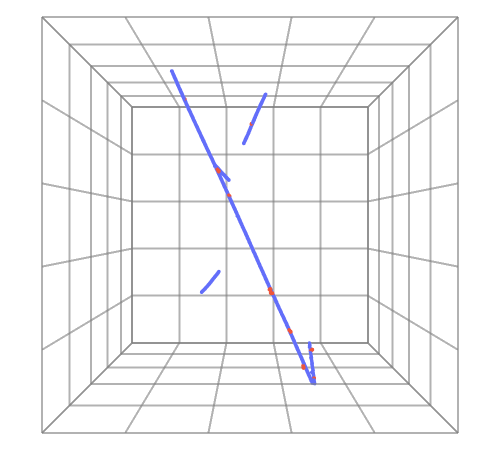}&
    \includegraphics[width=0.135\linewidth]{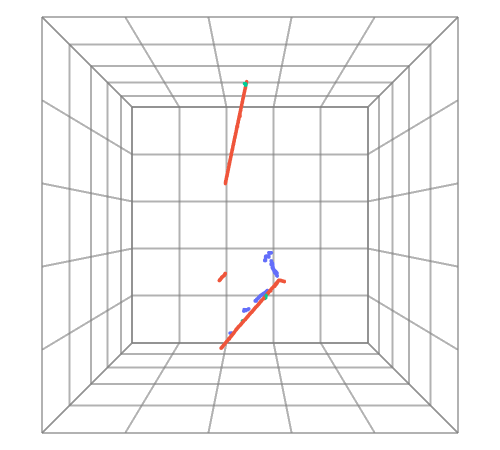} &
    \includegraphics[width=0.135\linewidth]{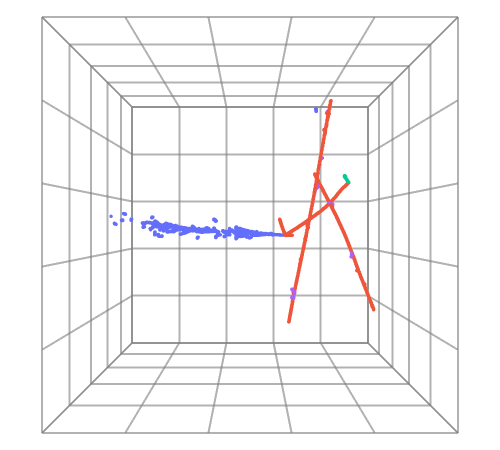} &
    \includegraphics[width=0.135\linewidth]{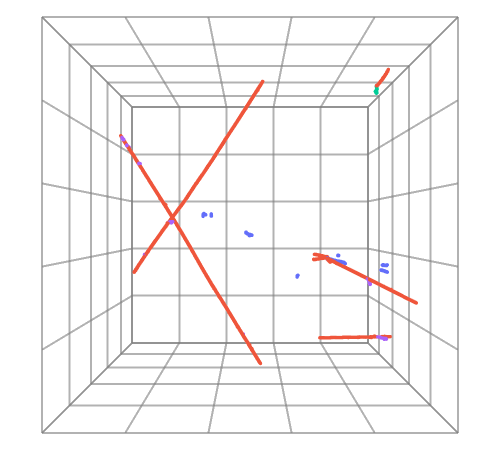} &
    \includegraphics[width=0.135\linewidth]{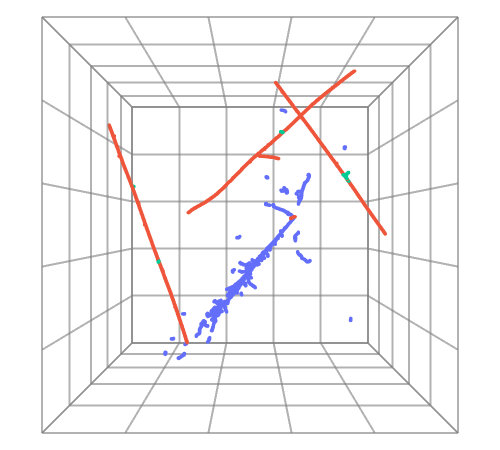} &
    \includegraphics[width=0.135\linewidth]{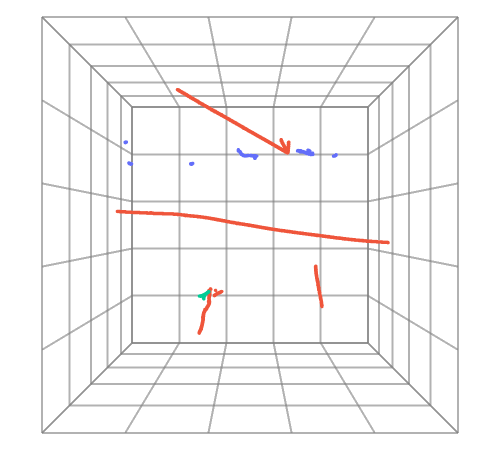} &
    \includegraphics[width=0.135\linewidth]{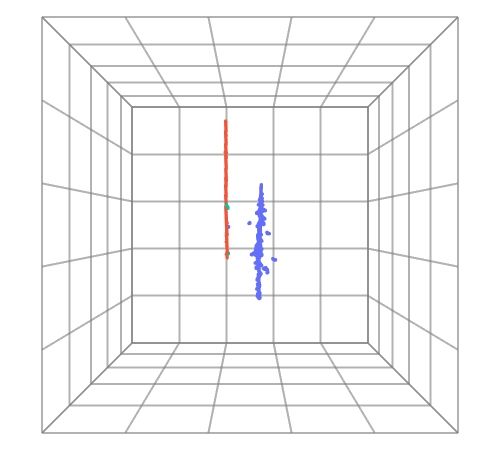} \\
\end{tabular}
    \caption{\textit{Visual depictions of 7 PILArNet-M events.} For each event, voxels are colored according to which integer identifiers they share, for each label type (fragment / cluster ID, group ID, interaction ID, and semantic ID). For ease of visualization, we remove low energy depositions, which would greatly degrade the ability to visualize what each identifier means.}
    \label{fig:ids}
\end{figure*}

The PILArNet-M dataset contains 5.2B points containing energy deposited from 28,251,859 individual particle trajectories, over 1,210,080 individual events. As described Sec.~\ref{sec:data} of the main text, each voxel contains information corresponding to not just the energy deposited, but also a fragment ID, group ID, interaction ID, and semantic type. Though we just use the semantic type in this work, we foresee the other IDs being quite useful for other tasks, like, i.e., latent clustering for individual particle (i.e., group ID) identification. Dataset-wide statistics are shown in Figure \ref{fig:statistics}. We also offer visualizations of individual events, where voxels with the same colors correspond to sharing different IDs, in Figure \ref{fig:ids}. See \cite{adams2020pilarnetpublicdatasetparticle} for detailed information about this dataset.

\newpage
\section{C-NMS}
\label{app:cnms}

Non-Maximum Suppression (NMS) algorithms are widely used to resolve redundancies in overlapping spatial proposals. In object detection, for instance, NMS iteratively selects the bounding box with the highest confidence score and suppresses neighboring proposals that exceed a predefined overlap threshold (e.g., Intersection-over-Union $>$ 0.5). This ensures that only the most relevant detections are retained, balancing precision and recall. 

Our \textbf{Centrality-based NMS (C-NMS)} adapts this philosophy to 3D point cloud grouping. Instead of bounding boxes, we process spherical groups centered on points selected through farthest point sampling (FPS). The algorithm prioritizes spheres whose centroids are most central within their local neighborhoods, iteratively suppressing overlapping candidates based on a tunable \textit{overlap factor} \( f \in [0, 1] \). Two spheres are considered overlapping if the distance between their centers is less than \( 2rf \), where \( r \) is the sphere radius. Lower values of \( f \) tolerate more overlap (yielding denser groups), while \( f \to 1 \) enforces strict non-overlap. 

Implementationally, C-NMS leverages CUDA-accelerated spatial queries via \texttt{pytorch3d}'s ball tree for efficient neighborhood searches, combined with a simple custom kernel to batch-process the iterative suppression across events. This balances computational efficiency with flexibility, allowing dynamic adjustment of group density without predefined cluster counts. 

The subsequent sections compare C-NMS+ball query against conventional FPS+\{\(k\)-NN, ball query\} methods, evaluating their trade-offs in coverage (\% missed points), and redundancy (\% duplicated points). By explicitly optimizing for minimal overlap, C-NMS addresses a limitation of prior grouping strategies when applied to irregular 3D geometries like particle trajectories.

The C-NMS algorithm is described in Algorithm \ref{alg:cnms}.

\renewcommand{\algorithmicrequire}{\textbf{Input:}}
\renewcommand{\algorithmicensure}{\textbf{Output:}}
\begin{algorithm}[H]
\caption{Centrality-Based Non-Maximum Suppression (C-NMS)}
\label{alg:cnms}
\begin{algorithmic}[1]
\REQUIRE $C \in \mathbb{R}^{P \times 3}$: Set of $P$ centroid coordinates, $r \in \mathbb{R}^{+}$: Sphere radius, $f \in [0, 1]$: Overlap factor.
\ENSURE $C' \in \mathbb{R}^{P' \times 3}$: Culled set of centroids, where $P' \leq P$.

\STATE \algcomment{Minimum overlapping distance}
\STATE $r_q \gets 2 \cdot r \cdot f$ 
\STATE \algcomment{Find neighbors within distance $r_q$}
\STATE $N \gets \emptyset$
\FOR{$c_i \in C$}
    \STATE $N_i \gets \{c_j \in C \mid \|c_i - c_j\| \leq r_q\}$
    \STATE $N \gets N \cup \{N_i\}$
\ENDFOR

\STATE \algcomment{Centrality scores for each centroid}
\STATE $S \gets \{|N_i| \;\forall i \in \{1, \dots, P\}\}$

\STATE \algcomment{Sort centroids by scores}
\STATE $\text{indices} \gets \text{argsort}(S, \text{descending=True})$
\STATE $C_{\text{sorted}} \gets C[\text{indices}]$

\STATE \algcomment{Perform iterative suppression}
\STATE $C' \gets \emptyset$
\STATE $\text{suppressed} \gets \text{zeros}(P)$ 
\FOR{$c_i \in C_{\text{sorted}}$}
    \IF{$\text{suppressed}[i] = 0$}
        \STATE $C' \gets C' \cup \{c_i\}$
        \FOR{$c_j \in N_i$ \textbf{and} $c_j \neq c_i$}
            \STATE $\text{suppressed}[j] \gets 1$ 
        \ENDFOR
    \ENDIF
\ENDFOR
\end{algorithmic}
\end{algorithm}

\subsection{Comparison of Grouping Strategies}

\subsubsection{Pareto Frontier of Grouping Methodologies.} 
\label{app:pareto}
A central challenge in adapting Masked Autoencoders (MAEs) to 3D particle trajectory reconstruction is optimizing the trade-off between two competing objectives: \textbf{comprehensive coverage} (minimizing ungrouped points) and \textbf{group distinctness} (minimizing overlap). We formalize this through two metrics:

\vspace{0.2cm}

\noindent
\textit{1. Missed Points Ratio}  
\begin{equation}
M = \left(1 - \frac{|\mathcal{X}_{\text{grouped}} \cap \mathcal{X}|}{|\mathcal{X}|}\right), 
\end{equation}  
where $\mathcal{X}$ is the full set of points in an event, and $\mathcal{X}_{\text{grouped}} \subseteq \mathcal{X}$ represents points assigned to any group.  

\vspace{0.2cm}

\noindent
\textit{2. Duplicated Points Ratio}  
\begin{equation}
D = \frac{|\{x \in \mathcal{X} \mid x \text{ appears in }> 1\; \text{group}\}|}{|\mathcal{X}|}. 
\end{equation}

\vspace{0.2cm}

\noindent
These metrics address critical failure modes in MAE pre-training:  
\begin{itemize}
    \vspace{-0.2cm}
    \item \textit{High $M$}: Fails to include trajectory segments in any group, degrading reconstruction quality.
    \vspace{-0.2cm}
    \item \textit{High $D$}: Allows leakage between masked and visible regions, artificially simplifying the task (unlike image MAEs, where masks are strictly non-overlapping).  
\end{itemize}

\noindent
Prior work mitigates leakage by using large masking ratios (greater than 60\%), but this only \textit{reduces} overlap probability rather than eliminating it. For particle trajectory datasets—where spatial correlations are strong—even minor overlaps risk exposing the model to trivial reconstruction shortcuts. Our \textbf{C-NMS} strategy directly optimizes the Pareto frontier of $M$ vs. $D$ (see Figure \ref{fig:pareto} in the main text), ensuring groups are both \textit{complete} (low missed points) and \textit{isolated} (low duplication). This mirrors the non-overlapping structure of successful 2D image masking while adapting to the irregular geometry of particle clouds.

\begin{figure}[ht!!!]
    \centering
    \includegraphics[width=1\linewidth]{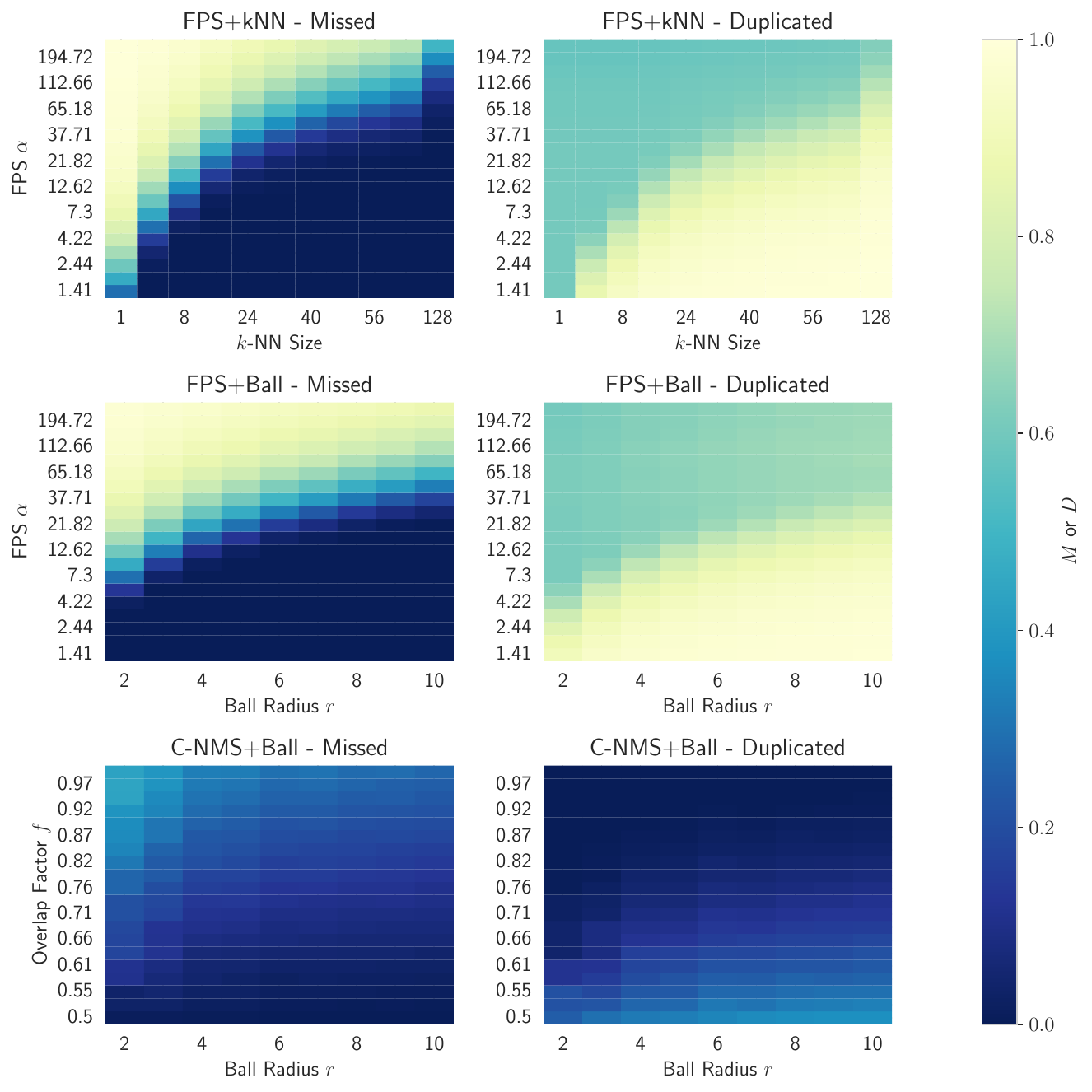}
    \vspace{-10pt}
    \caption{\textit{Detailed grouping hyperparameter sweep.} To find the pareto frontier of grouping methodologies, we sweep over an array of reasonable grouping parameters for each method, reporting the missed ratio $M$ and duplicated ratio $D$ heatmaps for each hyperparameter pair. Clearly, the na\"ive approaches result in an extreme number of duplicated points in any configuration, while the C-NMS-based grouping method empirically guarantees much smaller amounts of each quantity.}
    \label{fig:hyperparam_sweep}
\end{figure}

To systematically compare grouping strategies, we perform parameter sweeps for each method. Initial group centers are sampled via FPS (2048 per event). These parameter ranges are described in Table \ref{tab:param_sweep}. For FPS-based methods, the number of samples per event are defined as $\lceil |\mathcal{X}| / \alpha \rceil$ where $|\mathcal{X}|$ is the number of points in the event. Sweeps generate $80-220$ configurations per method, with Pareto frontiers computed from the missed/duplicated ratio metrics defined in App. \ref{app:pareto}. Figure \ref{fig:hyperparam_sweep} provides array of figures showing results for each metric across these sweeps.

\begin{table*}[t]
\centering
\caption{Parameter configurations for grouping method sweeps}
\label{tab:param_sweep}
\begin{tabular}{ll}
\toprule
\textbf{Method} & \textbf{Parameters} \\ \midrule
C-NMS + ball query & 
\begin{tabular}[t]{@{}l@{}}
Sphere radius: $r \in \text{arange}(2, 11, 1)$ voxels (1 voxel = 3~mm\textsuperscript{3}) \\
Overlap factor: $f \in \text{linspace}(0.5,\ 1.0,\ 20)$
\end{tabular} \\ \midrule 

FPS + $k$-NN & 
\begin{tabular}[t]{@{}l@{}}
Group count: $\alpha \in \text{logspace}(0.5,\ 8,\ 20,\ \text{base}=2)$ \\
Neighborhood size: $k \in \{1,\ 4,\ 8,\ 16,\ 24,\ 32,\ 40,\ 48,\ 56,\ 64,\ 128\}$
\end{tabular} \\ \midrule

FPS + ball query & 
\begin{tabular}[t]{@{}l@{}}
Group count: $\alpha \in \text{logspace}(0.5,\ 8,\ 20,\ \text{base}=2)$ \\
Query radius: $r \in \text{arange}(2,\ 11,\ 1)$ voxels \\
\end{tabular} \\
\bottomrule
\end{tabular}
\vskip -0.1in
\end{table*}

\subsubsection{Ablation study for pre-training with different grouping strategies}

We perform an ablation study on various grouping strategies to quantitatively understand the negative effects that result from using prior grouping methods. We pre-train PoLAr-MAE for 100,000 steps ($\sim$12 epochs) using FPS+{$k$-NN, ball query} and using C-NMS. We further show ablations on the C-NMS method itself by using extreme overlap factors (0.1, 1.0) and ablating the initial FPS step. We select optimal parameters for each setup using the Pareto frontier study in Appendix~\ref{app:pareto}.

Table \ref{tab:patch_ablation} provides per-class and overall $F_1$ scores from training linear SVMs on the token outputs, as described in Section~\ref{sec:eval} of the main paper. Clearly, using volumetric tokenization greatly increases the effectiveness of the representations learned. Additionally, we also show that it is not necessary to use FPS before C-NMS plus ball query, and not doing so seems to increase representation power by some amount. We do not supply further tests on these pre-trained models on downstream tasks, as we believe the main results in the paper show that the SVM evaluation is generally a good predictor of downstream performance on semantic segmentation fine-tuning.

\begin{table*}[t]
\centering
\caption{\textit{Ablation study on patch grouping strategies.}  
We report the overall $F_{1}$ score and per-class $F_{1}$ on the validation set.}
\label{tab:patch_ablation}
\begin{tabular}{llccccc}
\toprule
& & & \multicolumn{4}{c}{Per-class $F_{1}$}\\
\cmidrule(lr){3-7}
Method & Steps & $F_1$ & Track & Shower & Delta & Michel \\
& & \colorsquare{track} & \colorsquare{shower} & \colorsquare{delta} & \colorsquare{michel}\\
\midrule
FPS + kNN & 100k & 0.469 &  0.840 & 0.678 & 0.339 & 0.022 \\
FPS + Ball query & 100k & 0.555 & 0.943 & 0.797 & 0.417 & 0.061 \\
(Ours) FPS + C-NMS + Ball query & 100k & 0.687 & 0.988 & 0.966 & 0.449 & 0.344 \\
\cmidrule(lr){1-7}
$\rightarrow$ Overlap $f{=}0.1$ & 100k & \multicolumn{5}{c}{\textit{OOM}\footnote{OOM: Out of memory.}} \\
$\rightarrow$ Overlap $f{=}1.0$ & 100k & 0.669 & 0.989 & 0.967 & 0.381 & 0.339 \\
$\rightarrow$ C-NMS + Ball query & 100k & 0.698 & 0.989 & 0.968 & 0.454 & 0.383 \\
\bottomrule
\end{tabular}
\end{table*}

\subsubsection{Runtime comparison with different grouping strategies}

\begin{figure*}
    \centering
    \includegraphics[width=1\linewidth]{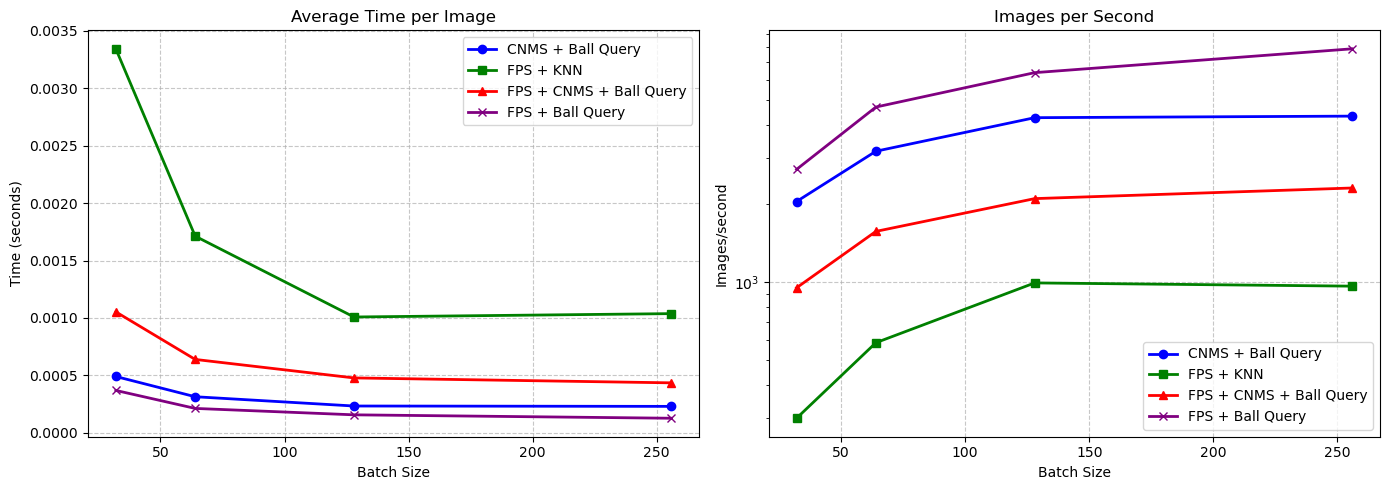}
    \caption{\textit{Computational efficiency of grouping methods.} \textbf{(Left.)} Average time spent to group a single image on the GPU. \textbf{(Right.)} Average number of images able to be grouped per second on the GPU. Our grouping method (C-NMS) is placed between the ball query-based (best) and $k$-NN-based (worst) methods.}
    \label{fig:efficiency}
\end{figure*}

Computational efficiency is an important part of foundation model architectures, since low efficiency can considerably increase the pre-training time. We test the efficiency of the C-NMS algorithm compared to prior methods by comparing runtimes of grouping a different number of events. We use the optimal parameters found in the hyperparameter search in App.~\ref{app:pareto}. Fig.~\ref{fig:efficiency} shows the relative efficiency of each grouping strategy as a function of batch size. The C-NMS method sits between the efficiencies of the standard $k$-NN and ball query methods, with FPS plus ball query being the most efficient, at $\sim4,000$ images per second at a batch size of 128. C-NMS plus ball query and without/with the initial FPS step are the second and third most efficient, grouping $\sim 3,300$ and $\sim 2,000$ images per second with the same batch size. Finally, the FPS plus $k$-NN approach is the slowest, at $\sim 1,000$ images per second. Considering the heavy use of $k$-NN based approaches in 3D computer vision literature, we argue the additional computational burden of using C-NMS is small relative to the overall performance of the grouping methods. 

\section{Hyperparameters}
\label{app:hyperparameters}

We provide detailed hyperparameter information in the following sections. The full implementation of these models, as well as training code, can be found at \url{https://github.com/DeepLearnPhysics/PoLAr-MAE}.

\begin{table*}[h!!!]
\centering
\caption{Model Architecture}
\begin{tabular}{ll}
\toprule
\textbf{Component} & \textbf{Parameters} \\
\midrule
\textbf{Tokenizer \& Masking} & Transformer Encoder \\
Num Channels & 4 \\
Num Seed Groups & 2048 \\
Context Length & 512 \\
Group Max Points $K_\text{max}$ & 32 \\
Group Radius $r$ &  5 voxels (15 mm) \\
Overlap factor & 0.73 \\
Group Size Reduction Method & FPS \\
Masking Method & Random \\
Masking Ratio & 0.6 \\
\midrule
\textbf{Encoder} & Transformer Encoder \\
Base Architecture & ViT-S \\
Activation Layer & GELU \\
Normalization Layer & LayerNorm \\
Normalization Type & Pre + Post \\
Embed Dim & 384 \\
Depth & 12 \\
Num Heads& 6 \\
MLP Ratio & 4 \\
Final LayerNorm & False \\
Add Position Embeddings at Every Layer & True \\
Dropout Rate & 0.0 \\
MLP / Projection Dropout Rate & 0.0 \\
Attention Dropout & 0.05 \\
Stochastic Depth Rate & 0.25 \\
\midrule
\textbf{Decoder} & Transformer Decoder \\
Base Architecture & ViT-S \\
Activation Layer & GELU \\
Normalization Layer & LayerNorm \\
Normalization Type & Pre + Post \\
Embed Dim & 384 \\
Depth & 4 \\
Num Heads& 6 \\
MLP Ratio & 4 \\
Final LayerNorm & True \\
Add Position Embeddings at Every Layer & True \\
Dropout Rate & 0.0 \\
MLP / Projection Dropout Rate & 0.0 \\
Attention Dropout & 0.05 \\
Stochastic Depth Rate & 0.25 \\
\bottomrule
\end{tabular}
\end{table*}

\begin{table}[h!!!]
\centering
\caption{Optimizer \& Learning Rate Schedule}
\begin{tabular}{ll}
\toprule
\textbf{Parameter} & \textbf{Value} \\
\midrule
Optimizer & AdamW \\
Base Learning Rate & 1e-4 \\
Weight Decay & 0.05 \\
$\beta_1$ & 0.9 \\
$\beta_2$ & 0.99 \\
Distributed Training Strategy & DDP \\
Gradient Clip & 3.0 \\
LR Schedule & Linear Warmup + Cosine Decay \\
Total Steps & \{100k, 500k\} \\
Warmup Steps & 12,500 \\
Warmup Start LR & 8.6e-6 \\
Cosine Eta Min & 8.6e-6 \\
Scheduler Steps & Per Optimization Step \\
Effective Batch Size & 128 \\
Training precision & BF16-Mixed \\
\bottomrule
\end{tabular}
\end{table}

\subsection{Grouping strategy}

\paragraph{Overlap factor $f$. }We chose a group radius equal to 5 voxels (15 mm) such that a 1m track will contain ~67 mm, and because it is approximately the characteristic distance of trajectory morphologies. To find the best value for the overlap factor $f$, we run a sweep over values of $f$, and examine the missed points ratio $M$ and duplicated points ratio $D$ as defined in Appendix \ref{app:pareto}. Fig. \ref{fig:f_sweep} shows this trade-off for different values of $f$. The value of $f$ that jointly minimizes both these scores is $f=0.73,$ and hence we use this value when pretraining with this grouping strategy.

\begin{figure}
    \centering
    \includegraphics[width=0.6\linewidth]{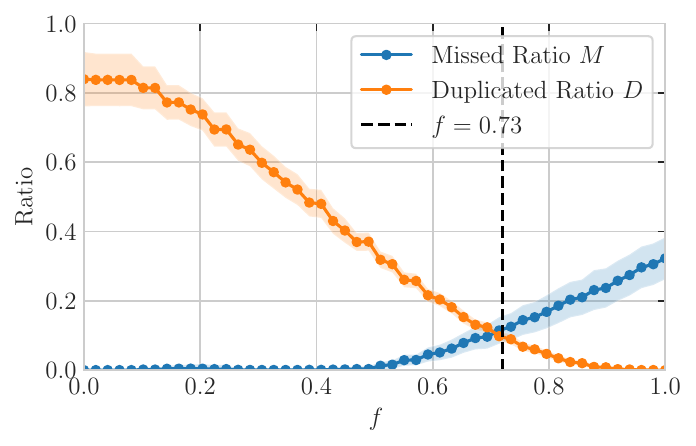}
    \vspace{-10pt}
    \caption{\textit{C-NMS overlap parameter $f$ sweep.} We find the optimal $f$ parameter for our group radius of 5 voxels by minimizing both the missed ratio $M$ and duplicated ratio $D$. We find $f=0.73$ to be approximately optimal. Shaded regions represent 1 standard deviation across 32 LArTPC events.}
    \label{fig:f_sweep}
\end{figure}

\begin{figure}
    \centering
    \includegraphics[width=0.6\linewidth]{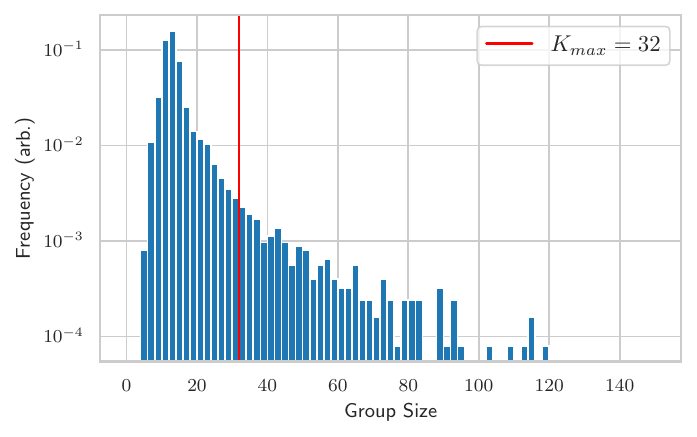}
    \vspace{-10pt}
    \caption{\textit{Group size statistics.} We show the number of points found in each group given a group radius of 5 voxels and $f=0.73$.}
    \label{fig:group_sizes}
\end{figure}

\paragraph{Group size.} We determine the number of points allowed per group based on the distribution of groups per event shown in Fig. \ref{fig:group_lens}, as well as our embedding size. This hyperparameter is extremely important, as our model is primarily bottlenecked by memory constraints from the point group embedding module. Within the mini-PointNet, each individual point is lifted to a large embedding dimension (384 for this work), and as such larger and larger group sizes greatly increase the memory budget required to train the model. Fig. \ref{fig:group_sizes} shows a distribution of group sizes across a number of LArTPC events. Here, the group sizes peak around 16 points, but there is a long tail that extends to $\sim~$150 points. To keep as many points as possible without running into memory issues, we set $K_\text{max}$, the maximum number of points to take within a group, to 32. For groups with more than 32 points, we randomly sample 32 points with farthest point sampling.

\section{Pretraining Results}
\label{app:pretraining}

\subsection{Spatial Debiasing of Features}
\label{app:spatial_debiasing}

Given a single event, we perform principal component analysis to cast each token's feature vector to three dimensions. We then normalize the final vectors to $[0,1]^3$ and interpret them as RGB channel values. Before PCA, however, we must remove spatial debiasing of individual vectors. Empirically, we found that without this debiasing procedure, tokens were mostly colored as a simple function of their 3-D position. Contrary to usual Vision Transformers, we inject positional encoding at each Transformer layer. This results in embeddings that are rich in spatial context -- as needed for the masked modeling pretext task, which heavily relies on the spatial position of masked tokens for reconstruction. However, we are interested in semantic differences between tokens. Thus, we train a simple linear model to predict the position $\textbf{x}=(x,y,z)$ of individual tokens given their feature vector $\textbf{z}$. We then compute the residual $R$ between the original and predicted feature vectors, and finally perform PCA. 

Concretely, let the dataset
\[
\mathcal{D}=\bigl\{(\mathbf{x}_{i},\mathbf{z}_{i})\bigr\}_{i=1}^{N},
\qquad 
\mathbf{x}_{i}\in\mathbb{R}^{3},\;
\mathbf{z}_{i}\in\mathbb{R}^{D_\text{embed}},\; D_\text{embed}=384.
\]
We model each latent vector as an affine function of its 3‑D position:
\[
\hat{\mathbf{z}}_{i}=W\mathbf{x}_{i}+\mathbf{b},
\]
where \(W\in\mathbb{R}^{D_\text{embed}\times 3}\) and \(\mathbf{b}\in\mathbb{R}^{D_\text{embed}}\) are learned parameters.
The optimal \((W^{\star},\mathbf{b}^{\star})\) are obtained by ordinary least squares:
\[
(W^{\star},\mathbf{b}^{\star})
=\arg\min_{W,\mathbf{b}}
\sum_{i=1}^{N}\bigl\lVert\mathbf{z}_{i}-W\mathbf{x}_{i}-\mathbf{b}\bigr\rVert_{2}^{2}.
\]
In matrix form, we augment the position matrix with a bias column:
\[
X=\begin{bmatrix}\mathbf{x}_{1}^{\!\top}\\[-2pt]\vdots\\\mathbf{x}_{N}^{\!\top}\end{bmatrix}\in\mathbb{R}^{N\times 3},
\quad
\tilde{X}=\begin{bmatrix}X & \mathbf{1}\end{bmatrix}\in\mathbb{R}^{N\times 4},
\quad
Z=\begin{bmatrix}\mathbf{z}_{1}^{\!\top}\\[-2pt]\vdots\\\mathbf{z}_{N}^{\!\top}\end{bmatrix}\in\mathbb{R}^{N\times D_\text{embed}}.
\]
Let \(\tilde{W}=\begin{bmatrix}W & \mathbf{b}\end{bmatrix}\in\mathbb{R}^{D_\text{embed}\times 4}\).
The closed‑form solution is
\[
\tilde{W}^{\star}=\bigl(\tilde{X}^{\top}\tilde{X}\bigr)^{-1}Z^{\top}\tilde{X}.
\]

For each sample, subtract the fitted component to obtain the residual
\[
\mathbf{r}_{i}
=\mathbf{z}_{i}-\hat{\mathbf{z}}_{i}
=\mathbf{z}_{i}-W^{\star}\mathbf{x}_{i}-\mathbf{b}^{\star}.
\]
Collecting all residuals,
\[
R=Z-\tilde{X}\tilde{W}^{\star\top}\in\mathbb{R}^{N\times D_\text{embed}},
\]
which can now be used for downstream analysis without linear spatial bias.

\subsection{Extra PCA Results}

\begin{figure*}[t]
    \centering
    \begin{tabular}{c@{\hspace{1mm}}*{8}{c}@{}} 
    & \textsc{Event 4} & \textsc{Event 5} & \textsc{Event 6} & \textsc{Event 7} & \textsc{Event 8} & \textsc{Event 9} & \textsc{Event 10} \\
    \rotatebox[origin=l]{90}{\textsc{~~~Random}} & 
    \includegraphics[width=0.125\linewidth]{figs/latent/latent_space_random_0}&
    \includegraphics[width=0.125\linewidth]{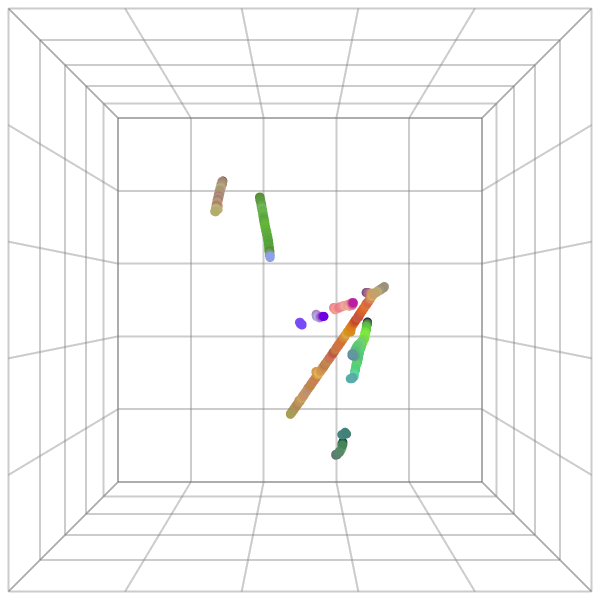} &
    \includegraphics[width=0.125\linewidth]{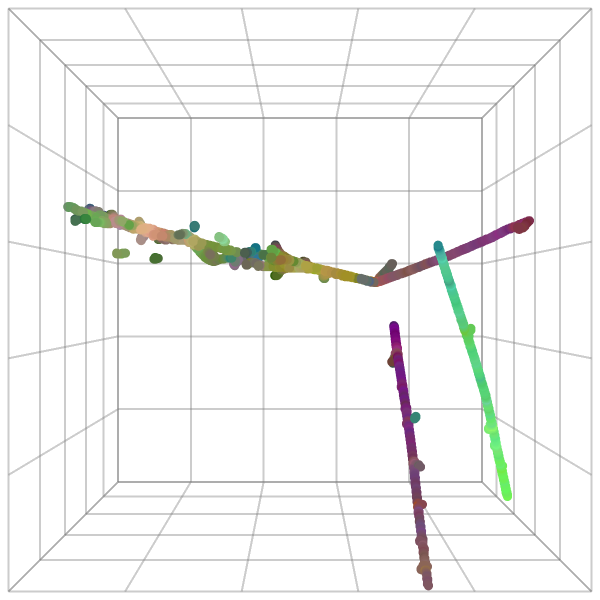} &
    \includegraphics[width=0.125\linewidth]{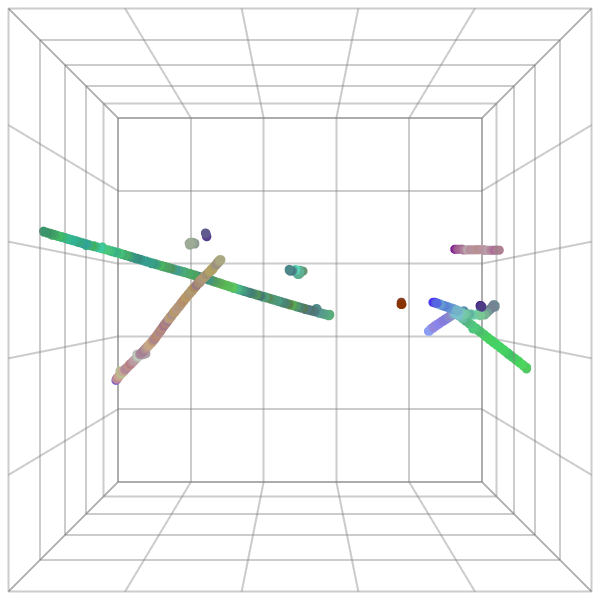} &
    \includegraphics[width=0.125\linewidth]{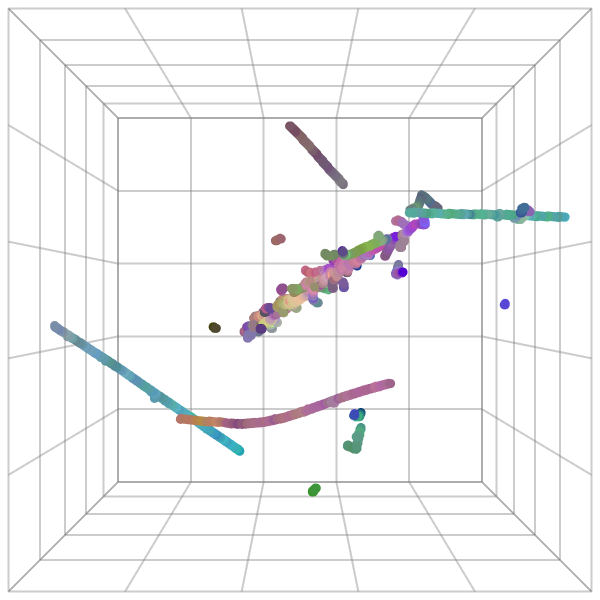} &
    \includegraphics[width=0.125\linewidth]{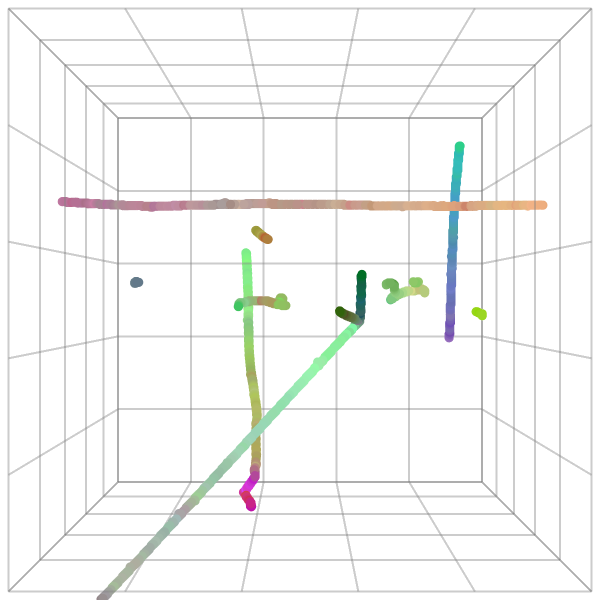} &
    \includegraphics[width=0.125\linewidth]{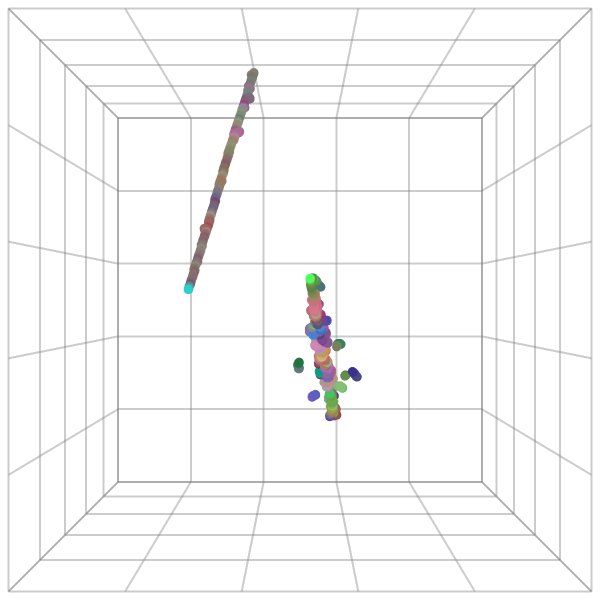} &

    \\
    
    \rotatebox[origin=l]{90}{\textsc{~Point-MAE}} & 
    \includegraphics[width=0.125\linewidth]{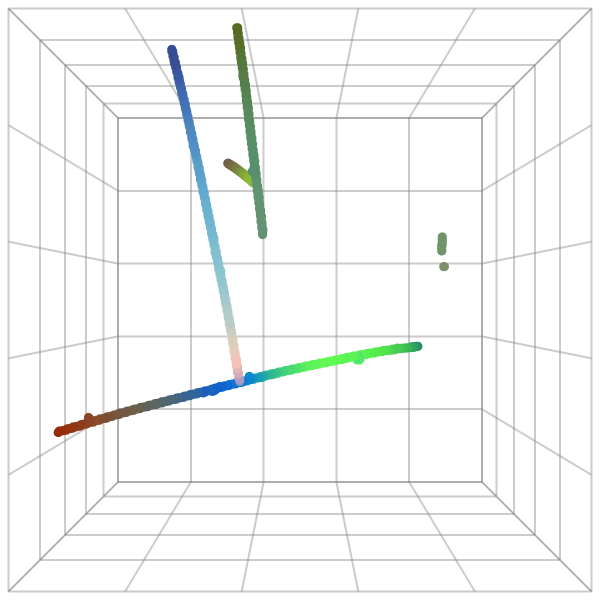}&
    \includegraphics[width=0.125\linewidth]{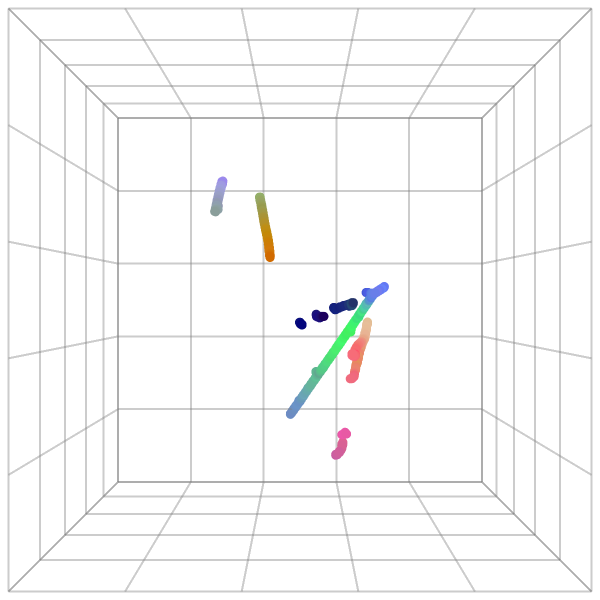} &
    \includegraphics[width=0.125\linewidth]{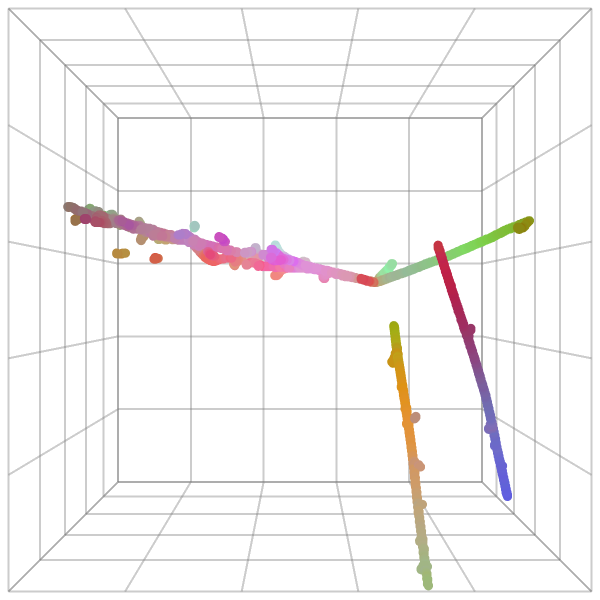} &
    \includegraphics[width=0.125\linewidth]{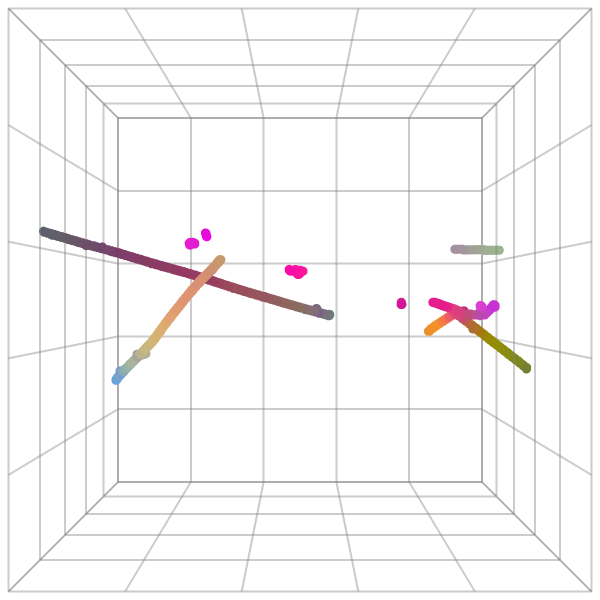} &
    \includegraphics[width=0.125\linewidth]{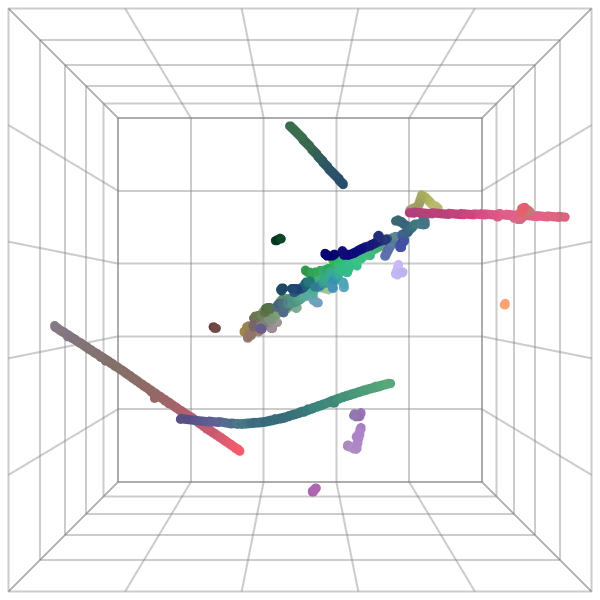} &
    \includegraphics[width=0.125\linewidth]{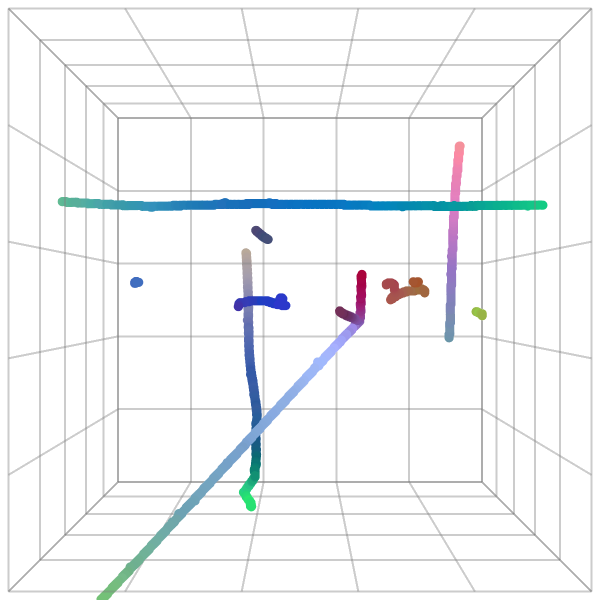} &
    \includegraphics[width=0.125\linewidth]{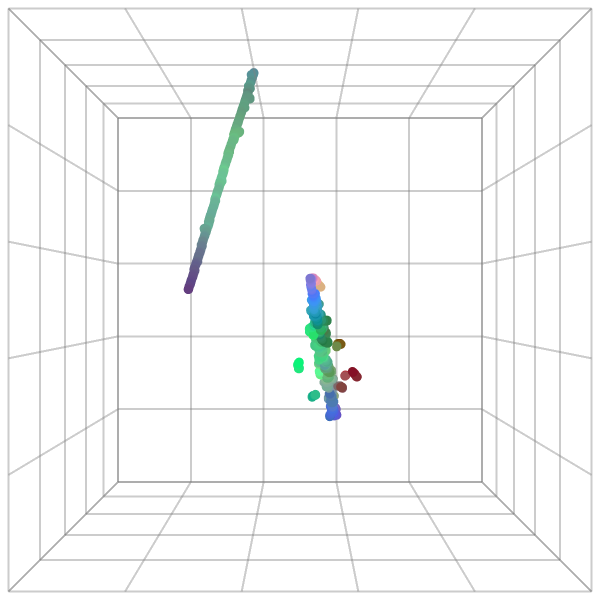} &

    \\
    \rotatebox[origin=l]{90}{\textsc{PoLAr-MAE}} & 
    \includegraphics[width=0.125\linewidth]{figs/latent/latent_space_multitask_0}&
    \includegraphics[width=0.125\linewidth]{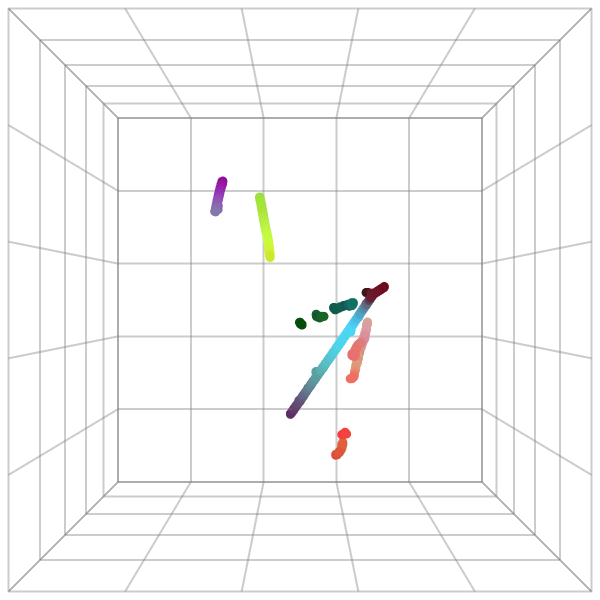} &
    \includegraphics[width=0.125\linewidth]{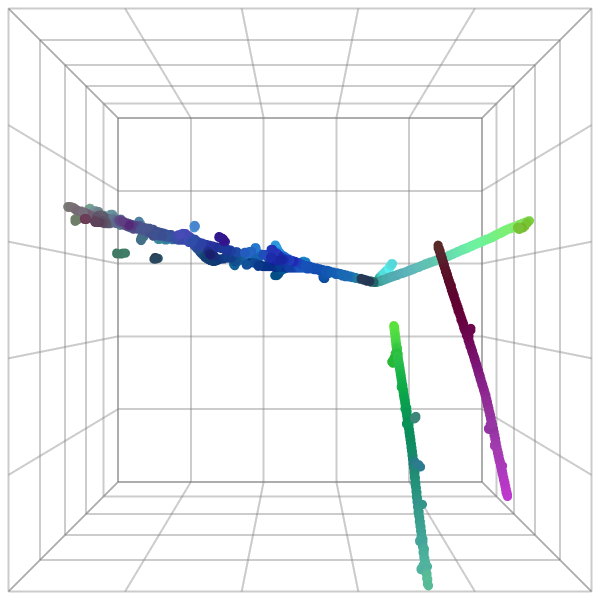} &
    \includegraphics[width=0.125\linewidth]{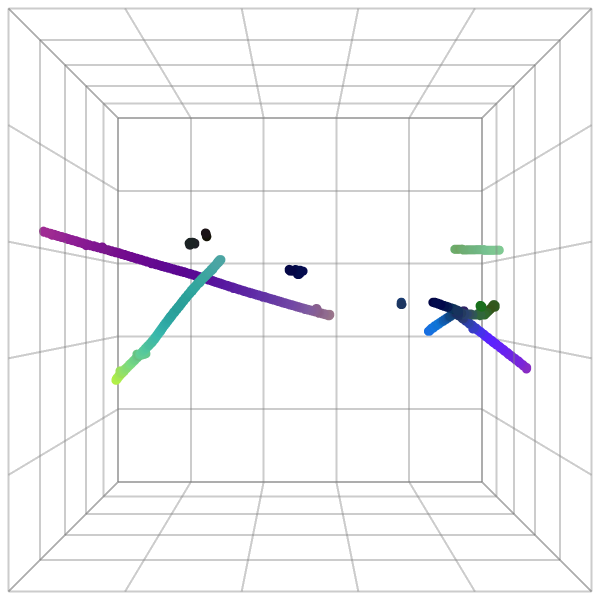} &
    \includegraphics[width=0.125\linewidth]{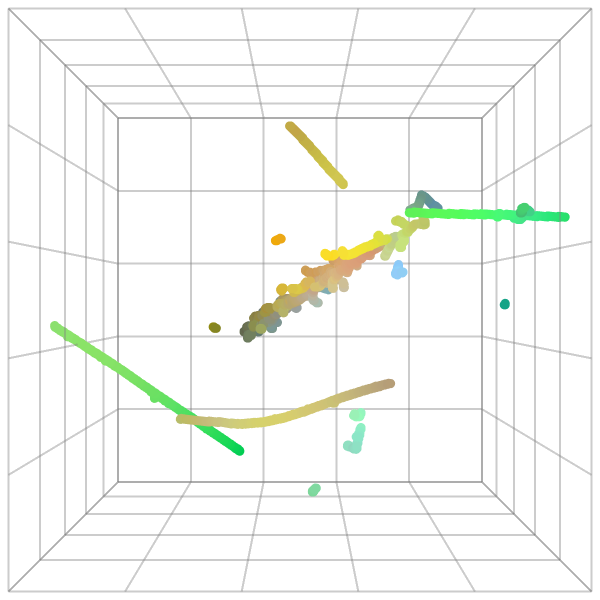} &
    \includegraphics[width=0.125\linewidth]{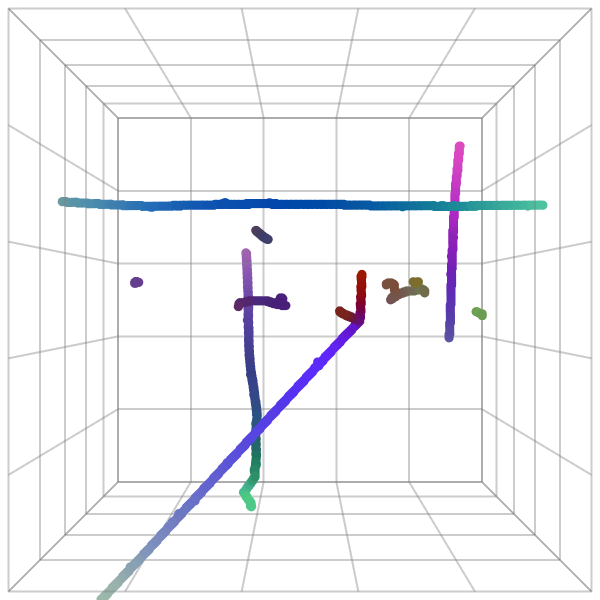} &
    \includegraphics[width=0.125\linewidth]{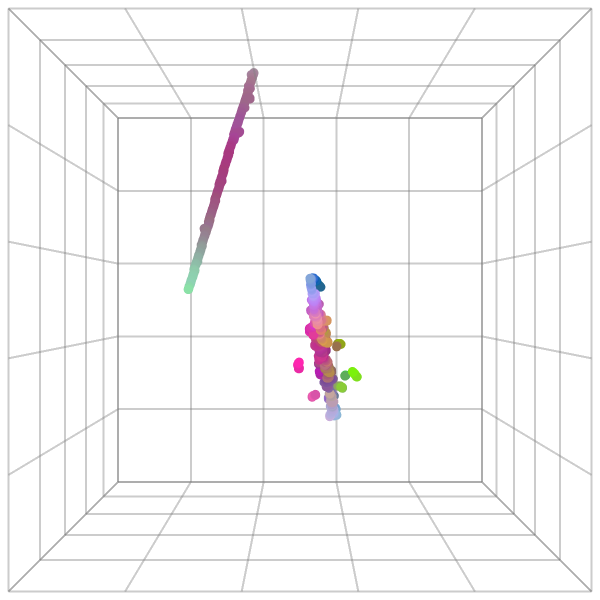} \\
\end{tabular}
    \caption{\textit{Additional visualizations of learned representations after pretraining.} See the caption of Fig. \ref{fig:latent_comparison} for an explanation.}
    \label{fig:latent_comparison_extras}
\end{figure*}

In Figure \ref{fig:latent_comparison_extras}, we show additional visualizations of representations learned by both Point-MAE and PoLAr-MAE.

\section{Segmentation Results}
\label{app:segsem}

\begin{figure*}[h]
    \centering
    \begin{tabular}{c@{\hspace{1mm}}*{8}{c}@{}} 
    & \textsc{Event 4} & \textsc{Event 5} & \textsc{Event 6} & \textsc{Event 7} & \textsc{Event 8} & \textsc{Event 9} & \textsc{Event 10} \\
    \rotatebox[origin=l]{90}{\textsc{~~~Truth}} & 
    \includegraphics[width=0.125\linewidth]{figs/segmentation/truth_0}&
    \includegraphics[width=0.125\linewidth]{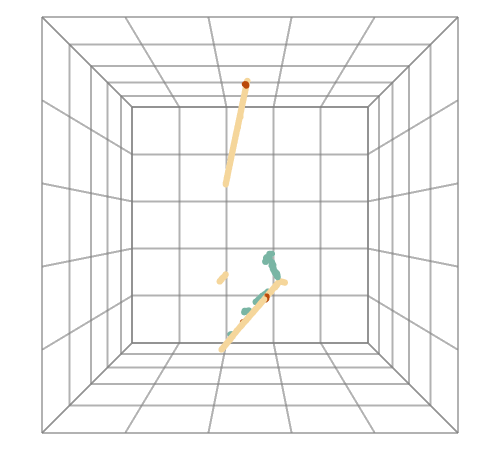} &
    \includegraphics[width=0.125\linewidth]{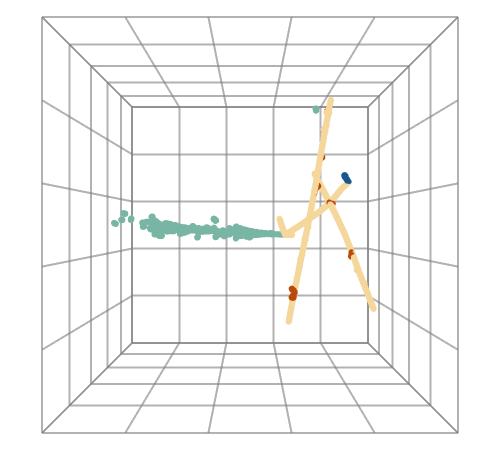} &
    \includegraphics[width=0.125\linewidth]{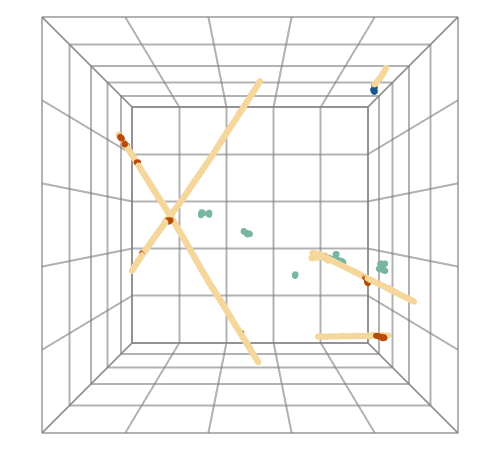} &
    \includegraphics[width=0.125\linewidth]{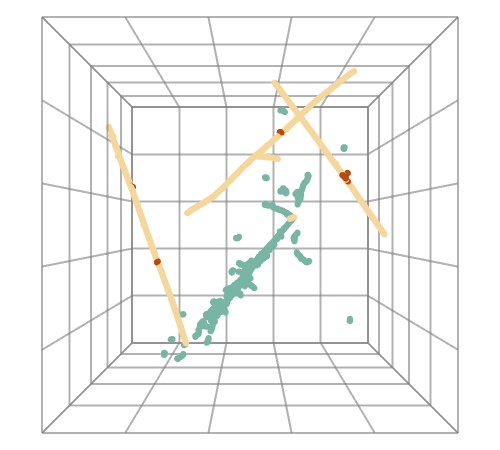} &
    \includegraphics[width=0.125\linewidth]{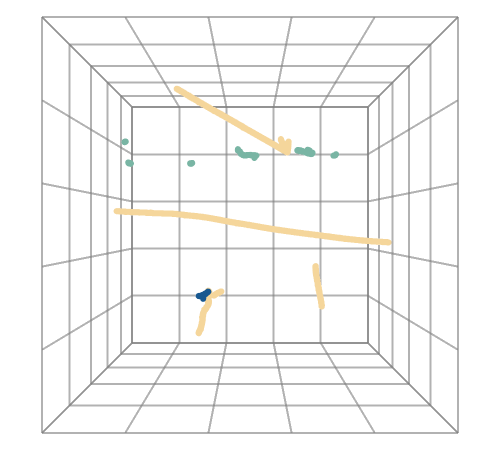} &
    \includegraphics[width=0.125\linewidth]{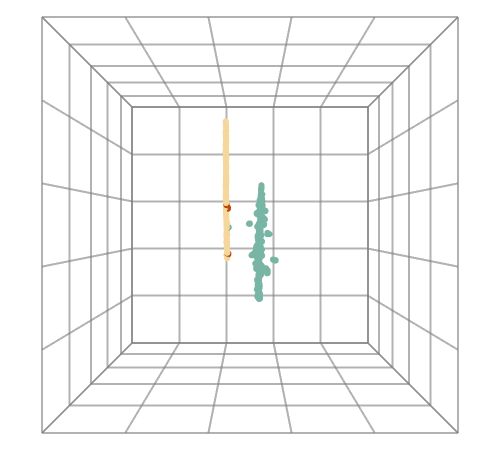} &

    \\
    
    \rotatebox[origin=l]{90}{\textsc{~~PL PEFT}} & 
    \includegraphics[width=0.125\linewidth]{figs/segmentation/mae_multitask_peft_pred_0}&
    \includegraphics[width=0.125\linewidth]{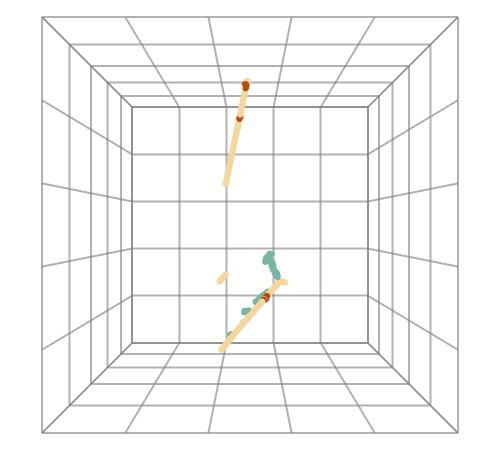} &
    \includegraphics[width=0.125\linewidth]{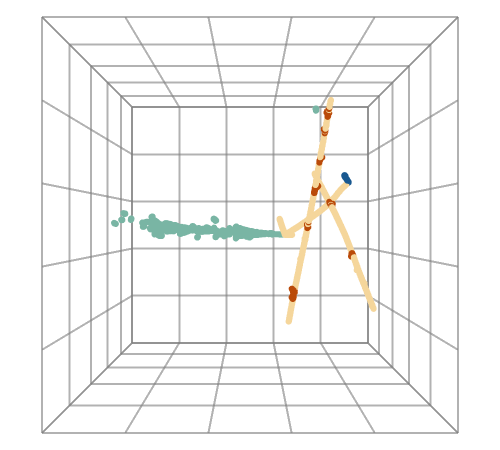} &
    \includegraphics[width=0.125\linewidth]{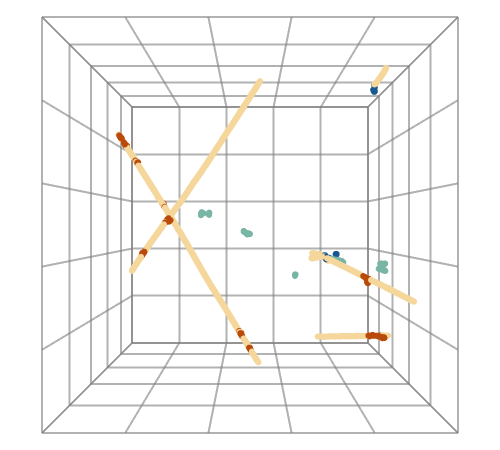} &
    \includegraphics[width=0.125\linewidth]{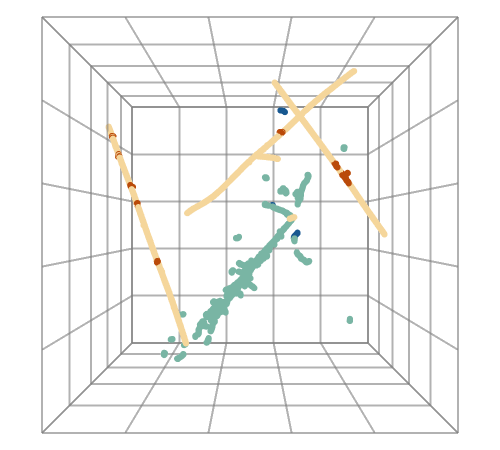} &
    \includegraphics[width=0.125\linewidth]{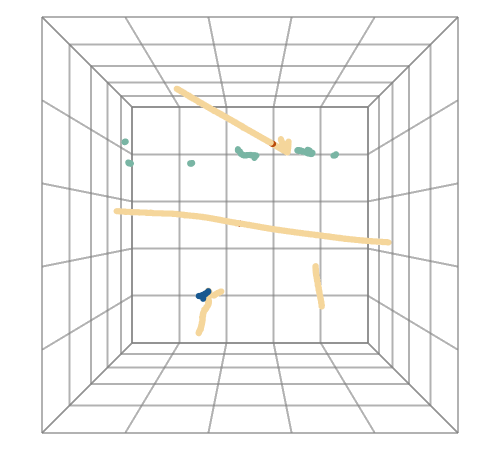} &
    \includegraphics[width=0.125\linewidth]{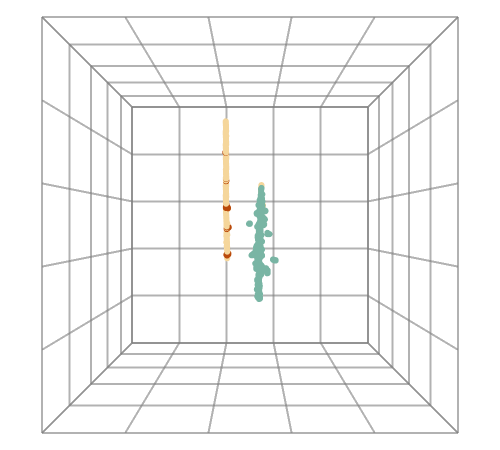} &

    \\
    \rotatebox[origin=l]{90}{\textsc{~PM PEFT}} & 
    \includegraphics[width=0.125\linewidth]{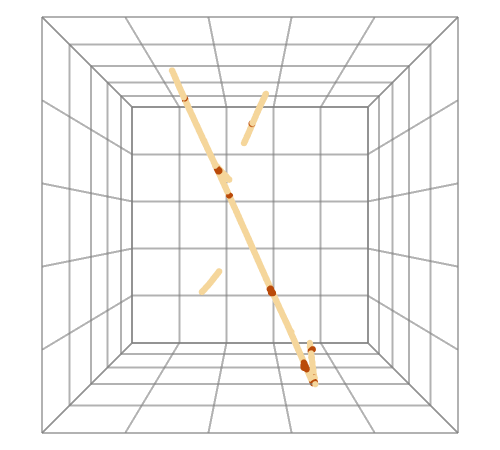}&
    \includegraphics[width=0.125\linewidth]{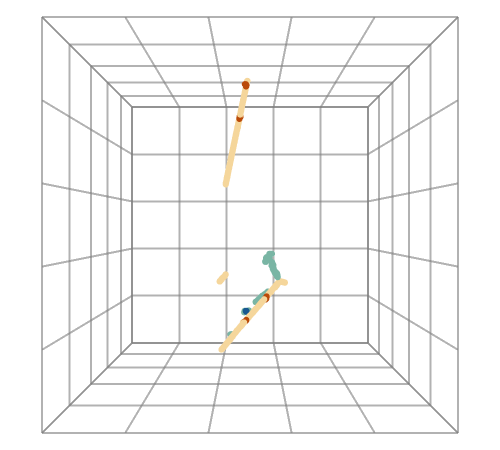} &
    \includegraphics[width=0.125\linewidth]{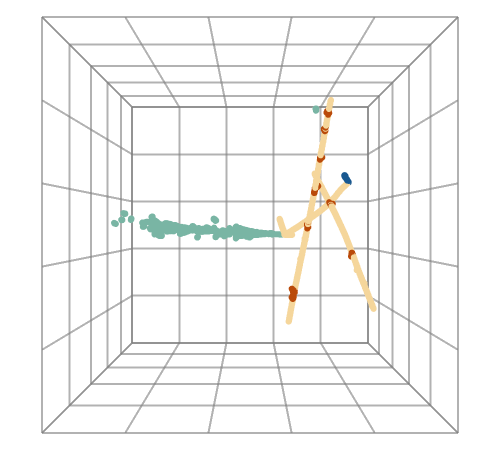} &
    \includegraphics[width=0.125\linewidth]{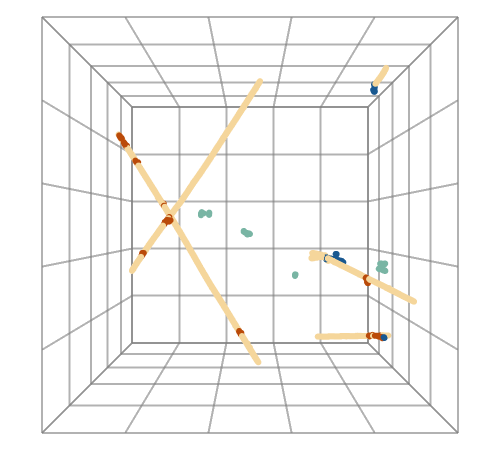} &
    \includegraphics[width=0.125\linewidth]{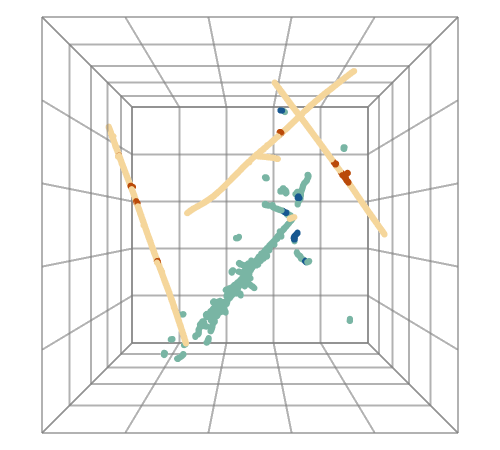} &
    \includegraphics[width=0.125\linewidth]{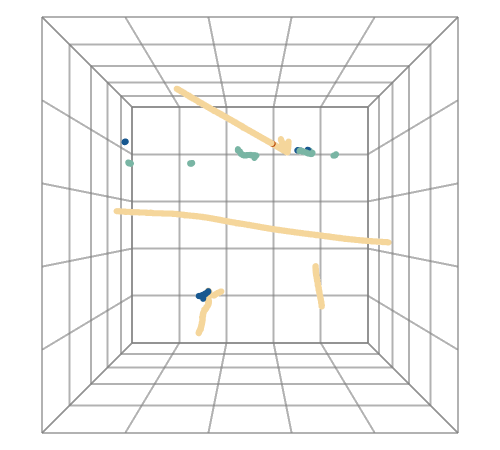} &
    \includegraphics[width=0.125\linewidth]{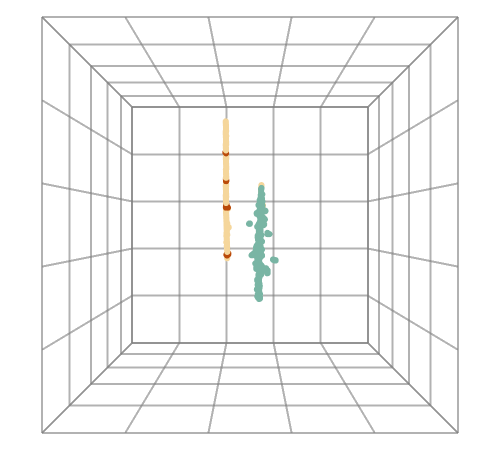} &

    \\
    \rotatebox[origin=l]{90}{\textsc{~~PL FFT}} & 
    \includegraphics[width=0.125\linewidth]{figs/segmentation/mae_multitask_fft_pred_0}&
    \includegraphics[width=0.125\linewidth]{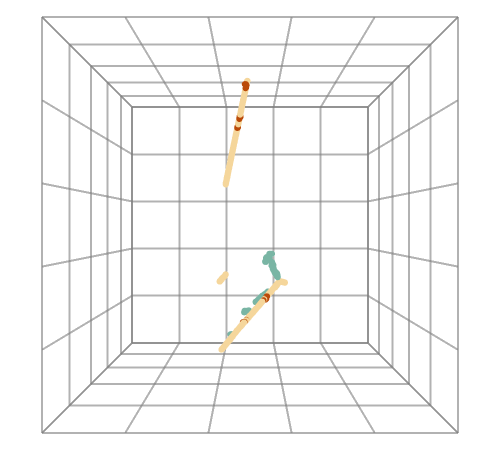} &
    \includegraphics[width=0.125\linewidth]{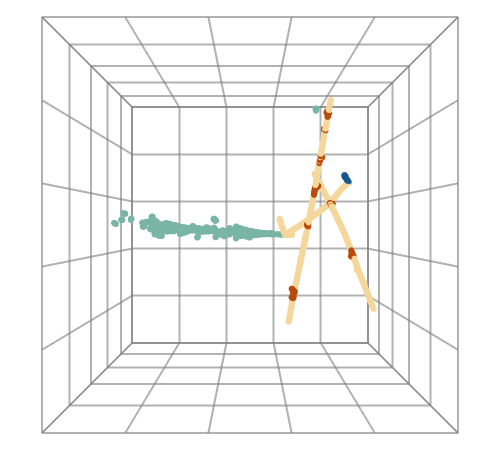} &
    \includegraphics[width=0.125\linewidth]{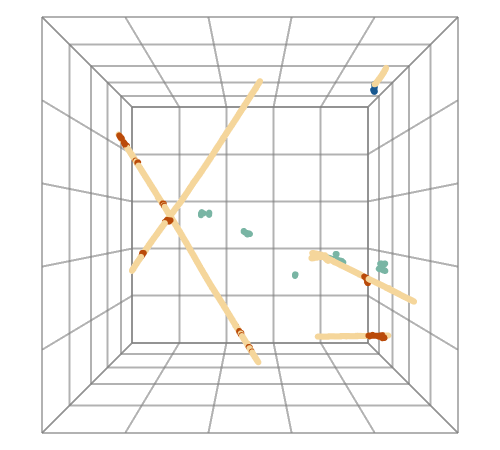} &
    \includegraphics[width=0.125\linewidth]{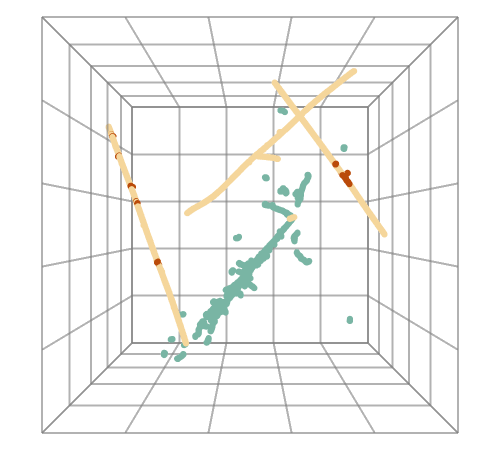} &
    \includegraphics[width=0.125\linewidth]{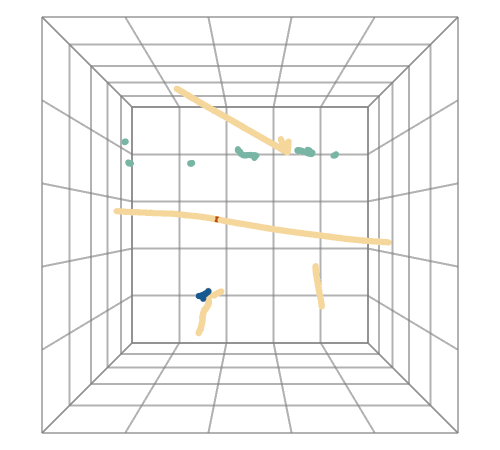} &
    \includegraphics[width=0.125\linewidth]{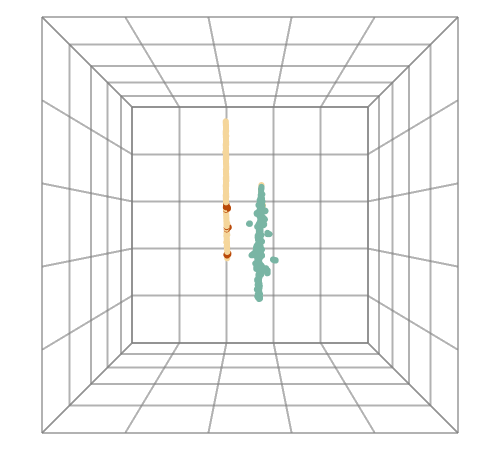} &
    \\
    \rotatebox[origin=l]{90}{\textsc{~~PM FFT}} & 
    \includegraphics[width=0.125\linewidth]{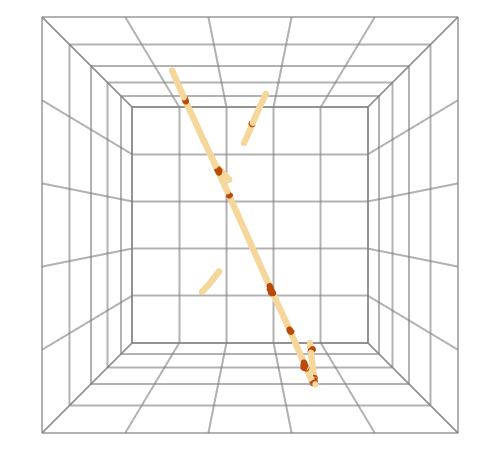}&
    \includegraphics[width=0.125\linewidth]{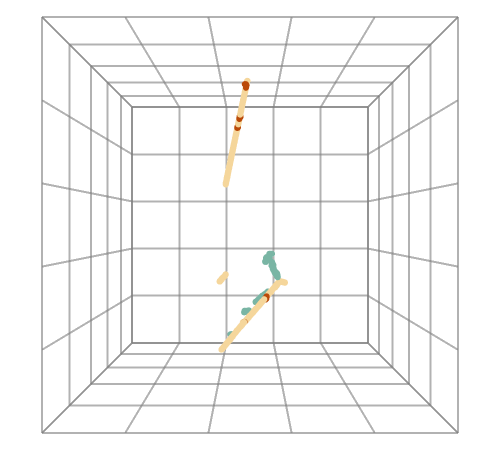} &
    \includegraphics[width=0.125\linewidth]{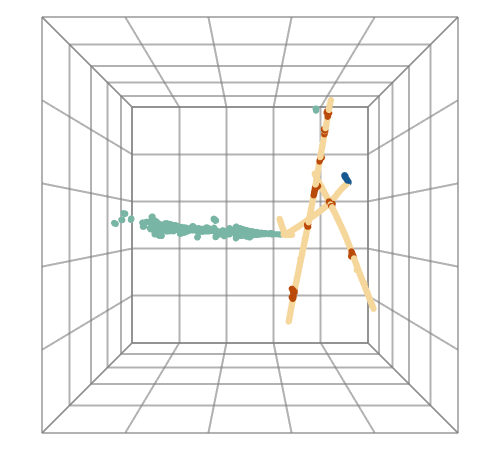} &
    \includegraphics[width=0.125\linewidth]{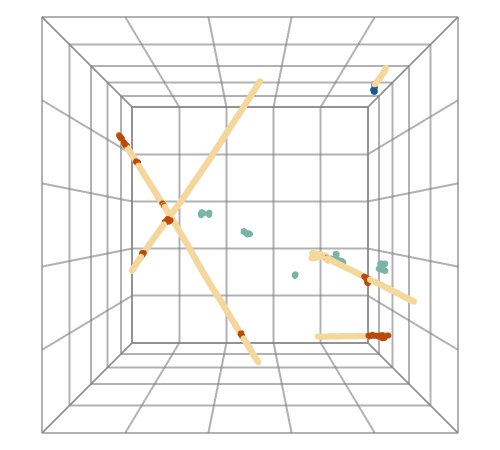} &
    \includegraphics[width=0.125\linewidth]{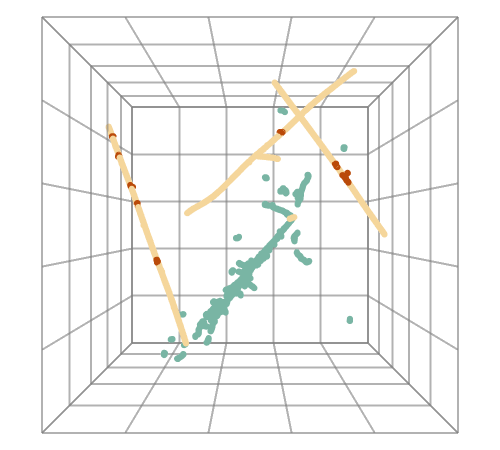} &
    \includegraphics[width=0.125\linewidth]{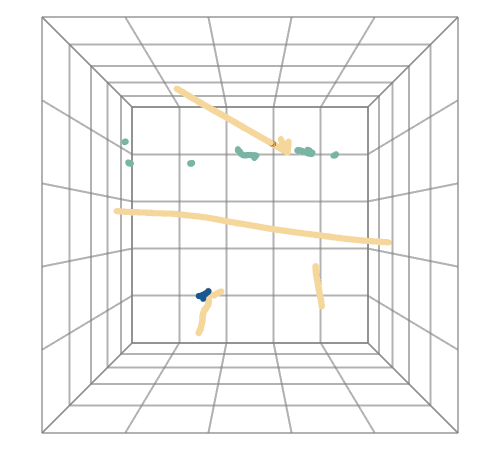} &
    \includegraphics[width=0.125\linewidth]{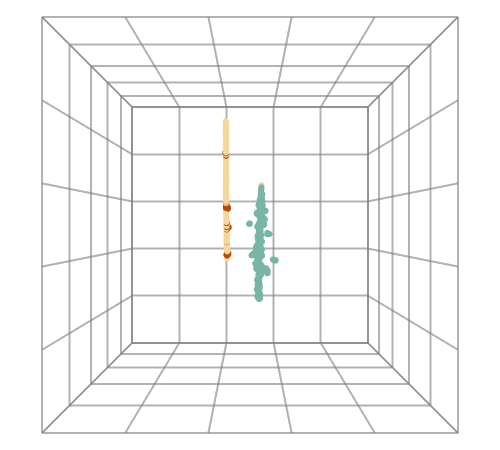} \\
\end{tabular}
    \makebox[\linewidth][r]{%
        \includegraphics[width=0.25\linewidth]{figs/legend.pdf}}
    \caption{\textit{Further examples of dense classification.} Model types are abbreviated with PL meaning PoLAr-MAE, PM meaning Point-MAE, PEFT meaning parameter-efficient fine-tuning, and FFT meaning full fine-tuning. Best viewed zoomed in.}
    \label{fig:segsem_comparison}
\end{figure*}

\begin{figure*}[h]
\centering
\begin{tabular}{ccc}
    & \textsc{Point-MAE} & \textsc{PoLAr-MAE} \\
    \rotatebox[origin=l]{90}{~~~~~~~~~~~~~~~~~~~~~~~~~~~~~~\textsc{PEFT}} &
    \includegraphics[width=0.4\linewidth]{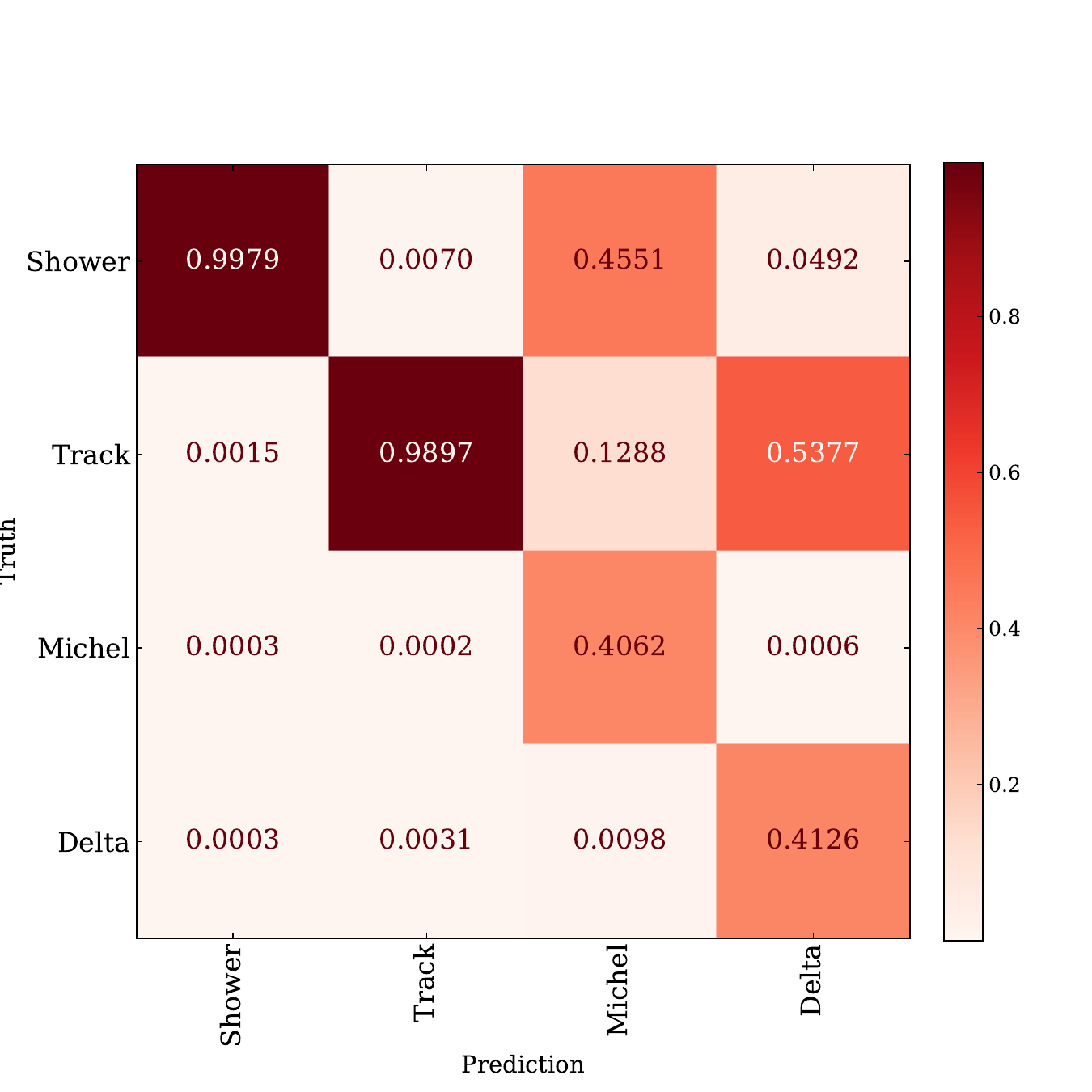} &
    \includegraphics[width=0.4\linewidth]{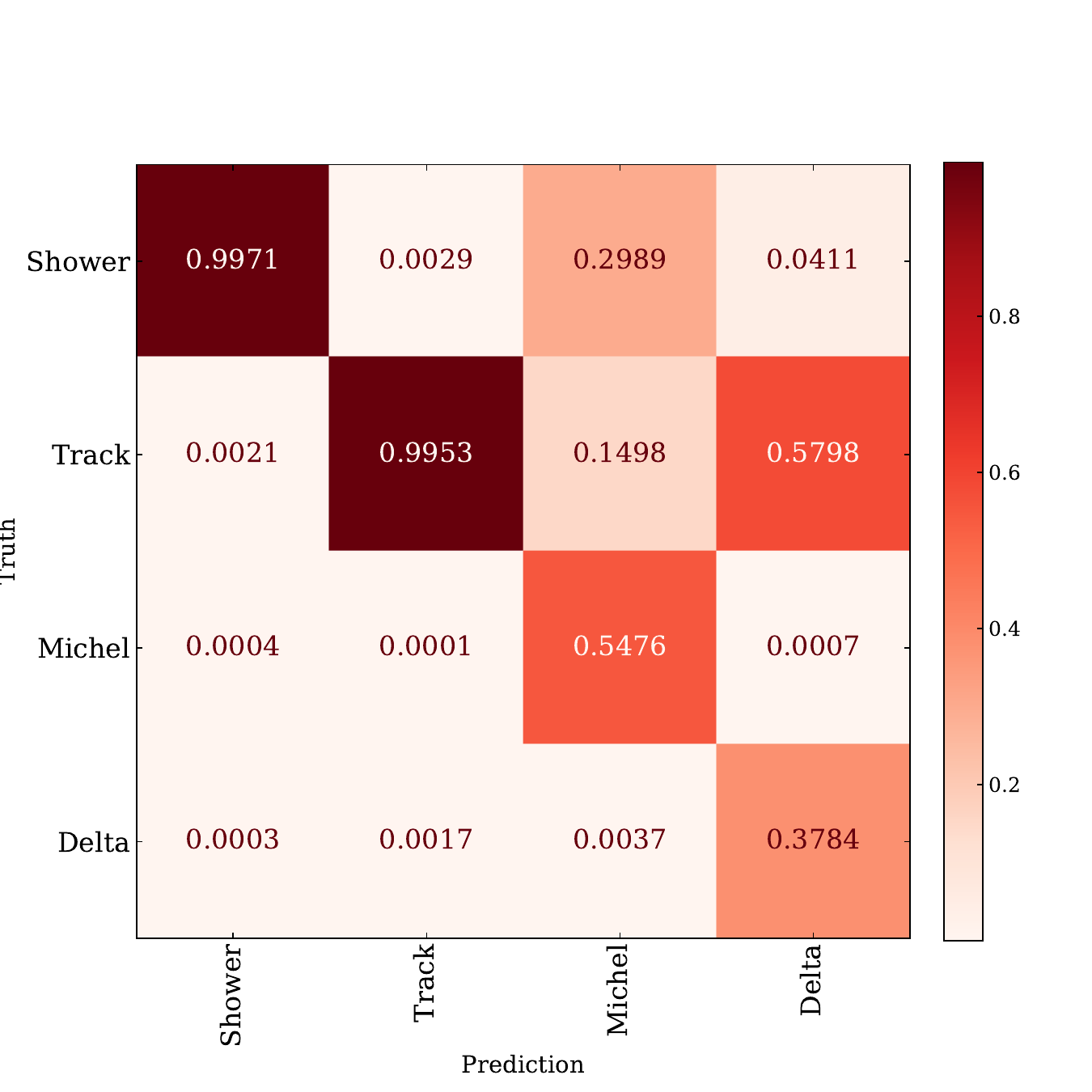} \\[1em]
    \rotatebox[origin=l]{90}{~~~~~~~~~~~~~~~~~~~~~~~~~~~~~\textsc{FFT}} &
    \includegraphics[width=0.4\linewidth]{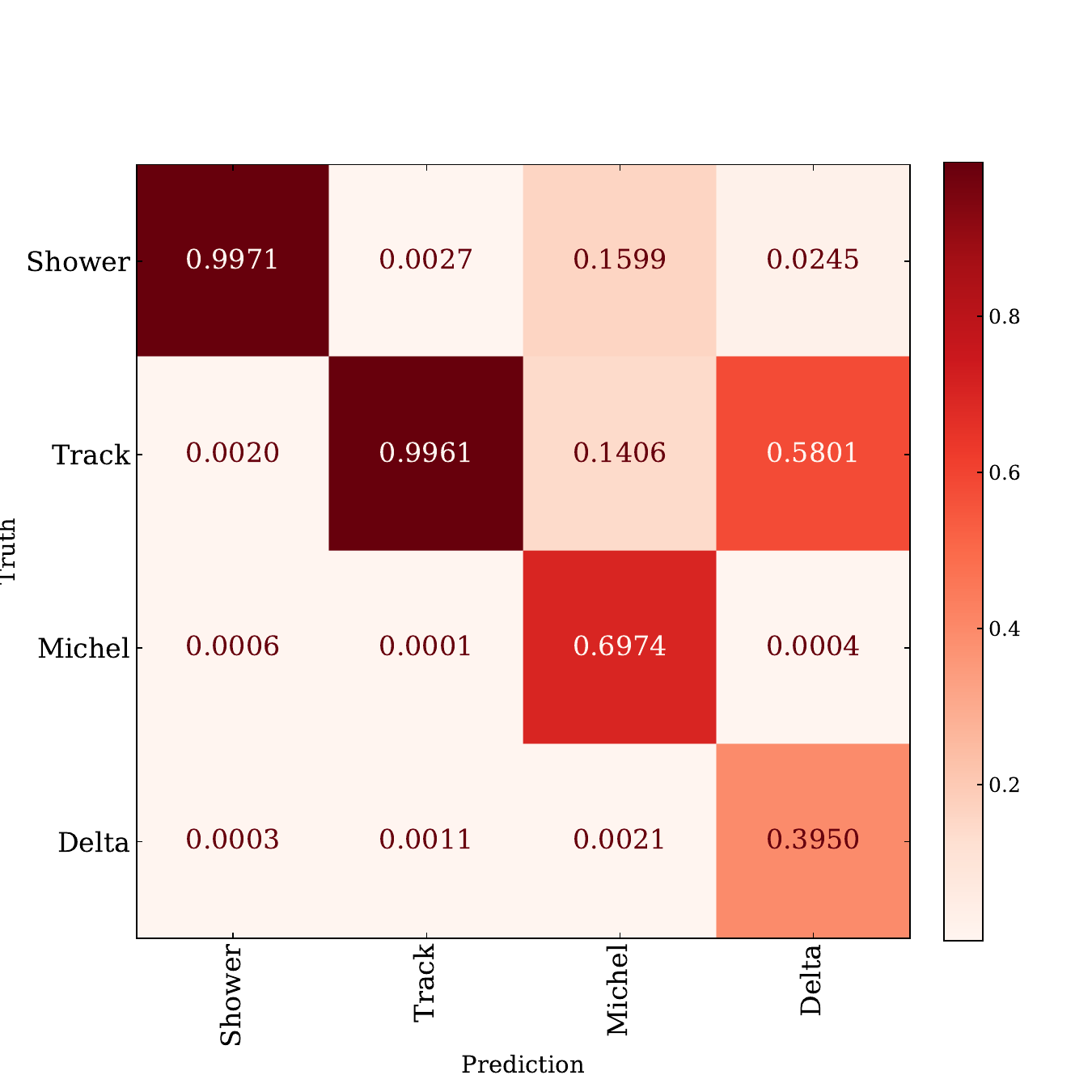} &
    \includegraphics[width=0.4\linewidth]{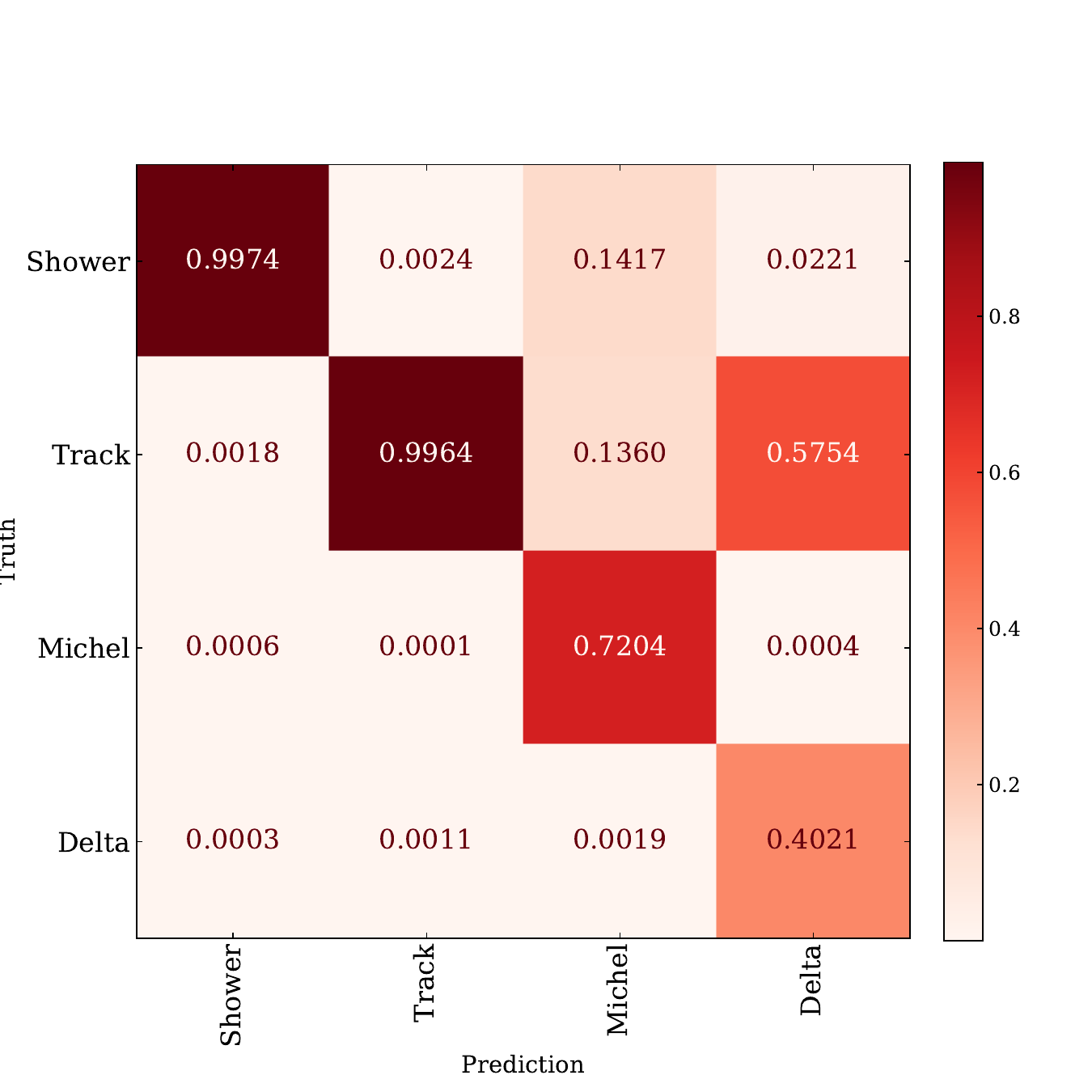} \\
\end{tabular}
\caption{\textit{Semantic segmentation confusion matrices.} Rows denote fine-tuning strategies: PEFT and FFT. Columns denote the pre-trained model used: Point-MAE and PoLAr-MAE. Each column sums to 1, and the diagonal values represent precision metrics for each class.}
\label{fig:confusion}
\end{figure*}

In Figure \ref{fig:segsem_comparison}, we show additional examples of semantic segmentation results for both Point-MAE and PoLAr-MAE fine-tuned models. In Table \ref{tab:extended_seg}, we provide additional evaluation metrics for the fine-tuned Point-MAE and PoLAr-MAE models. We also supply confusion matrices for each model, normalized across predictions, in Figure \ref{fig:confusion}.

\begin{table*}[th!!]
\centering\hspace{-20pt}
\caption{\textit{Semantic segmentation results (extended).} We present detailed metrics including mean F\textsubscript{1}, precision, recall, and per-class scores. The best results are \besto{highlighted}.}
\vskip 0.15in
\label{tab:extended_seg}
\begin{tabular}{l *{3}{c c c c c}}
\toprule
& \multicolumn{5}{c}{F\textsubscript{1}} & \multicolumn{5}{c}{Precision} & \multicolumn{5}{c}{Recall} \\
\cmidrule(lr){2-6} \cmidrule(lr){7-11} \cmidrule(lr){12-16}
Model & Mean & Track & Shower & Delta & Michel & Mean & Track & Shower & Delta & Michel & Mean & Track & Shower & Delta & Michel\\
& & \colorsquare{track} & \colorsquare{shower} & \colorsquare{delta} & \colorsquare{michel} & & \colorsquare{track} & \colorsquare{shower} & \colorsquare{delta} & \colorsquare{michel} & & \colorsquare{track} & \colorsquare{shower} & \colorsquare{delta} & \colorsquare{michel} \\
\midrule
\textsc{PoLAr-MAE PEFT}
& 0.798 &	0.961 &	0.990 &	0.542 &	0.698
& 0.730 &	0.995 &	0.997 &	0.378 &	0.548
& 0.958 &	0.929 &	0.983 &	0.956 & 0.963 \\

\textsc{Point-MAE PEFT}
& 0.772 &	0.965 &	0.983 &	\best{0.569} &	0.572
& 0.702 &	0.990 &	\best{0.998} &	\best{0.413} &	0.406
& 0.948 &	\best{0.941} &	0.968 &	0.919 &	\best{0.967} \\

\textsc{Point-MAE FFT}
& 0.831 &	0.963 &	\best{0.994} &	0.561 &	0.807
& 0.771 &	\best{0.996} &	0.997 &	0.395 &	0.697
& 0.962 &	0.932 &	0.990 &	0.969 &	0.957 \\

\textsc{PoLAr-MAE FFT}
& \best{0.837} &	\best{0.964} &	\best{0.994} &	\best{0.569} &	\best{0.823}
&\best{0.779} &	\best{0.996} &	0.997 &	0.402 &	\best{0.720}
& \best{0.964} &	0.934 &	\best{0.992} &	\best{0.970} &	0.959 \\
\bottomrule
\end{tabular}
\end{table*}


\newpage
\section{Failures}
\label{app:failures}

In the spirit of the computer vision community's recent increasing efforts in portraying both positive and negative results, we document the directions we explored to improve representations that ultimately resulted in either worse or similar performance.

\begin{itemize}
    \item \textbf{Energy Normalization and Embedding.} We attempted to apply centering and scaling group normalizations to the energy values and added an additional ``center energy" embedding to the tokens, similar to the positional embeddings used for coordinates. However, this approach did not yield any noticeable improvement.

    \item \textbf{Handling Stochasticity in LArTPC Events.} LArTPC events are inherently stochastic, as particle energy depositions result from a complex chain of Markov processes. Some particle types, such as tracks, are easier to predict due to their structured nature, while others, like delta rays and electromagnetic (EM) showers, are highly stochastic and challenging to predict. For example, predicting masked portions of delta rays or EM showers is nearly impossible because their presence and trajectories depend on probabilistic physical interactions. To address this, we conditioned the mask decoder with embeddings of diffused masked groups, following DiffPMAE \cite{li2024diffpmaediffusionmaskedautoencoders}. We tested both permutation-equivariant and regular PointNet-based per-point reconstruction methods. Despite these efforts, no meaningful improvements in semantic understanding were observed.

    \item \textbf{Enforcing Sub-Token Semantics.} To address the limitations in sub-token semantic representation, we added an autoencoder task on visible tokens, forcing the encoder to simultaneously encode both the original points and global context within all tokens. However, this approach introduced confusion in the encoder, resulting in degraded representations and worse performance.
\end{itemize}

\end{document}